\documentclass[aps, prd,  superscriptaddress, showpacs, longbibliography,notitlepage, floatfix, nofootinbib, 10pt]{revtex4-1}

\usepackage{bm, amsmath, amssymb, relsize, amsfonts, mathrsfs, multirow, braket, siunitx, color, booktabs, arydshln}
\usepackage{graphicx}
\graphicspath{{./figures/}}

\DeclareSIUnit \parsec {pc}

\usepackage[dvipsnames]{xcolor}
\usepackage[unicode]{hyperref}
\hypersetup{
  colorlinks=true,
  citecolor=RoyalBlue,
  linkcolor=blue,
  urlcolor=blue
}

\usepackage{array}
\usepackage{arydshln}
\usepackage{cancel}
\usepackage{graphicx}
\usepackage{booktabs}
\usepackage{dcolumn}
\usepackage{bm}
\usepackage{paralist}
\usepackage{epsfig}
\usepackage{placeins}
\usepackage{mathrsfs}
\usepackage{comment}
\usepackage{color, colortbl}
\usepackage[inline]{enumitem}
\usepackage{soul}
\usepackage{adjustbox}
\usepackage{amssymb}
\usepackage{caption}
\usepackage{subcaption}
\usepackage{float}
\usepackage{cancel}
\usepackage{verbatim}
\usepackage{amsmath}
\usepackage{xcolor}
\usepackage[toc,page]{appendix}
\usepackage{graphicx}

\DeclareMathAlphabet{\mathpzc}{OT1}{pzc}{m}{it}
	
\definecolor{LightCyan}{rgb}{0.88,1,1}
\definecolor{lightgray}{gray}{0.9}
\definecolor{lightorange}{RGB}{255, 230, 204}

\def \IITGn     {Indian Institute of Technology Gandhinagar, Gujarat 382055, India.\vspace*{4pt}}
\def \HRI    {\mbox{Regional Centre for Accelerator-based Particle Physics, Harish-Chandra Research Institute, Prayagraj (Allahabad) 211019, India}\vspace*{2pt}}
\def \SP   {Instituto de F\'isica, Universidade de S\~ao Paulo, R. do Mat\~ao 1371, 05508-090 S\~ao Paulo, Brazil}

\begin{document}

\title{Optimizing The Cut And Count Method In Phenomenological Studies}

\author{\textsc{Baradhwaj Coleppa}}
\email{baradhwaj@iitgn.ac.in }
\affiliation{\IITGn}

\author{\textsc{Gokul B. Krishna \footnote{corresponding author}}}
\email{gokulb@iitgn.ac.in}
\affiliation{\IITGn}

\author{\textsc{Agnivo Sarkar}} 
\email{agnivosarkar@hri.res.in}
\affiliation{\HRI}

\author{\textsc{Sujay Shil}} 
\email{sujayshil1@gmail.com}
\affiliation{\SP}

\begin{abstract}
We introduce an optimization technique to discriminate signal and background in any phenomenological study based on the cut and count-based method. The core ideas behind this technique are the introduction of a ranking scheme that can quantitatively assess the relative importance of various observables involved in a new physics process, and a more methodical way of choosing what cuts to impose. The technique is an iterative process that works with the help of the MadAnalysis5 interface. Working in the context of a BSM (Beyond Standard Model) scenario where we carry out a signal search of singly charged Higgs in the context of the Two Higgs Doublet Model (2HDM), we demonstrate how automating the cut and count process in this specific way results in an enhanced discovery potential compared with the more traditional way of imposing cuts.
\end{abstract}
\maketitle
\section{Introduction}
\label{sec:intro}
It is by now firmly established that the Standard Model (SM) \cite{Glashow:1961tr,Weinberg:1967tq,Salam:1968rm} of particle physics, while an enormously successful theory whose predictions have been tested to very good accuracy in various colliders over the years, needs to be supplanted with new dynamics at the TeV scale. With the advent of the Large Hadron Collider (LHC), we now have a testing ground for many of these so-called ``Beyond the SM" (BSM) avenues. Typically one expects some new resonances at the TeV scale to show up in high energy collisions - however, such a discovery has thus far eluded us. This has prompted both experimantalists and theorists to employ increasingly sophisticated approaches toward phenomenological analyses with various machine learning (ML) and deep learning algorithms \cite{Guest:2018yhq,deOliveira:2015xxd} now at the forefront of particle physics research. The reason for this is of course clear - the traditional ``bump hunting" searches, while successful in the past, might need to be replaced for a more thorough search. In this work, we offer an alternative strategy - while still employing the cut-based methodology, we aim to demonstrate that suitably automating this process can lead to significant enhancements in our search efficiency and thus might lead to a better discovery potential of new physics. 

The study of BSM physics has been structured and scrutinized in many ways over the decades. The standard practice amongst phenomenologists now is a simulation tool such as $\mathtt{MadGraph\_aMC@NLO}$ \cite{Alwall:2014hca} for event generation which is compatible with  $\mathtt{FeynRules}$~\cite{Alloul:2013bka} for new physics model implementation. Also, to consider realistic collider situations, one has to cascade the parton level process from $\mathtt{MadGraph\_aMC@NLO}$ to PYTHIA8 \cite{Sjostrand:2014zea}  for jet showering and $\mathtt{Delphes3}$~\citep{deFavereau:2013fsa} for detector-level simulations. The analysis of collision events proceeds via $\mathtt{MadAnalysis5}$~\cite{Conte:2012fm} which is a framework for phenomenological analysis that handles data structures and facilitates effective analysis of particle physics scattering experiments through a user-friendly interface. which provides the user with normalized observable distributions. The ``cut and count" strategy then relies on looking at the distributions and  imposing suitable cuts to improve the signal over background significance. However, this practice of the cut-and-count method is ignorant of \emph{how implementing a cut on a particular distribution will affect the remaining observable distributions}. While this might not be a huge disadvantage in say a $2\to 2$ process with not a lot of SM background, for more complicated decay chains, one would need to understand the interplay between the various kinematic quantities.  As we will demonstrate, we can optimize the cut and count methodology by formulating an iterative process with a suitably designed set of techniques to optimize the cut-flow.

While high-performing deep learning algorithms display superior capabilities in discrimination tasks \cite{Cranmer:2015bka,Baldi:2016fzo,Metodiev:2017vrx}, the intricate structure of these data models is notably complex. This complexity arises from many tunable parameters and the numeric nature of the optimization process, posing challenges to a comprehensive understanding of its physical nature. Some studies have directed their attention towards exploring the phenomenologically relevant characteristics of neural networks in terms of interpretability \cite{Chakraborty:2019imr,Faucett:2020vbu,Shlomi:2020gdn,Vatellis:2024vjl}, and there are issues regarding uncertainty quantification \cite{Ghosh:2021roe,Bollweg:2019skg,CMS:2025cwy,Benato:2025rgo} as well. Our present study is not intended as a substitution for ML techniques~\cite{Feickert:2021ajf}; it is presented as a methodology which gives phenomenologically interpretable results and improves the sequential cut and count techniques for challenging final states at the LHC where traditional methods might not isolate the signals efficiently. Although our work closely resembles a Decision tree (DT) classifier~\cite{Xia:2018cfz} or Boosted Decision Tree (BDT)~\cite{Coadou:2022nsh,Roe:2004na} in general, the algorithm is fundamentally different from a DT because of its implementation and optimisation schema. More details regarding this can be found in the upcoming sections along with comparision of both our algorithm and BDT.

This paper is organized as follows: we begin by briefly explaining the BSM model that we will use as an example for illustrating the methodology in \autoref{sec:sec2} and outline the data domain that we are going to use to implement the algorithm. In \autoref{sec:sec4}, we detail the development of the algorithm - this forms the crux of the paper wherein we present and explain a flowchart representing the execution of the various steps involved in the method. Then we continue with DT and BDT Analysis to demonstrate the comparison of our algorithm with DT and BDT and to examine how observables are associated in \autoref{sec:comparision}. Finally, the benefits of adopting such a method - and its shortcomings - are discussed in \autoref{sec:Conclusion}.

\section{Set-Up and Example}
\label{sec:sec2}

Our goal in this section is to lay out the example that we will use to illustrate our methodology and the underlying model details. Specifically, we choose to study charged Higgs pair production in the Type III\footnote{In the case of Type III 2HDM, the quark sector receives mass after both the scalar doublets $\Phi_{1}$ and $\Phi_{2}$ acquire \emph{vev}s. As a consequence, this model can potentially generate flavour changing neutral currents at the tree level. A detailed discussion this can be found in Ref.\cite{Branco:2011iw}. We point out at this stage that in the upcoming phenomenology sections, we have chosen the benchmark point for our study that is compatible with flavour constraints.} Two Higgs Doublet Model (2HDM) \cite{Branco:2011iw}. In the 2HDM, the SM scalar sector is extended by an additional scalar doublet $\Phi$, which is charged under the electroweak gauge group $SU(2)_{L}\times U(1)_{Y}$ with hypercharge $Y = 1$. We lay out the basics of the scalar sector of the model that is directly relevant to our analysis and proceed to the phenomenological process of interest.

\subsection{Model Set Up}
\noindent
In this section, we will present a brief review of the 2HDM model. For a more detailed description of the model, the interested reader is invited to consult Refs.\cite{Branco:2011iw,Karan:2023ufg}. For our purpose here, we will concentrate on the scalar sector of the model. We begin, in~\autoref{eq:pot1}, with the most general scalar potential of the theory which remains invariant under the SM gauge group $SU(2)_L\times U(1)_Y$. 
\begin{equation}
\begin{aligned}
V_{H} = & m_{11}^{2} \Phi_{1}^{\dagger} \Phi_{1}+m_{22}^{2} \Phi_{2}^{\dagger} \Phi_{2}-\left(m_{12}^{2} \Phi_{1}^{\dagger} \Phi_{2}+\tt {h.c.}\right)  + \frac{1}{2} \lambda_{1}\left(\Phi_{1}^{\dagger} \Phi_{1}\right)^{2}+\frac{1}{2} \lambda_{2}\left(\Phi_{2}^{\dagger} \Phi_{2}\right)^{2}\\
& + \lambda_{3}\left(\Phi_{1}^{\dagger} \Phi_{1}\right)\left(\Phi_{2}^{\dagger} \Phi_{2}\right) +\lambda_{4}\left(\Phi_{1}^{\dagger} \Phi_{2}\right)\left(\Phi_{2}^{\dagger} \Phi_{1}\right) \\
& + \left( \frac{1}{2} \lambda_{5}\left(\Phi_{1}^{\dagger} \Phi_{2}\right)^{2} + \tt{h.c.}\right) + \left( \lambda_{6}\left(\Phi_{1}^{\dagger} \Phi_{1}\right)\left(\Phi_{1}^{\dagger} \Phi_{2}\right)+\lambda_{7}\left(\Phi_{2}^{\dagger} \Phi_{2}\right)\left(\Phi_{1}^{\dagger} \Phi_{2}\right)+\tt {h.c.}\right),
\end{aligned}
\label{eq:pot1}
\end{equation}
\noindent
where we have denoted the two $SU(2)_L$ doublets by $\Phi_1$ and $\Phi_2$. Implementing the hermiticity condition, one can realize that $m^{2}_{11}$, $m^{2}_{22}$, and $\lambda_{i}$ ($i =$ 1 to 4) are real parameters. In contrast, $m^{2}_{12}$ and $\lambda_{i}$ ($i = $ 5 to 7) can, in general, be complex parameters. The presence of complex parameters can give rise to CP as well as charge violating minima which in turn hint upon interesting new physics scenarios. However, for the present purpose we will ignore these possibilities and impose a few conditions to simplify the above potential. We begin with the assumption that CP is conserved in the scalar sector. As a result, all the quatric terms which are odd powers of either of the scalar doublet drop out. Furthermore, we will assume all the scalar parameters in the potential are real. Under these conditions,~\autoref{eq:pot1} takes the following form:
\begin{equation}
\begin{aligned}
V_{H} = & m_{11}^{2} \Phi_{1}^{\dagger} \Phi_{1}+m_{22}^{2} \Phi_{2}^{\dagger} \Phi_{2}-\left(m_{12}^{2} \Phi_{1}^{\dagger} \Phi_{2}+\tt{h.c.}\right)  + \frac{1}{2} \lambda_{1}\left(\Phi_{1}^{\dagger} \Phi_{1}\right)^{2}+\frac{1}{2} \lambda_{2}\left(\Phi_{2}^{\dagger} \Phi_{2}\right)^{2}\\
& + \lambda_{3}\left(\Phi_{1}^{\dagger} \Phi_{1}\right)\left(\Phi_{2}^{\dagger} \Phi_{2}\right) +\lambda_{4}\left(\Phi_{1}^{\dagger} \Phi_{2}\right)\left(\Phi_{2}^{\dagger} \Phi_{1}\right) + \left(\frac{1}{2} \lambda_{5}\left(\Phi_{1}^{\dagger} \Phi_{2}\right)^{2} +\tt{h.c.}\right).
\end{aligned}
\label{eq:pot2}
\end{equation}

Both these doublets acquire vacuum expectation values (\emph{vev}s) given by $\langle\Phi_{1}\rangle = v_{1}$ and $\langle\Phi_{2}\rangle = v_{2}$, and thus engineer the electroweak symmetry breaking (EWSB) $SU(2)_L\times U(1)_Y\to U(1)_{\text{EM}}$. The explicit form of these doublets can be written as: 
\begin{equation}
\Phi_i = \left( \begin{array}{c} \phi_i^+ \\ \frac{(\rho_i + v_1 + i \eta_1)}{\sqrt{2}} \end{array} \right)~~~~~~~i= 1,2.
\label{Eq:phiform}
\end{equation} 
One can see by direct inspection that both these doublets contain total eight total degrees of freedom (\emph{d.o.f}) ($\phi_i^{\pm},\rho_i$, and $\eta_i$). After EWSB, three linear combinations of three of these \emph{d.o.f}'s - the massless Goldstone modes - get absorbed by the SM electroweak gauge bosons $W^{\pm}$ and $Z$ rendering them massive. Thus, the scalar spectrum contains five physical Higgs bosons - a pair of charged Higgs bosons $H^{\pm}$ (with mass $m_{H^{\pm}}$), a CP-odd scalar $A$ (with mass $m_{A}$), and two CP-even scalars $\{h, H\}$ (with mass $m_{h}$ and $m_{H}$ respectively). We can see the spectrum emerge in the following way: the mass matrix corresponding to the charged scalars can be read off from $V_H$ by collecting all the quadratic terms in fields and takes the following form:
\begin{equation}
    \mathcal{L}_{\text{charged}} = \left(m^{2}_{12} \left(\lambda_{4} + \lambda_{5}\right)v_{1}v_{2}\right)\begin{bmatrix} \phi^{-}_{1} & \phi^{-} _{2} 
    \end{bmatrix}
    \begin{bmatrix}
       \frac{v_2}{v_1} & -1 \\
       -1 & \frac{v_1}{v_2}
    \end{bmatrix} 
    \begin{bmatrix} \phi^{-}_{1} \\
     \phi^{-} _{2} 
    \end{bmatrix}.
    \label{Eq:mHcmatrix}
\end{equation}
The determinant of the above $2\times2$ square matrix is zero. This signifies that at least one of the eigenvalues correspond to that matrix must be zero. The eigenvector correspond to the zero eigenvalue is identified as the charged Goldstone mode $G^{\pm}$ which gets eaten up by the $W^{\pm}$ boson. On other hand, the physical state correspond to the non zero eigenvalue is identified as $H^{\pm}$ state with mass 
\begin{equation}
  m_{H^\pm} = \left[\frac{m^{2}_{12}}{v_{1}v_{2}} - \lambda_{4} -\lambda_{5}\right]^\frac{1}{2}\sqrt{\left(v^{2}_1 + v^{2}_{2}\right)}.
  \label{Eq:mHcmass}
\end{equation}
Similarly, collecting the relevant quadratic terms one can write down the mass matrix correspond to CP-odd scalars: 
\begin{equation}
    \mathcal{L}_{\text{CP-odd}} =\frac{m^{2}_{A}}{v^{2}_1 + v^{2}_2}\begin{bmatrix} \eta_{1} & \eta_{2} 
    \end{bmatrix}
    \begin{bmatrix}
       v^{2}_{2} & - v_{1}v_{2} \\
       - v_{1}v_{2} & v^{2}_{1}
    \end{bmatrix} 
    \begin{bmatrix} \eta_{1} \\
     \eta_{2} 
    \end{bmatrix}.
    \label{Eq:mAmatrix}
\end{equation}
Here also, one of the eigenvalue would be zero which is associated with the neutral goldstone mode $G^{0}$. The non zero eigenvalue represents the pseudo-scalar boson with mass 
\begin{equation}
 m_{A} = \left[\frac{m^{2}_{12}}{v_{1}v_{2}} - \lambda_{5}\right]^\frac{1}{2}\sqrt{\left(v^{2}_1 + v^{2}_{2}\right)}.
  \label{Eq:mAmass}
\end{equation} 
\noindent
Finally, we turn to the mass matrix for the CP-even scalars - this can again be written down by straightforward inspection of the potential collecting quadratic terms in the $\rho$ fields:
\begin{equation}
\begin{aligned}
    \mathcal{L}^{\text{CP-even}}_{mass} &= \begin{bmatrix} \rho_{1} & \rho_{2} 
    \end{bmatrix}
    \begin{bmatrix}
       \mathcal{M}_{11} & \mathcal{M}_{12} \\
       \mathcal{M}_{12} & \mathcal{M}_{22}
    \end{bmatrix} 
    \begin{bmatrix} \rho_{1} \\
     \rho_{2} 
    \end{bmatrix} \\
        &= \begin{bmatrix} \rho_{1} & \rho_{2} 
    \end{bmatrix}
    \begin{bmatrix}
       m^{2}_{12}\frac{v_{2}}{v_{1}} + \lambda_{1}v^{2}_{1} & -m^{2}_{12} +\lambda_{345}v_{1}v_{2} \\
       -m^{2}_{12} +\lambda_{345}v_{1}v_{2} & m^{2}_{12}\frac{v_{2}}{v_{1}} + \lambda_{2}v^{2}_{2}
    \end{bmatrix} 
    \begin{bmatrix} \rho_{1} \\
     \rho_{2} 
     \end{bmatrix}
    \end{aligned}
    \label{Eq:mHmhmass}
\end{equation} 
In general, the above mass matrix has two non-zero eigenvalues which can be expressed as
\begin{equation}
    \begin{aligned}
        m^{2}_{H} &=  \frac{1}{2}\left[\mathcal{M}_{11} + \mathcal{M}_{22} + \sqrt{\left(\mathcal{M}_{11} - \mathcal{M}_{22}\right)^{2} + \mathcal{M}^{2}_{12}}\right], \\
        m^{2}_{h}   &= \frac{1}{2}\left[\mathcal{M}_{11} + \mathcal{M}_{22} + \sqrt{\left(\mathcal{M}_{11} - \mathcal{M}_{22}\right)^{2} + \mathcal{M}^{2}_{12}}\right].
    \end{aligned}
    \label{Eq:CPmass}
\end{equation} 
\noindent
Hereafter we assume $m_{h} < M_{H}$ and the CP-even state $h$ with the mass $m_{h}$ can be identified with the SM-like 125 GeV Higgs boson\footnote{Even though a bit more constrained, 2HDM also allows the identification of the heavier of the two as the 125 GeV SM-like Higgs, thus making the other Higgs lighter. We do not explore this possibility in this paper.} discovered at the LHC. One can also define two parameters $\tan\beta$ and $\alpha$ using the \emph{vev}'s and the scalar parameters:
\begin{equation}
    \tan\beta =\frac{v_2}{v_1}, ~~ \tan2\alpha = \frac{2\mathcal{M}_{12}}{\mathcal{M}_{11} - \mathcal{M}_{22}}.
    \label{Eq:10}
\end{equation} 
With these parameters, one can write down the relations between the gauge basis and the mass basis of the scalar fields.  
\begin{equation}
        \begin{aligned}
    \begin{bmatrix} G^{\pm} \\
     H^{\pm} 
     \end{bmatrix}
      &=  \begin{bmatrix}
       \cos\beta & \sin\beta \\
       - \sin\beta & \cos\beta
    \end{bmatrix} 
    \begin{bmatrix} \phi^{\pm}_{1} \\
     \phi^{\pm} _{2} 
     \end{bmatrix} \\
     \begin{bmatrix} G^{0} \\
     A 
     \end{bmatrix}
      &=  \begin{bmatrix}
       \cos\beta & \sin\beta \\
       - \sin\beta & \cos\beta
    \end{bmatrix} 
    \begin{bmatrix} \eta_{1} \\
     \eta_{2} 
     \end{bmatrix} \\
    \begin{bmatrix}
        H \\
        h
    \end{bmatrix}
     & =  \begin{bmatrix}
       \cos\alpha & \sin\alpha \\
       - \sin\alpha & \cos\alpha
    \end{bmatrix} 
    \begin{bmatrix} \rho_{1} \\
     \rho_{2} 
     \end{bmatrix}
    \end{aligned}
    \label{Eq:11}
\end{equation} 
In computing all relevant couplings and studying the phenomenology, we use the above definitions of the mass eigenstates. Before closing this section, we present the relevant couplings in~\autoref{tab:vertices} that will play a central role in the signal topology we will soon discuss. We point out that the vertices responsible for the charged Higgs pair production is independent of any new physics parameters. On top of that, the coupling responsible for the $H^{\pm} \to W^{\pm} A$ decay strictly depends on the $SU(2)_{L}$ gauge coupling. As a result, the relevance of this signal is not only restricted to 2HDM but a larger class of scalar extended BSM scenarios which can in principle allow the kind of signal we will study. Furthermore, comparing~\autoref{Eq:mHcmass} and~\autoref{Eq:mAmass}, one can realise that the mass gap between the $m_{H^{\pm}}$ and $m_{A}$ is controlled by the scalar parameter $\lambda_{4}$. With the relevant pieces of the model in place, we now turn to the specific new physics process that we will use to illustrate out methodology. 

\begin{center}
\begin{table}
\begin{tabular}{c c} 
 \hline
 $ ZH^+H^-$ & $ -ie \cot{\theta_{W}} \bigl[P^{\mu}_{h^+} + P^{\mu}_{h^-} \bigr] $ \cite{Logan:2014jla}\\
\hline
$\gamma H^+H^-$ & $ -ie \bigl[P^{\mu}_{h^+} + P^{\mu}_{h^-} \bigr] $ \cite{Logan:2014jla} \\
\hline
 $ H^+W^{-}A$ & $-\frac{ig}{2} $ \cite{Hashemi:2023osd}\\
\hline
 $ Ab\bar{b}$ & $-i \frac{m_b}{v} \bigl[-\cot{\beta}\sin{(\beta - \alpha)}+\cos{(\beta-\alpha)}\bigr] $ \cite{Hashemi:2023osd}
\end{tabular}
\caption{The vertices in the 2HDM relevant for our study. Here, $\theta_{W}$ is the weak mixing angle, $P_{h^{\pm}}^{\mu}$ is the momentum of $h^{\pm}$, $m_{b}$ is the b-quark mass, and $g$ is the $SU(2)_{L}$ coupling.}
\label{tab:vertices}
\end{table}
\end{center}

\subsection{Illustrative Process}

In order to demonstrate the efficacy of our algorithm, we will study the pair production of charged Higgs bosons in the 2HDM with their subsequent decay into $AW^{\pm}$ with the $A$'s decaying to $\bar{b}b$ and the $W$'s decaying leptonically:
\begin{equation}
pp \rightarrow H^+ H^- \rightarrow W^+ W^- A A \rightarrow 4b + 2l + \cancel{E}_T \nonumber
\end{equation}
This signal is quite challenging to reconstruct, given its small cross-section compared to the SM backgrounds and a complex final state topology. For the purposes of this study, we fix the mass of the charged Higgs boson at 450~GeV. We have also assigned the masses of the CP-even Higgs $m_{H} = 250~\text{GeV}$ and the CP-odd pseudoscalar Higgs $m_{A} = 350~\text{GeV}$ for the purpose of this study, apart from fixing the charged Higgs mass $m_{H^{\pm}} = 450~\text{GeV}$. This choice of charged Higgs mass, \( H^\pm = 450~\text{GeV} \), remains viable in several popular beyond-Standard-Model (BSM) scenarios~\cite{Branco:2011iw}. This mass point has not yet been ruled out in the \( H^+ \rightarrow \tau^\pm \nu \) channel by ATLAS~\cite{ATLAS:2014otc} and CMS~\cite{CMS:2015lsf}, nor in the \( H^+ \rightarrow tb \) channel by ATLAS~\cite{ATLAS:2015nkq} and CMS~\cite{CMS:2015lsf}. Furthermore, updated information regarding the Two-Higgs-Doublet Model fits~\cite{Chowdhury:2017aav} supports this scenario. Moreover, this mass value lies within a range of particular interest for future colliders, such as the High-Luminosity LHC (HL-LHC).

We first present a conventional cut and count analysis with a luminosity of 3000 $fb^{-1}$. The initial preselection cuts include $N(l) \geq 2 $ and $N(b) \geq 4$ , and $\Delta R (ll)$ and $\Delta R (bb) > 0.4$ to effectively separate all pairs of leptons and b-quarks. Examining the final state topology, the major SM backgrounds considered for this process are $t\bar{t}+jets$, $VV+jets$, and those with subdominant contributions from $VVV+jets$. Moreover, we have also included $Vh$, $t\bar{t}h$ as essential part of the SM background. The signal of our interest and all relevant SM backgrounds are generated at a centre-of-mass energy of $\sqrt{s} = 14~\text{TeV}$, which corresponds to the design energy of the LHC and represents the operational energy for future runs, particularly in the HL-LHC phase. We have compiled a set of 29 observables for the analysis.\footnote{While this might be exhaustive for a traditional cut and count analysis, it will prove effective to start with a maximal set for the algorithm which will be designed in the next section.} which can, in principle, aid in discriminating signal from background. These are

\begin{enumerate*}

   \item $P_T(\ell_1)$,  \item $P_T(\ell_2)$, \item  $P_T(b_1)$, \item $P_T(b_2)$, \item $P_T(b_3)$, \item $P_T(b_4)$,
   \item $\eta(\ell_1)$,  \item $\eta(\ell_2)$, \item $\eta(b_1)$, \item $\eta(b_2)$,  \item $\eta(b_3)$, \item $\eta(b_4)$, 
   \item $\Delta R (\ell_1,\ell_2)$,  \item $\Delta R (b_1,b_2)$,  \item $\Delta R (b_1,b_3)$,  \item $\Delta R (b_1,b_4)$,  \item $\Delta R (b_2,b_3)$, \item $\Delta R (b_2,b_4)$, \item $\Delta R (b_3,b_4)$,
   \item $THT$,\item $\cancel{E}_T$,
   \item $M (\ell_1,\ell_2)$, \item $M (b_1,b_2)$, \item $M (b_1,b_3)$, \item $M (b_1,b_4)$,  \item $M (b_2,b_3)$, \item $M (b_2,b_4)$, \item $M (b_3,b_4)$,  
   
   \item $M (\ell_1,\ell_2,b_1,b_2,b_3,b_4)$.
    \end{enumerate*}

\begin{figure*}
   \centering
   \includegraphics[width=0.19\textwidth]{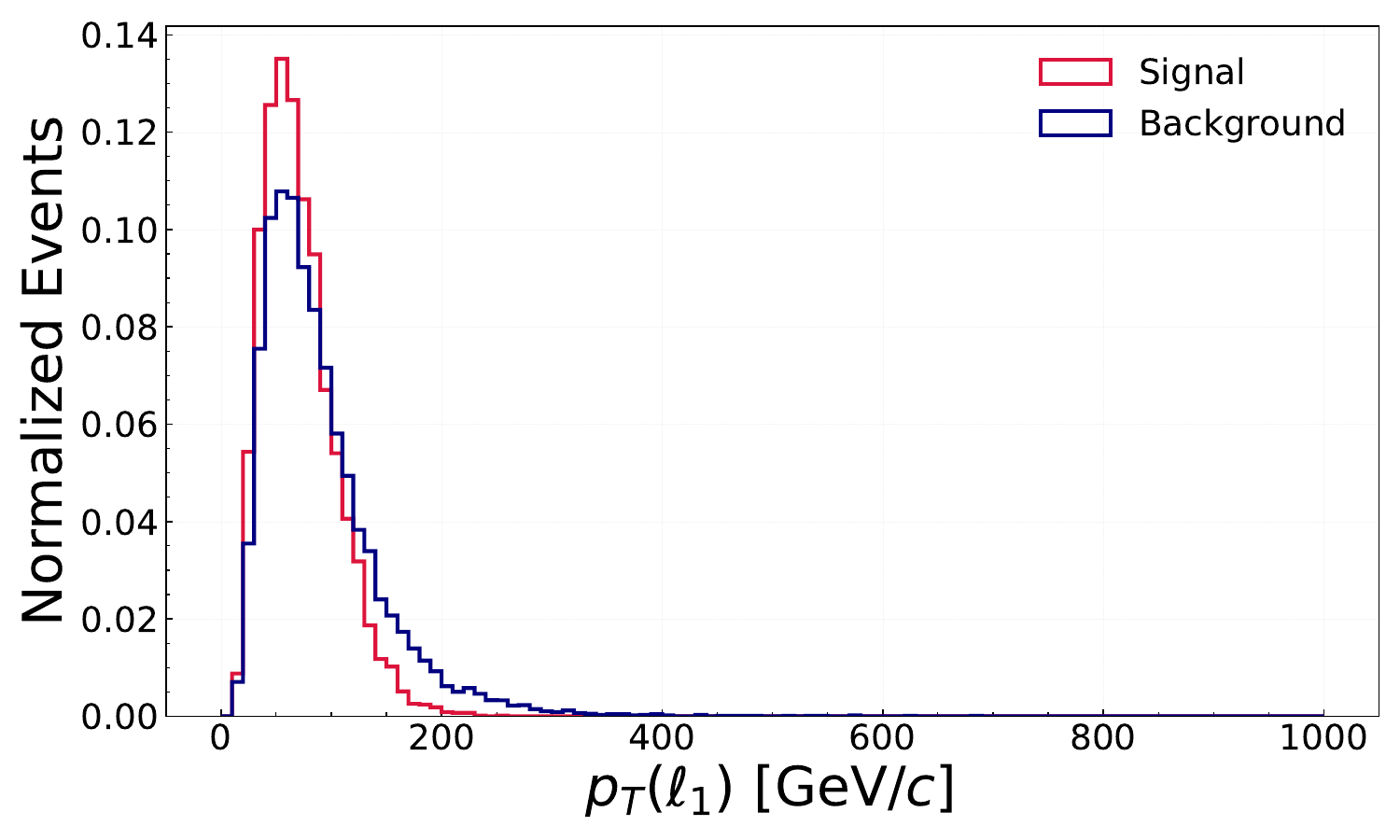}
   \includegraphics[width=0.19\textwidth]{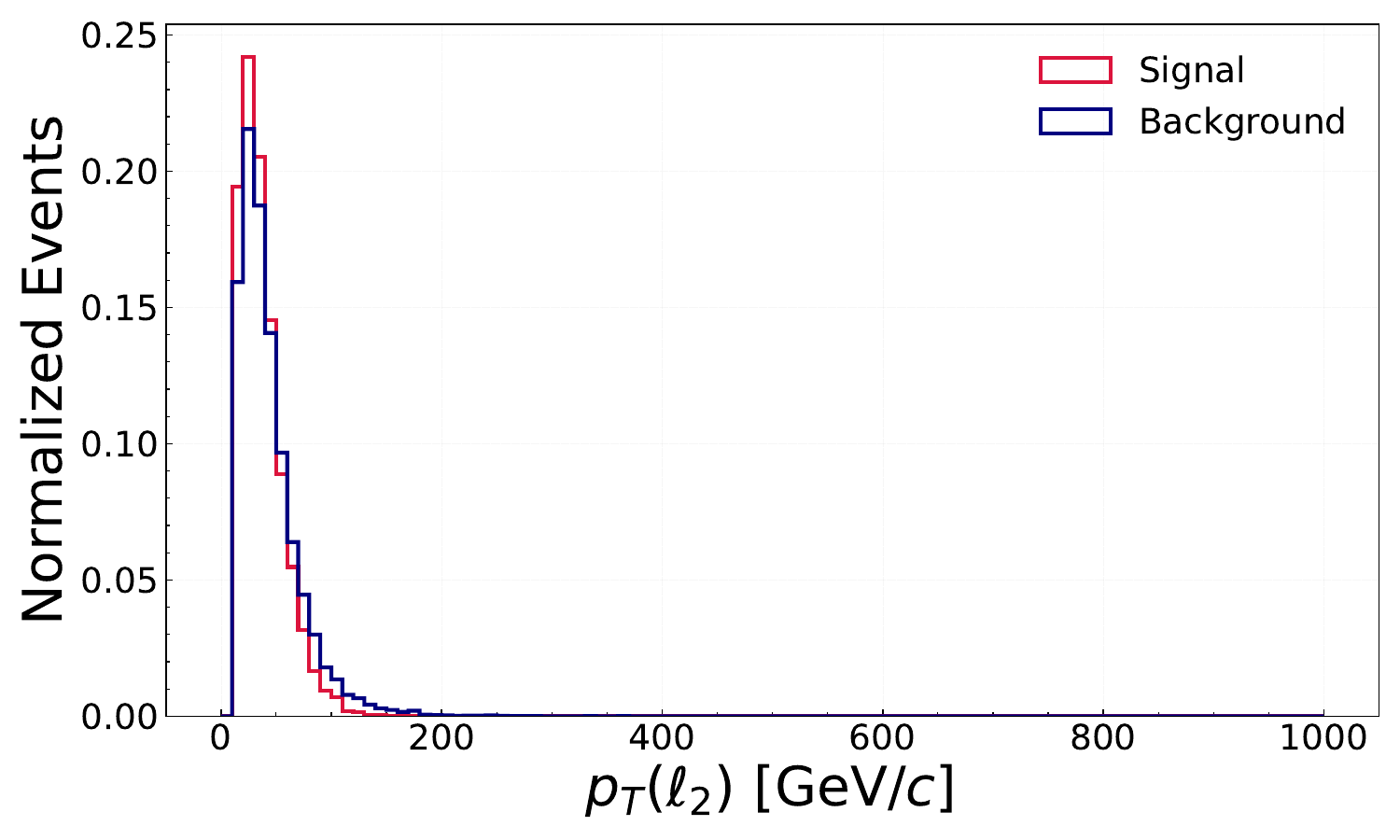}
   \includegraphics[width=0.19\textwidth]{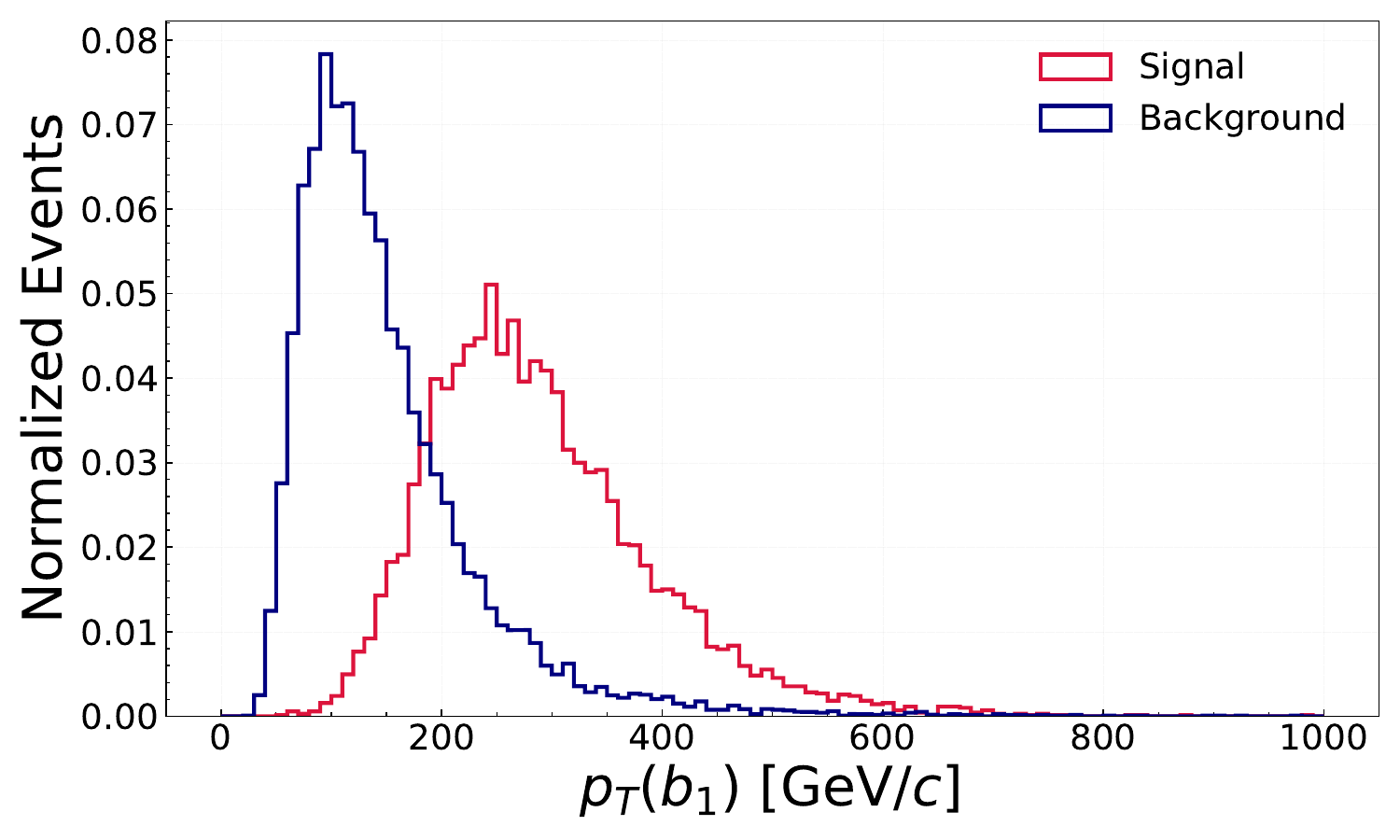}
   \includegraphics[width=0.19\textwidth]{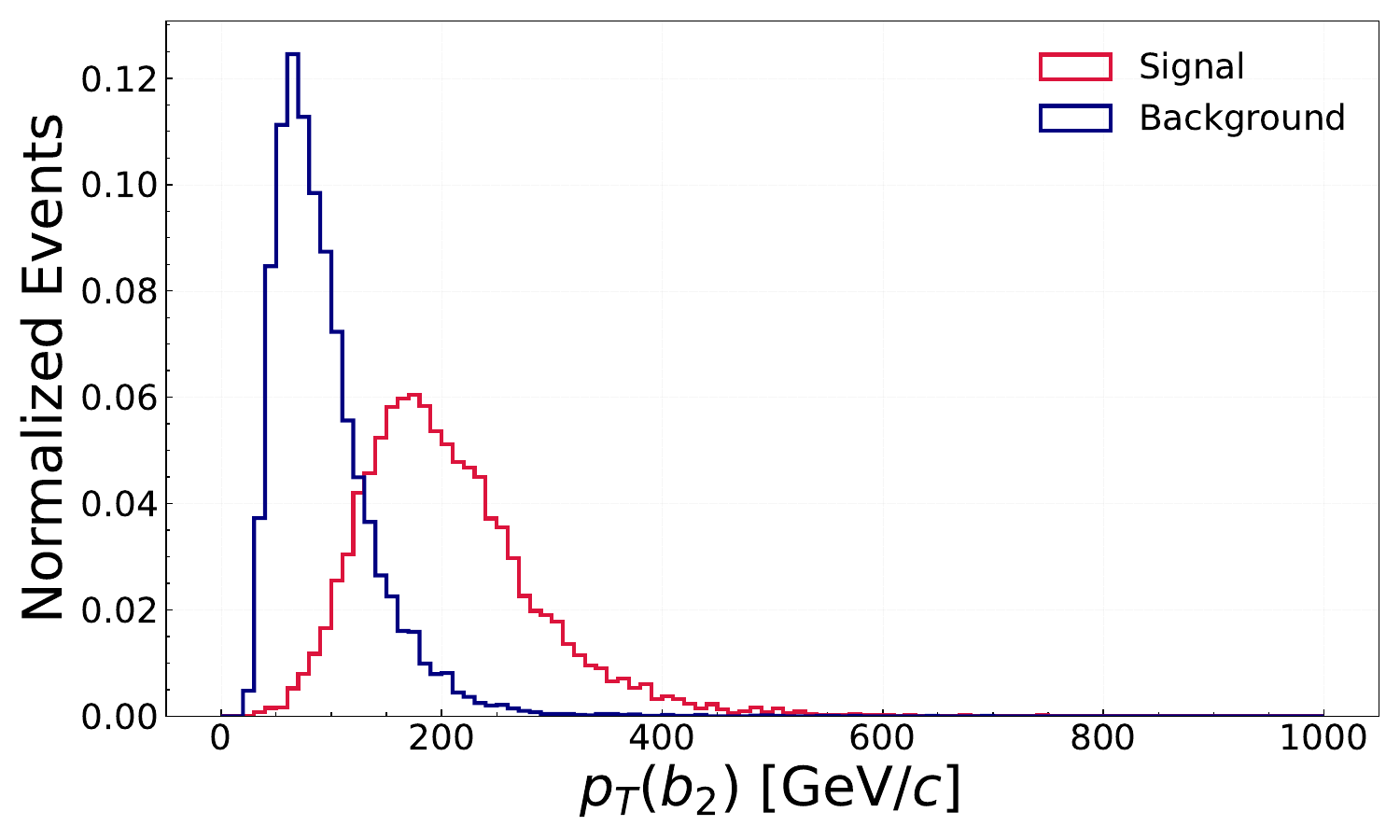}
   \includegraphics[width=0.19\textwidth]{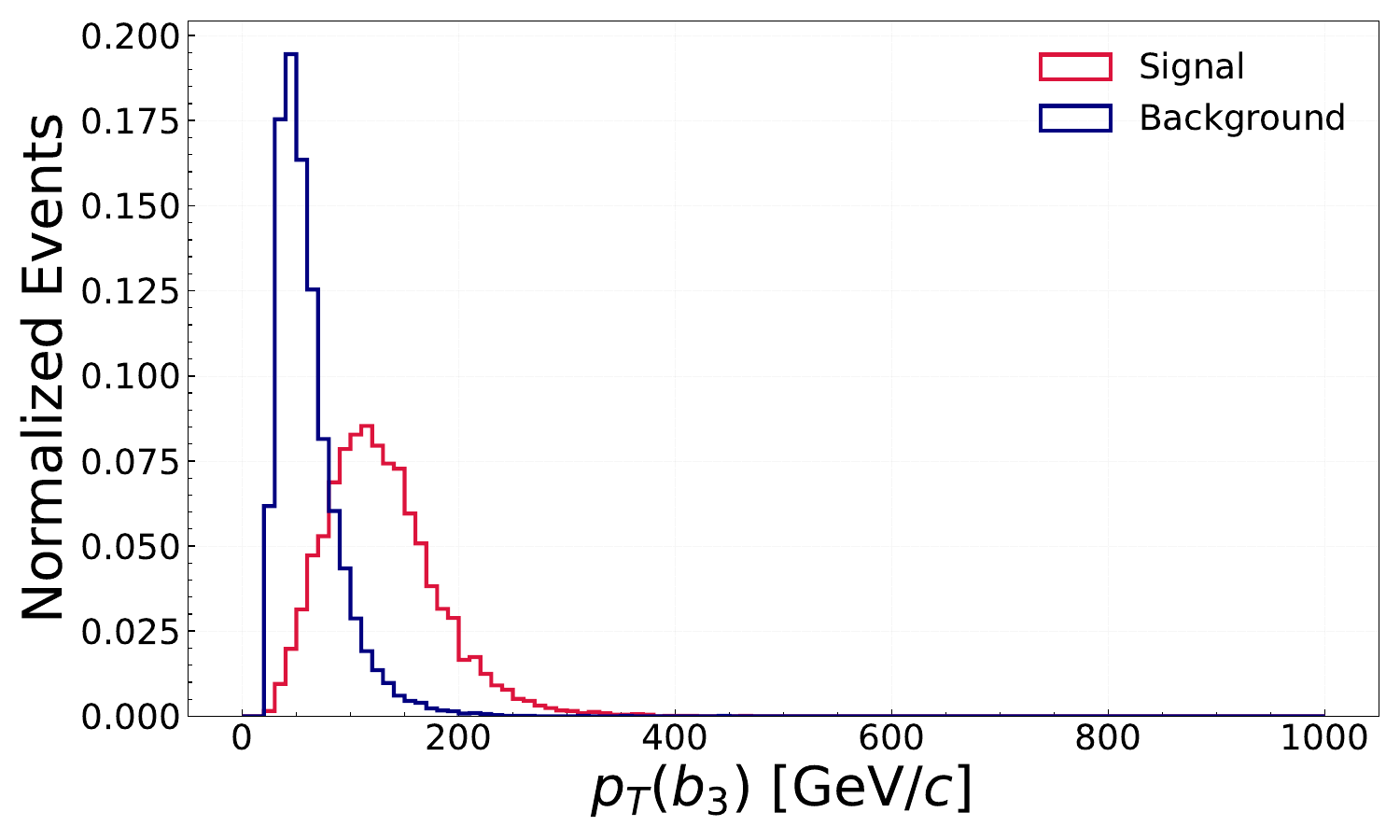}
   \includegraphics[width=0.19\textwidth]{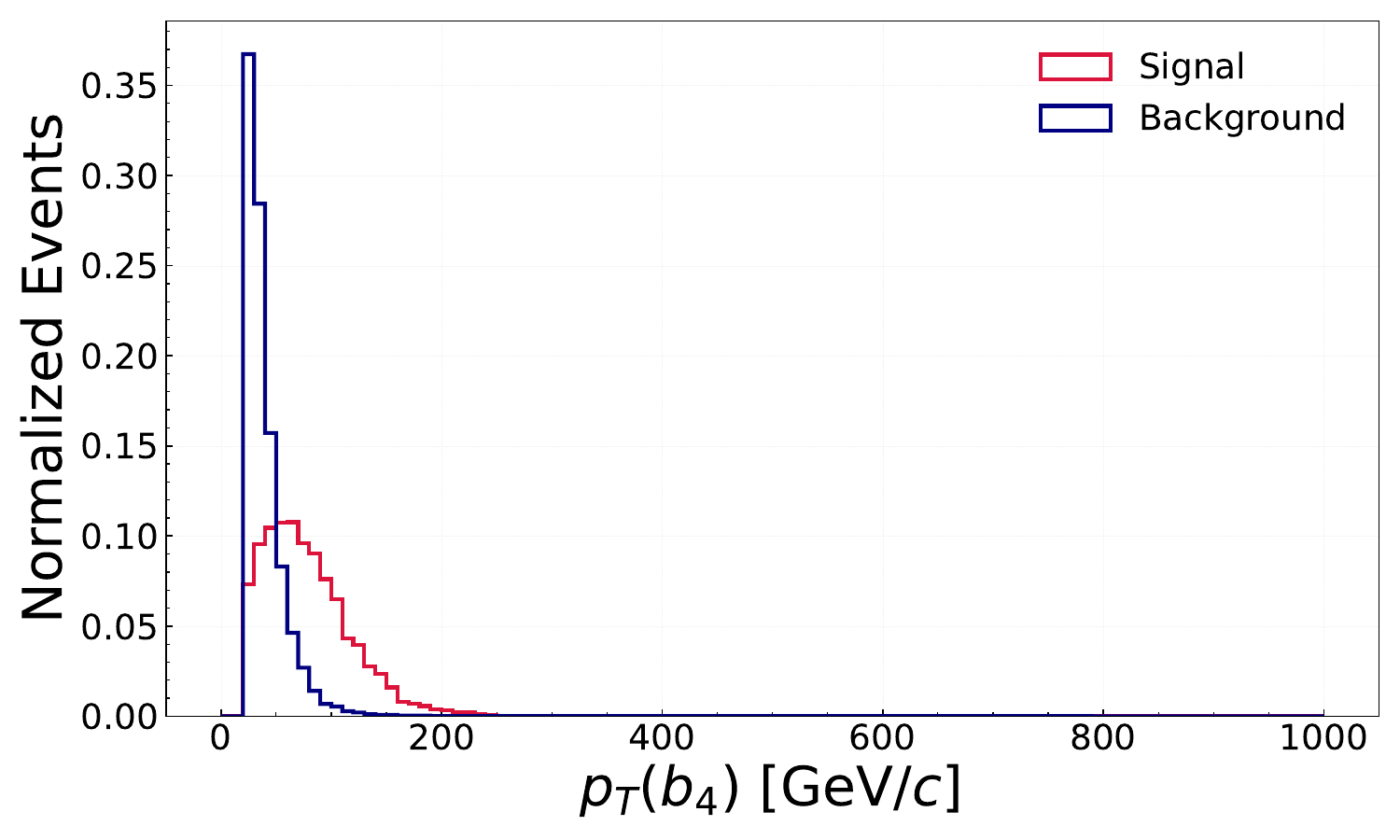}
   \includegraphics[width=0.19\textwidth]{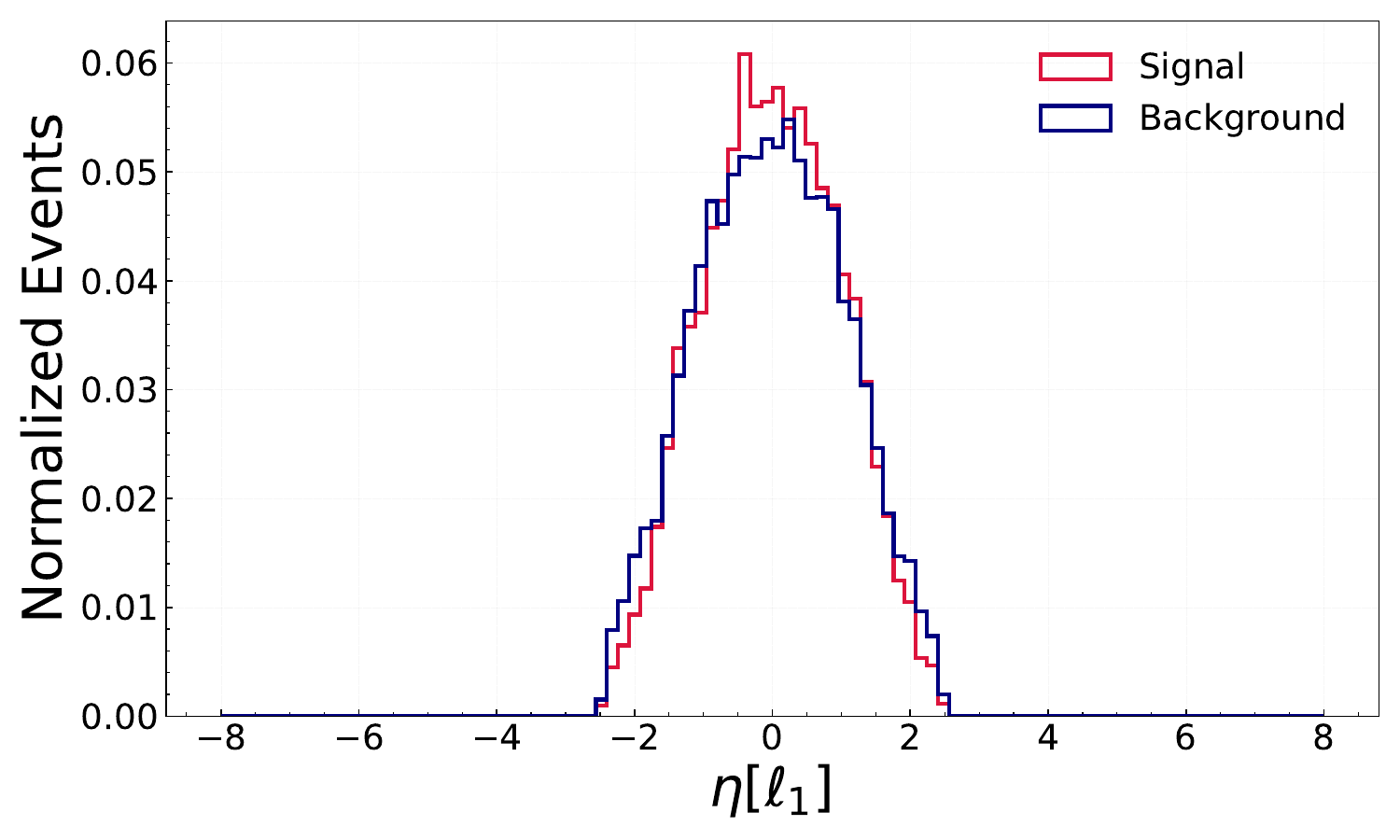}
   \includegraphics[width=0.19\textwidth]{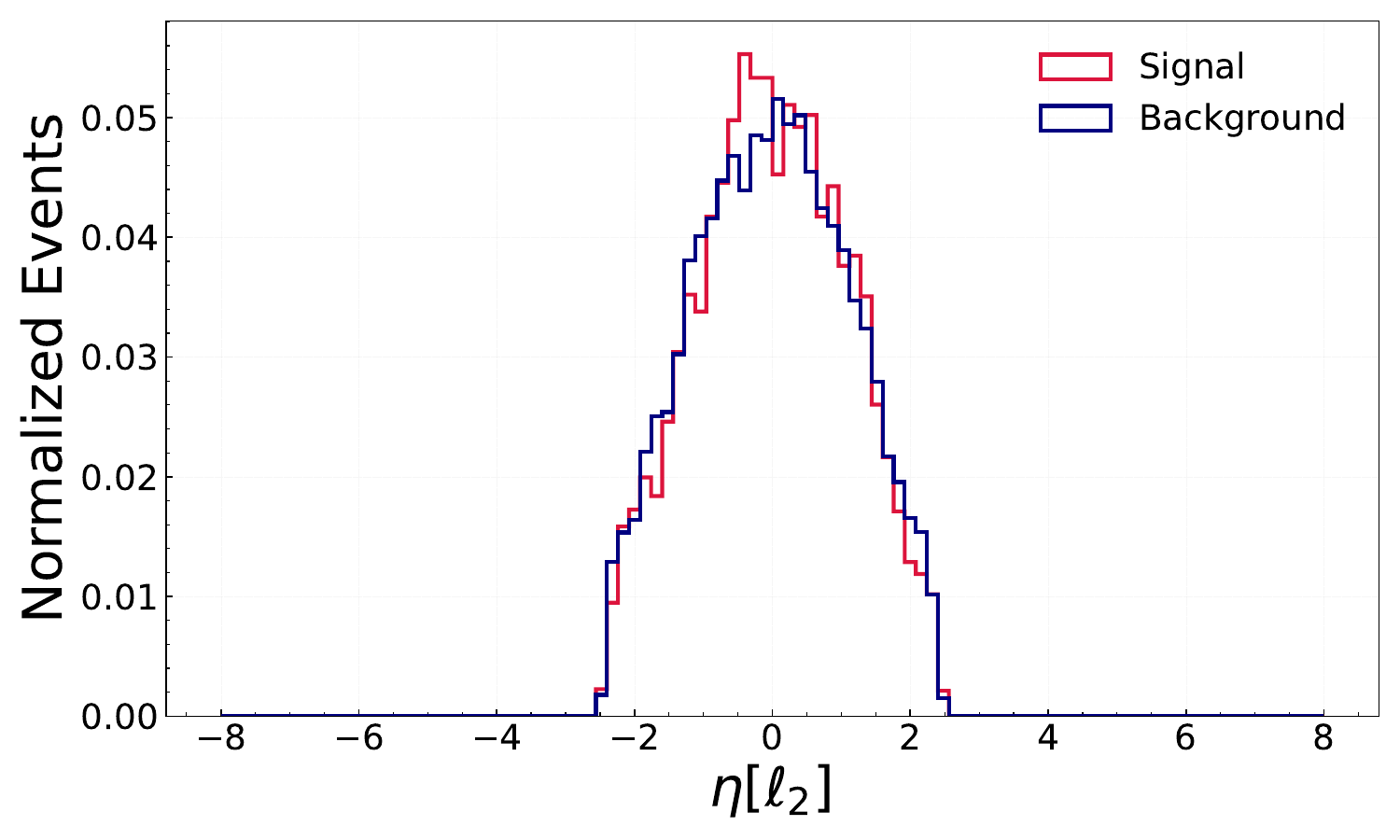}
   \includegraphics[width=0.19\textwidth]{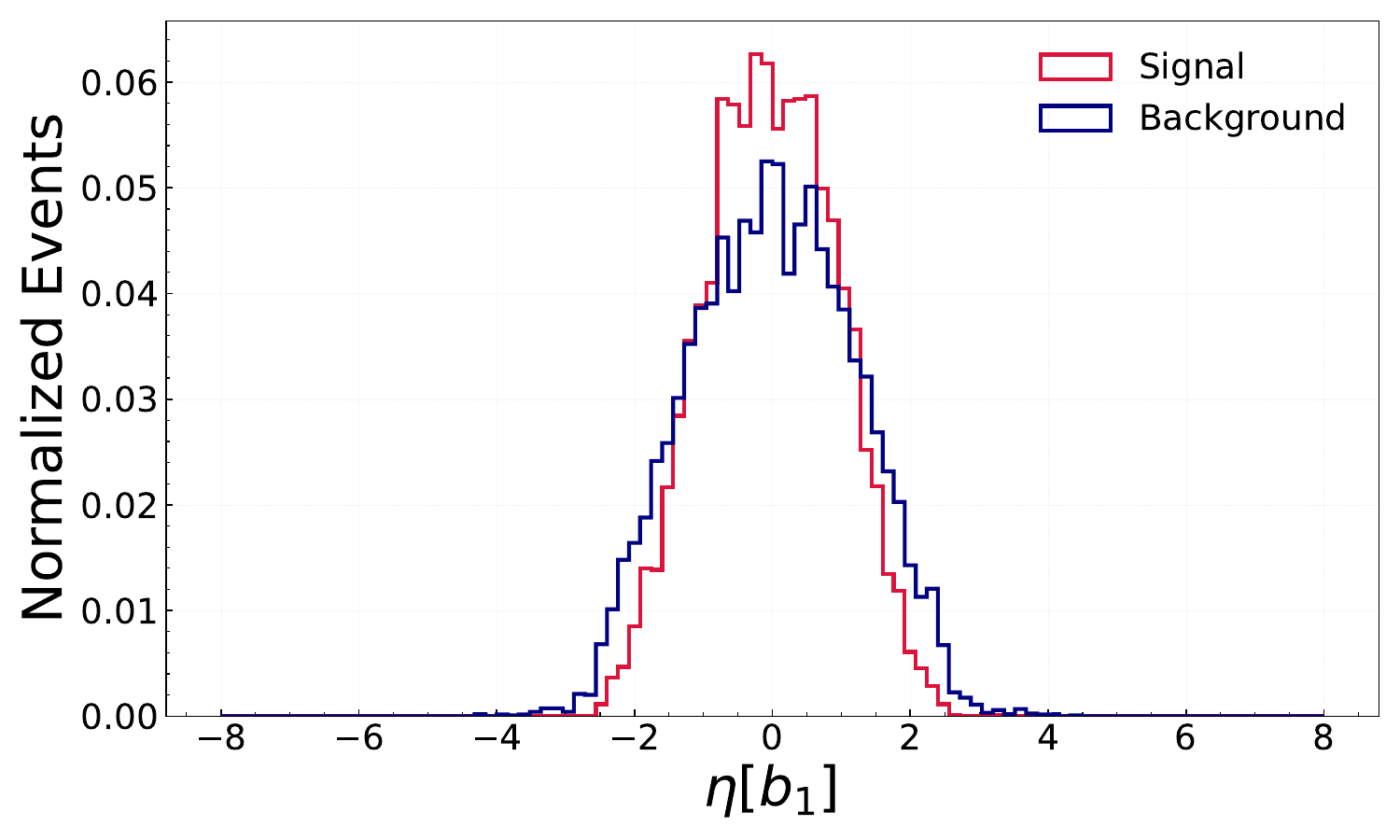}
 \includegraphics[width=0.19\textwidth]{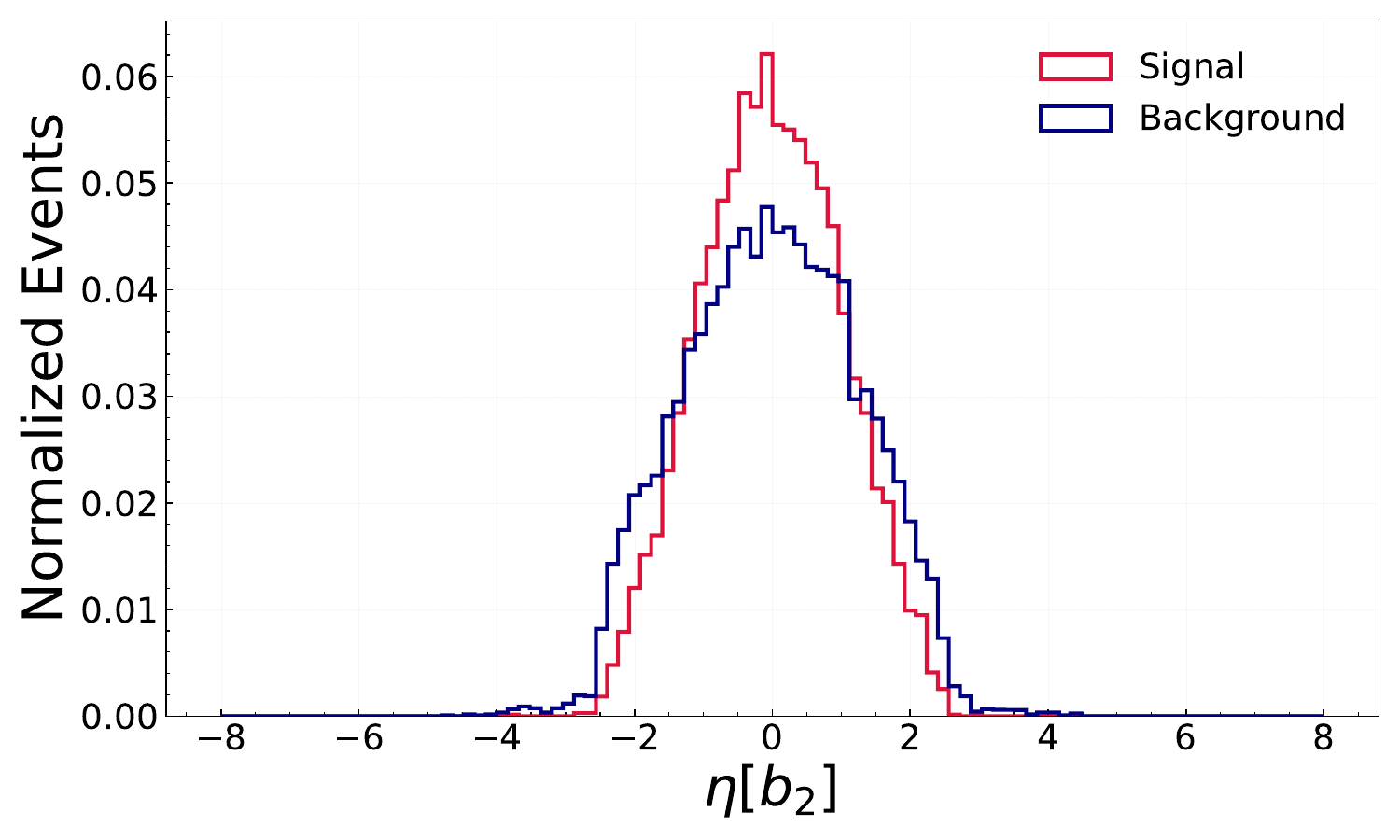}
\includegraphics[width=0.19\textwidth]{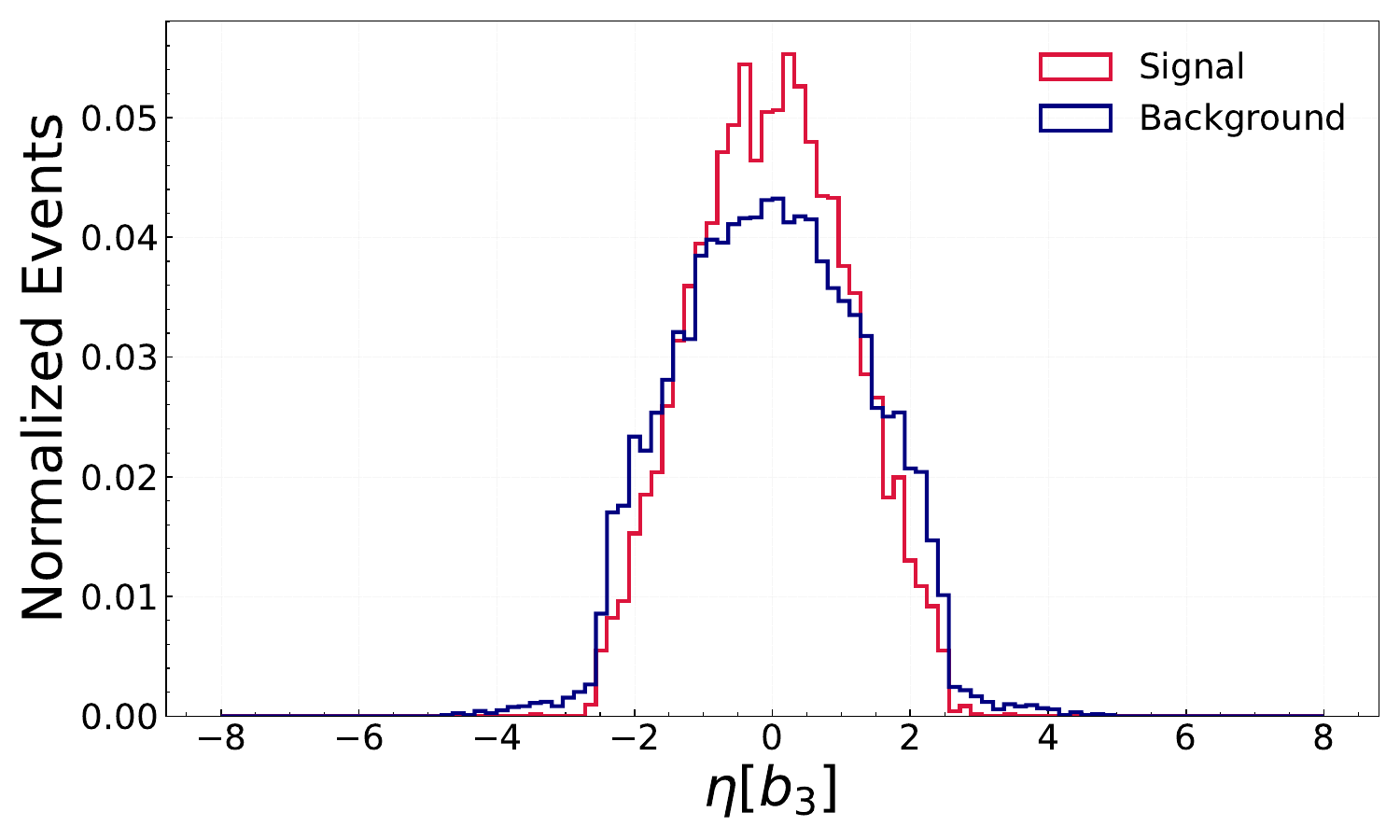}
\includegraphics[width=0.19\textwidth]{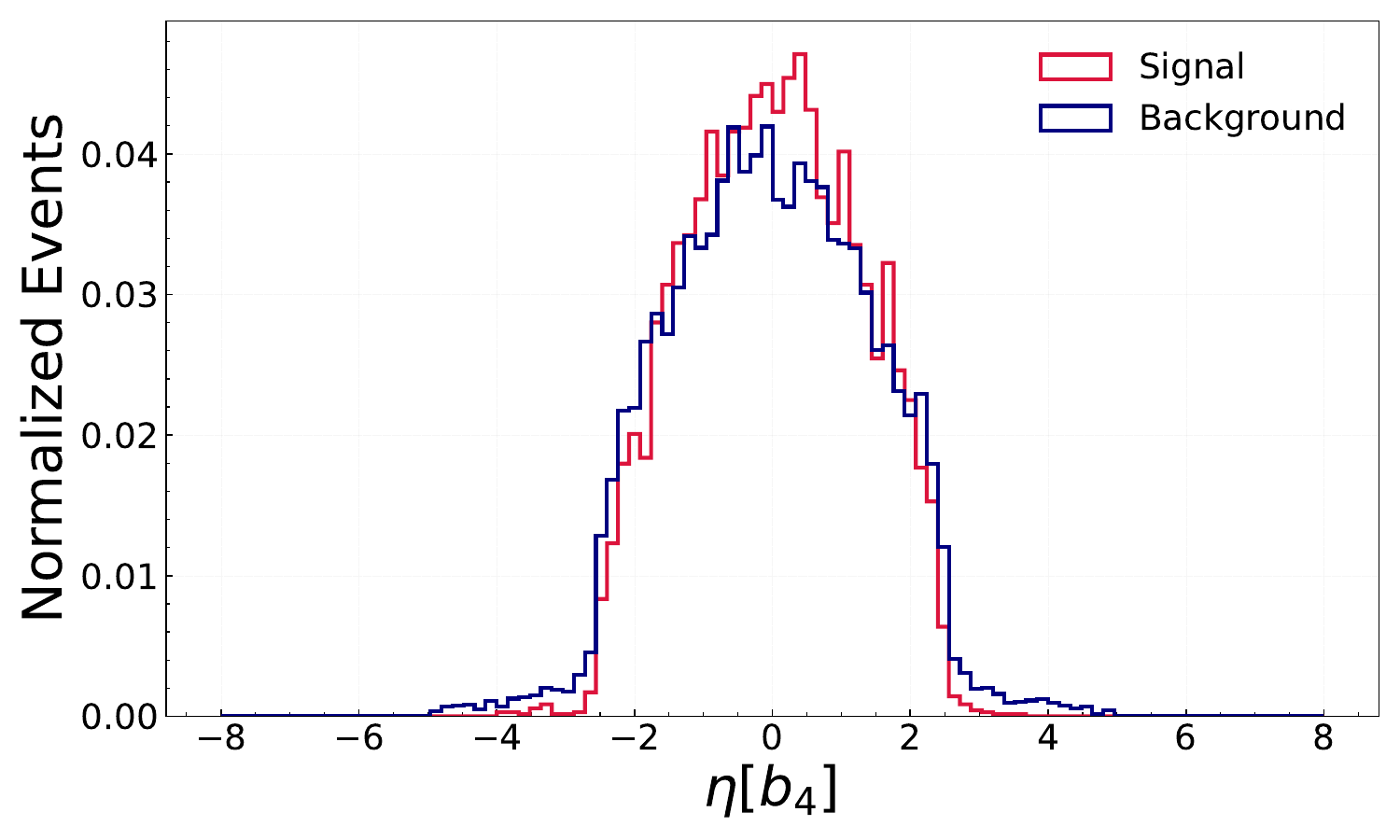}
 \includegraphics[width=0.19\textwidth]{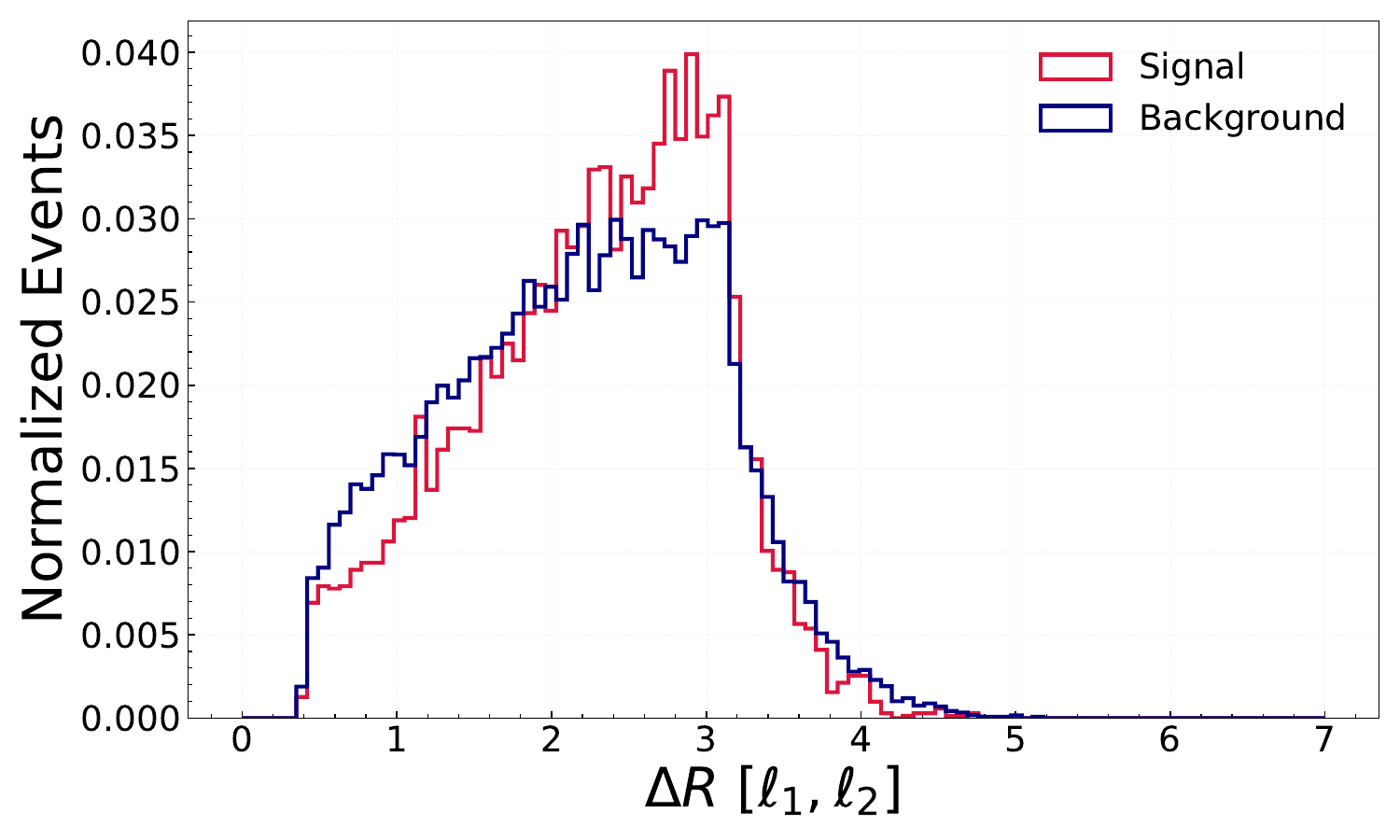}
 \includegraphics[width=0.19\textwidth]{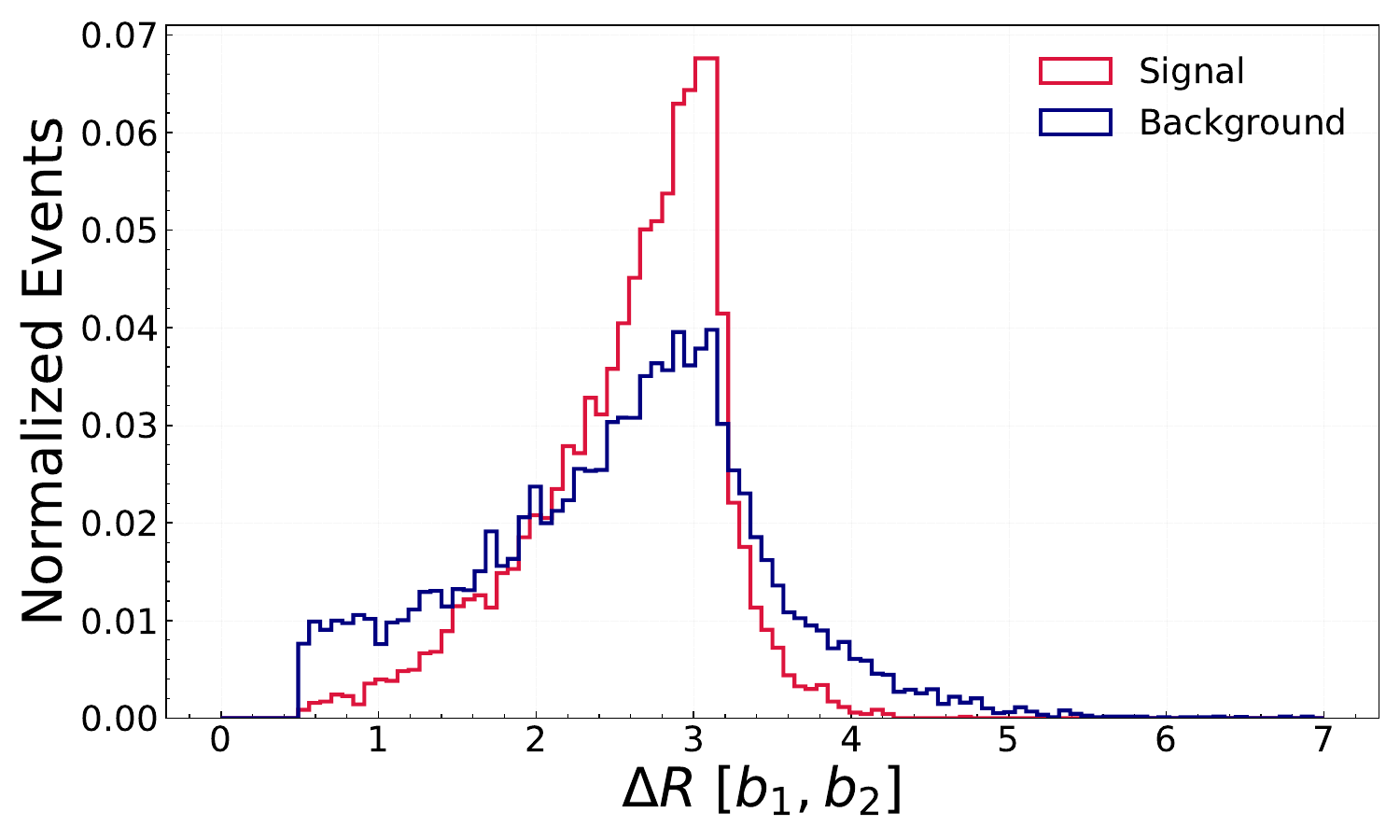}
\includegraphics[width=0.19\textwidth]{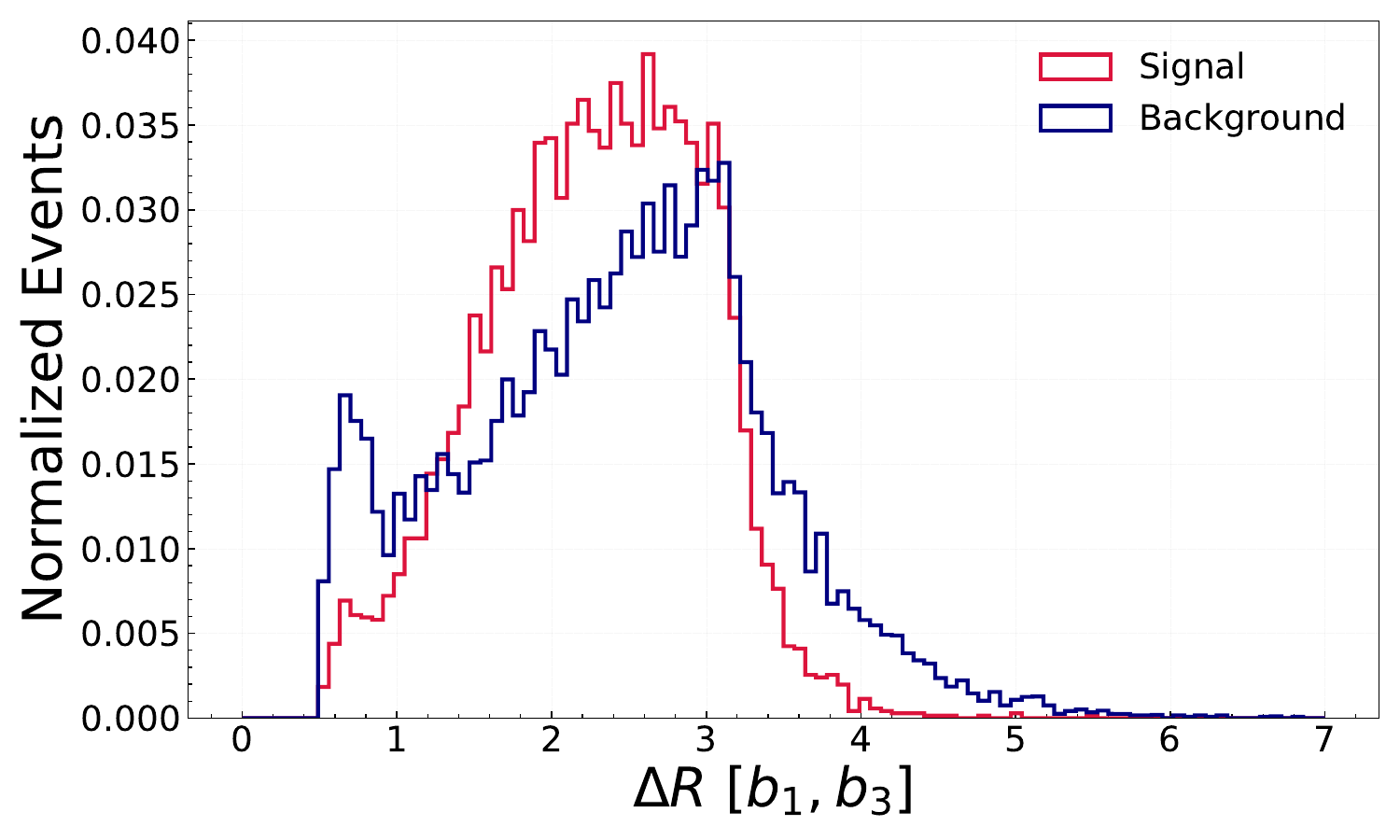}
 \includegraphics[width=0.19\textwidth]{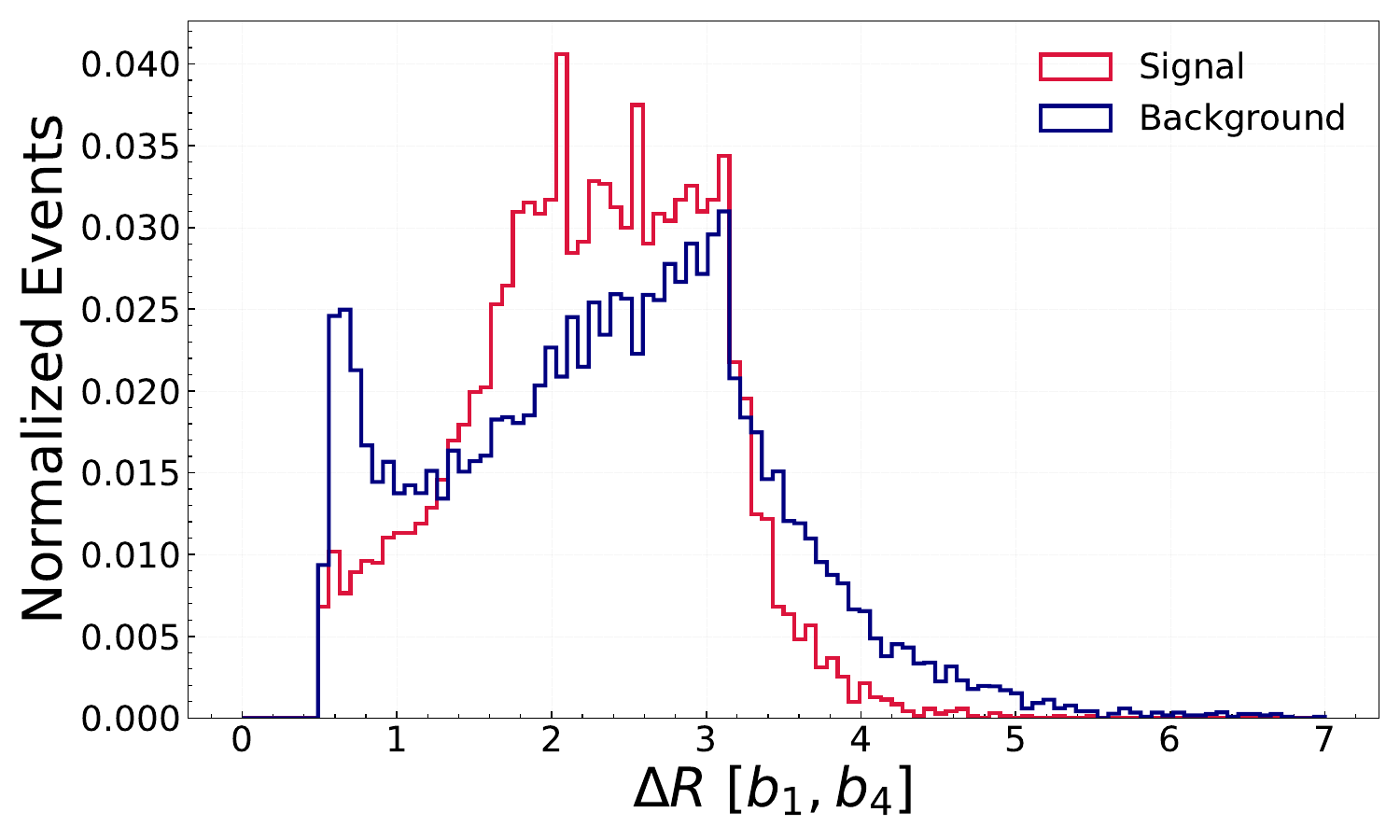}
  \includegraphics[width=0.19\textwidth]{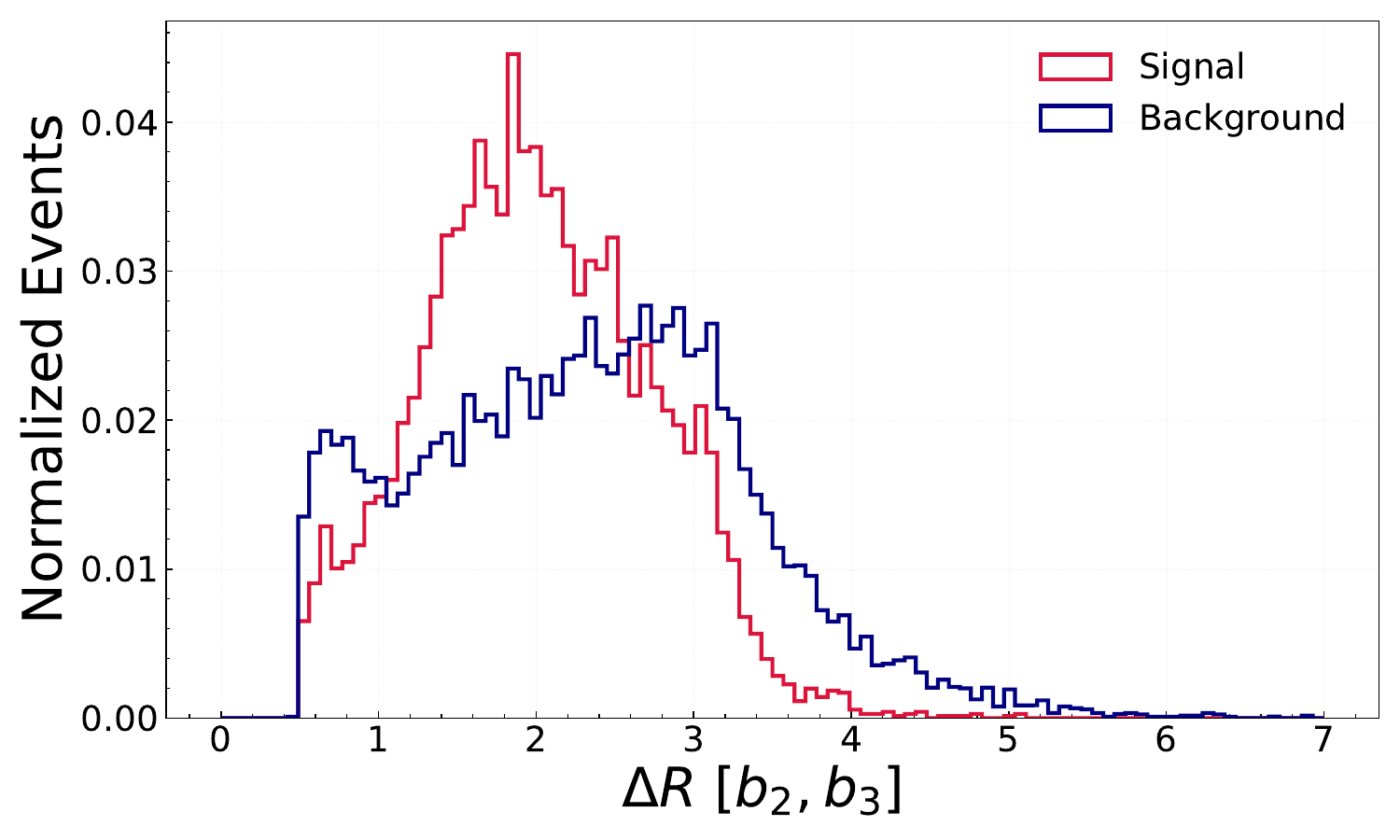}
 \includegraphics[width=0.19\textwidth]{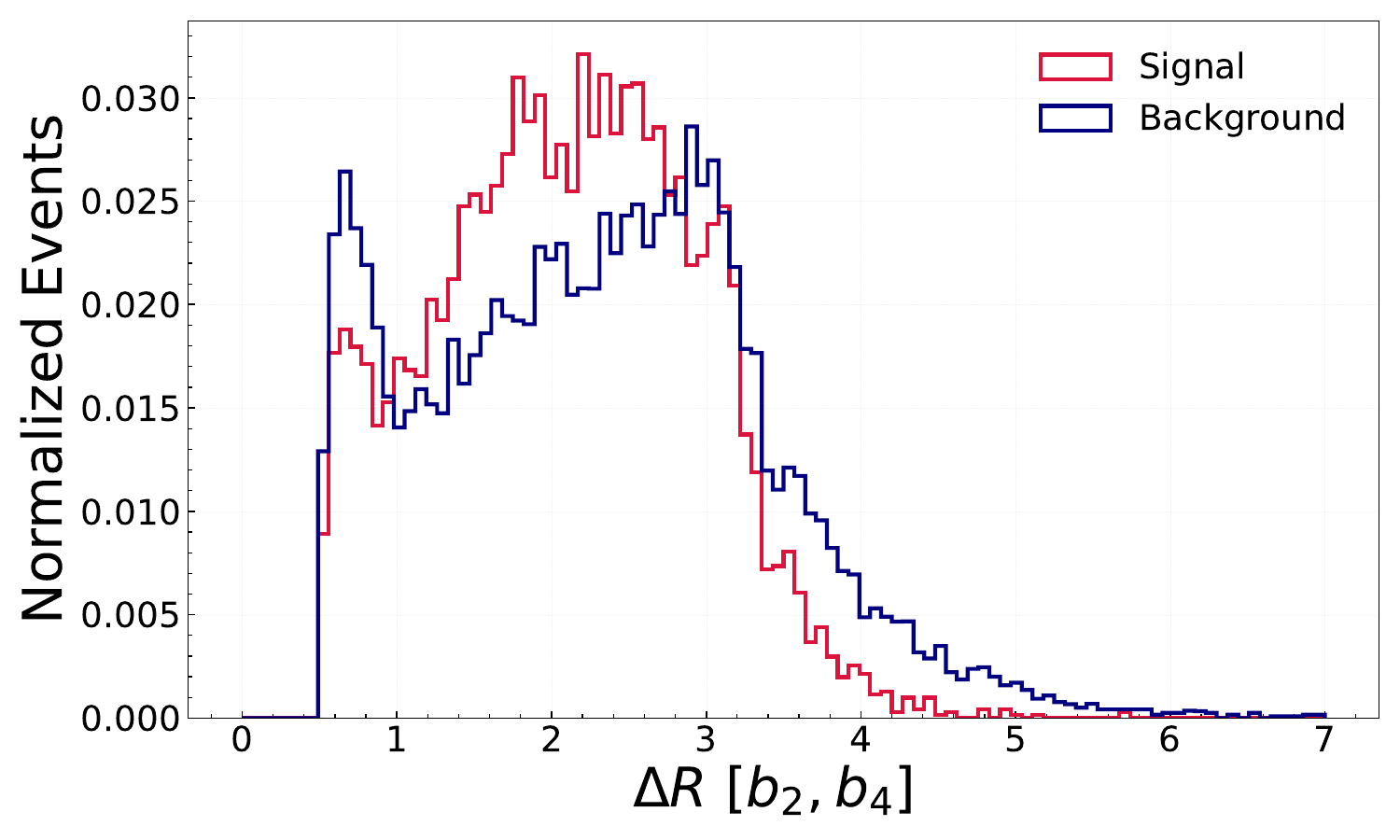}
\includegraphics[width=0.19\textwidth]{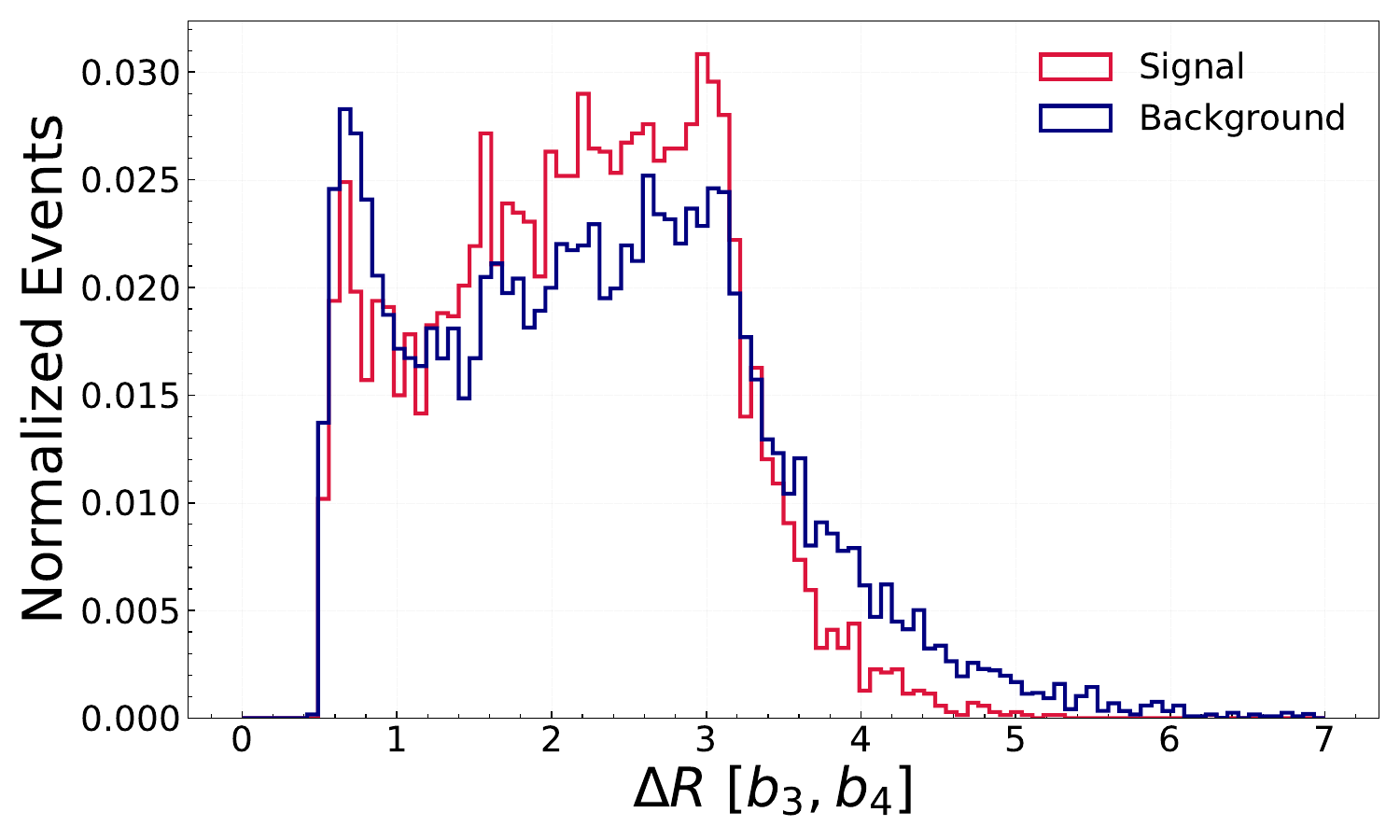}
\includegraphics[width=0.19\textwidth]{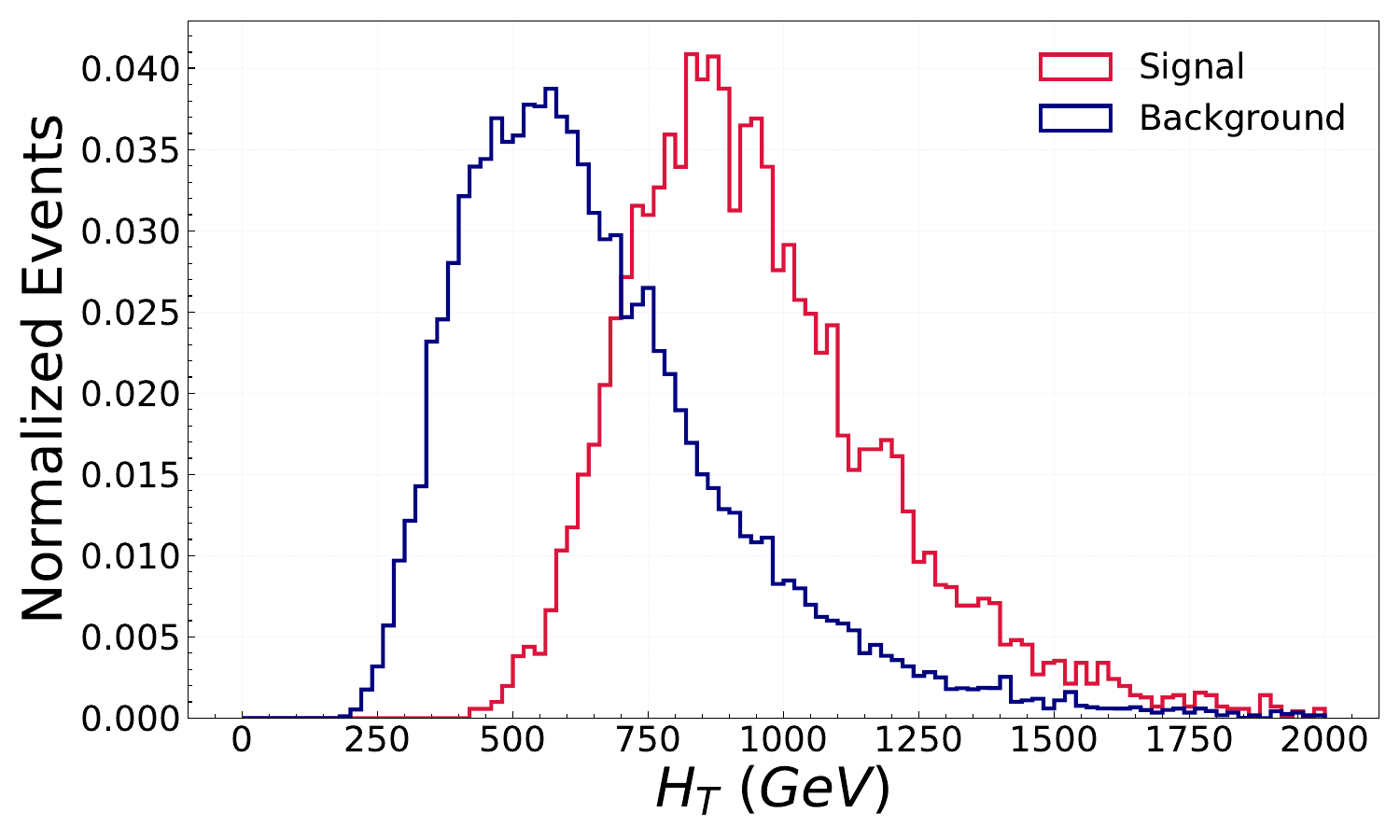}
\includegraphics[width=0.19\textwidth]{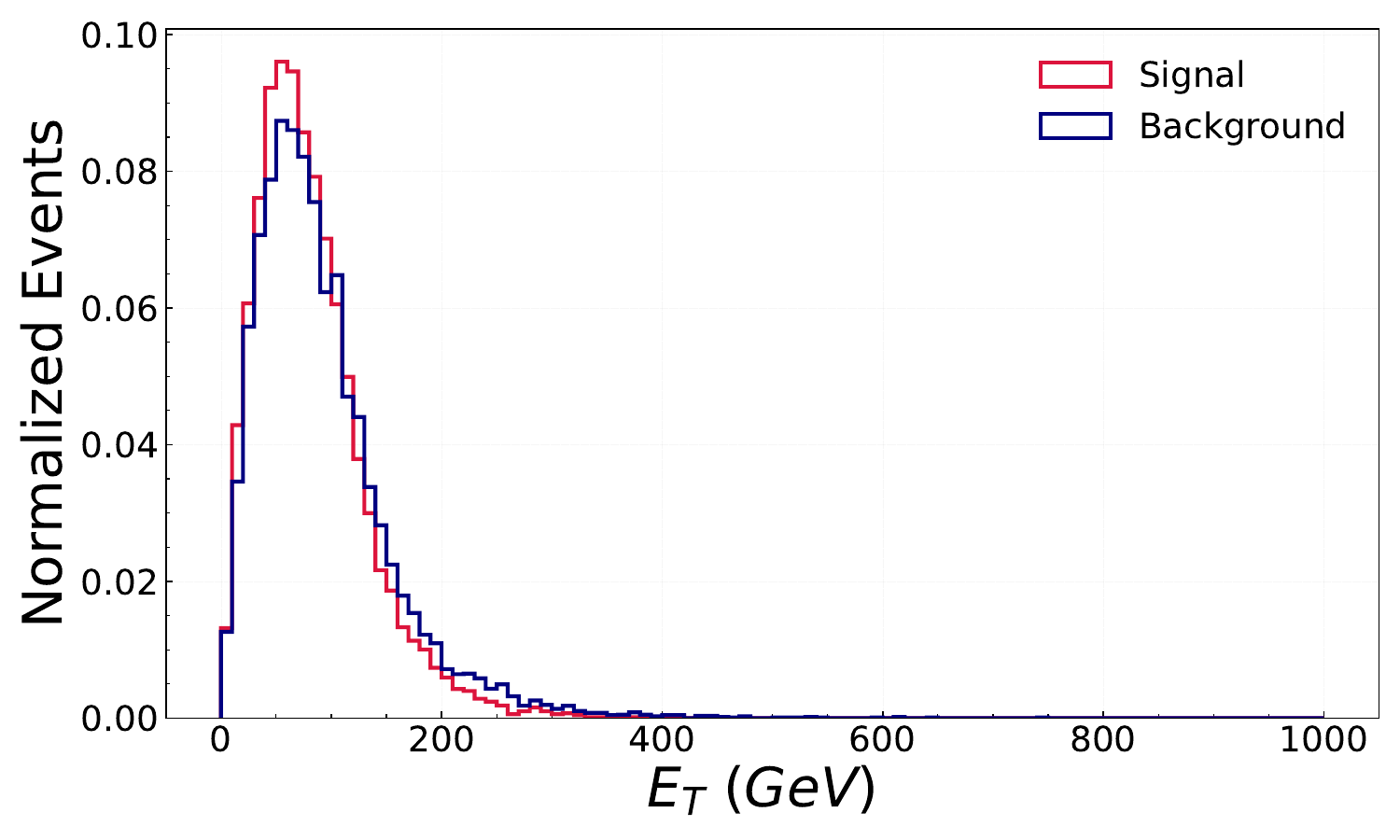}
\includegraphics[width=0.19\textwidth]{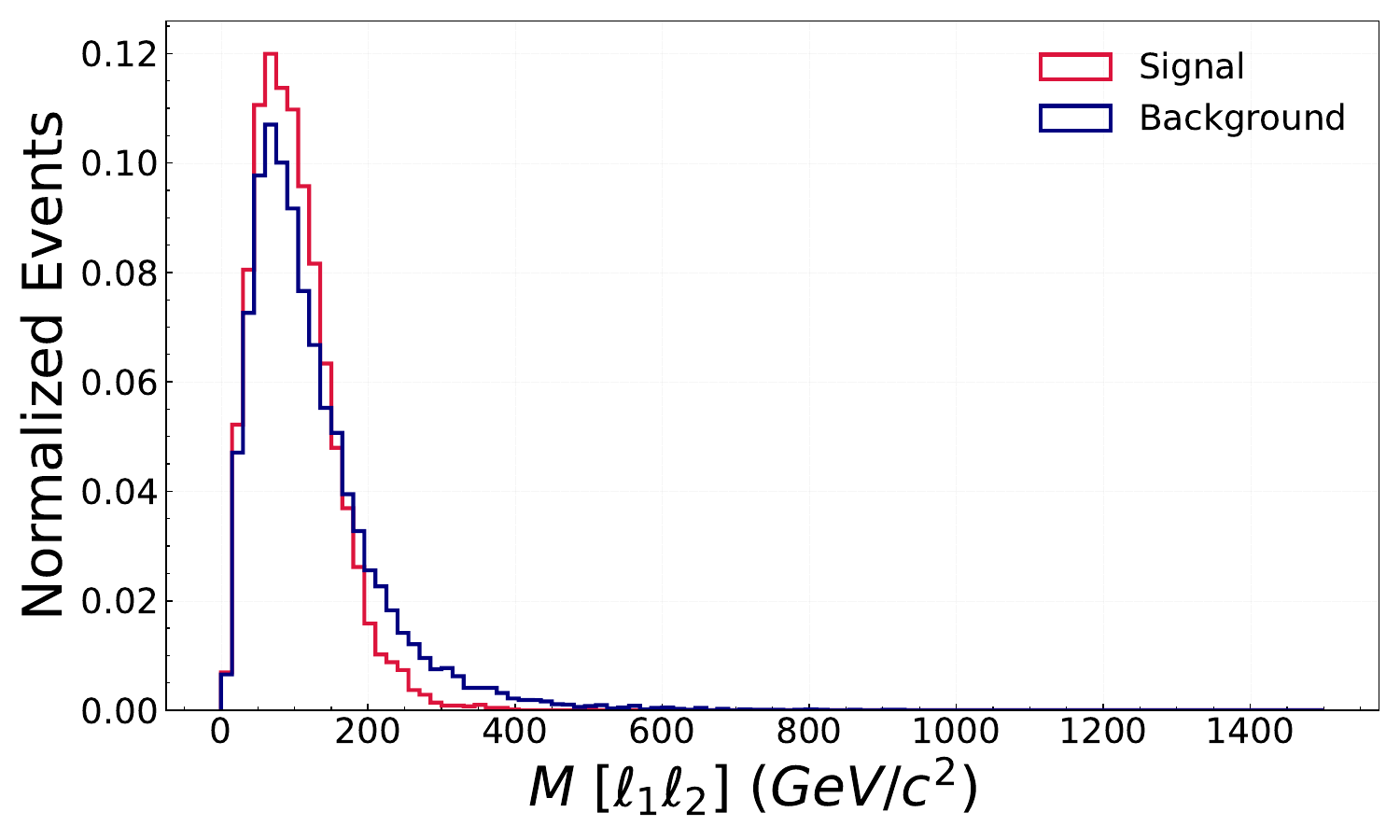}
\includegraphics[width=0.19\textwidth]{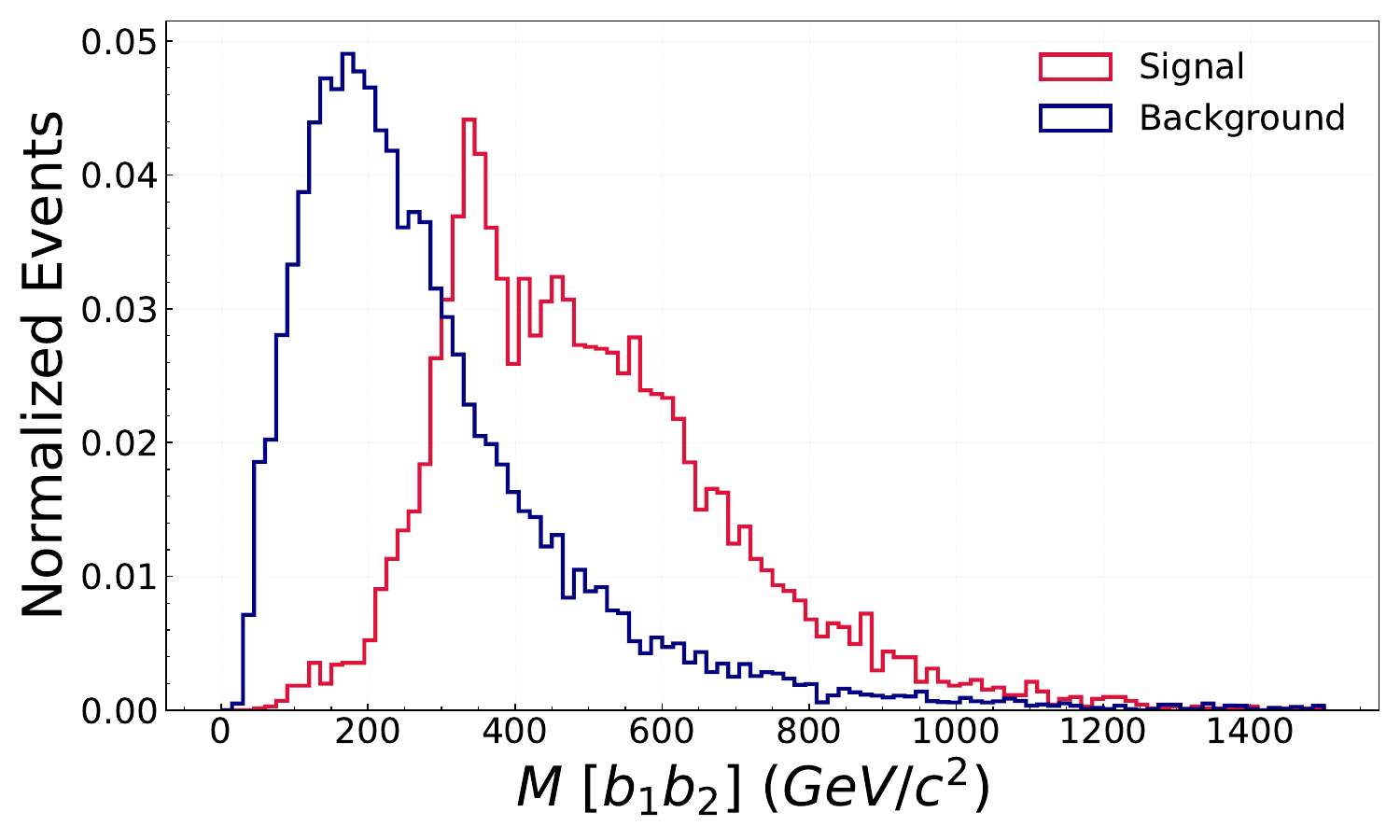}
\includegraphics[width=0.19\textwidth]{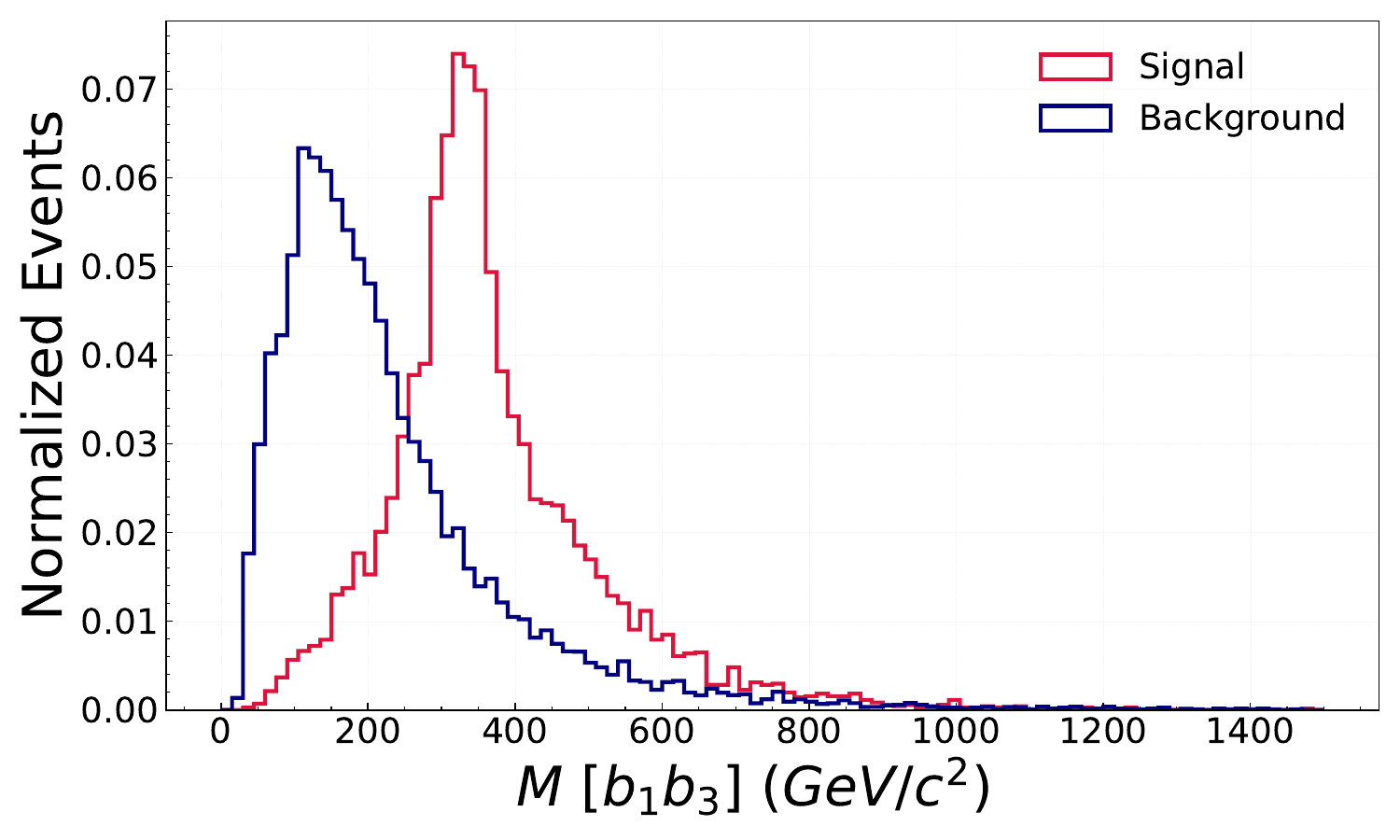}
\includegraphics[width=0.19\textwidth]{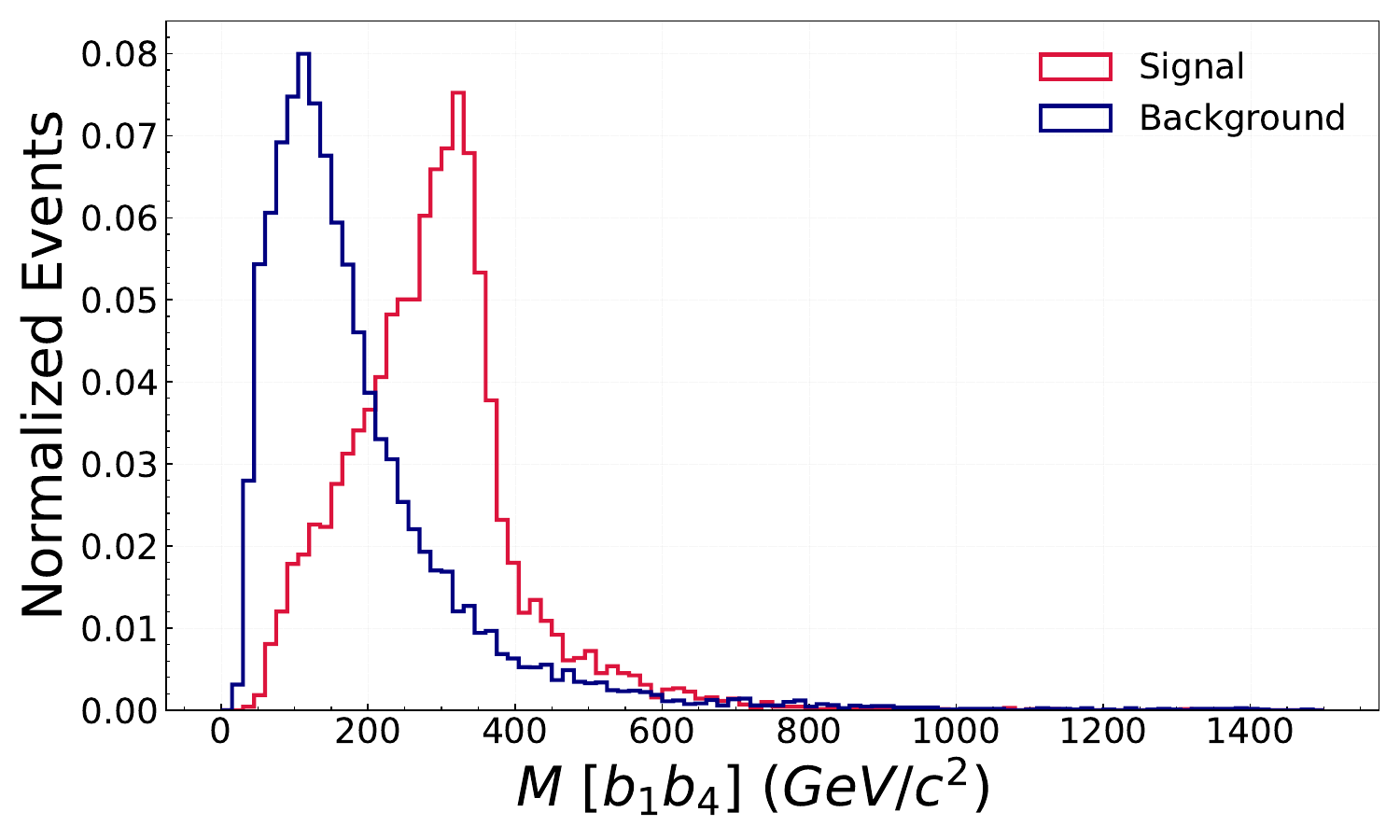}
\includegraphics[width=0.19\textwidth]{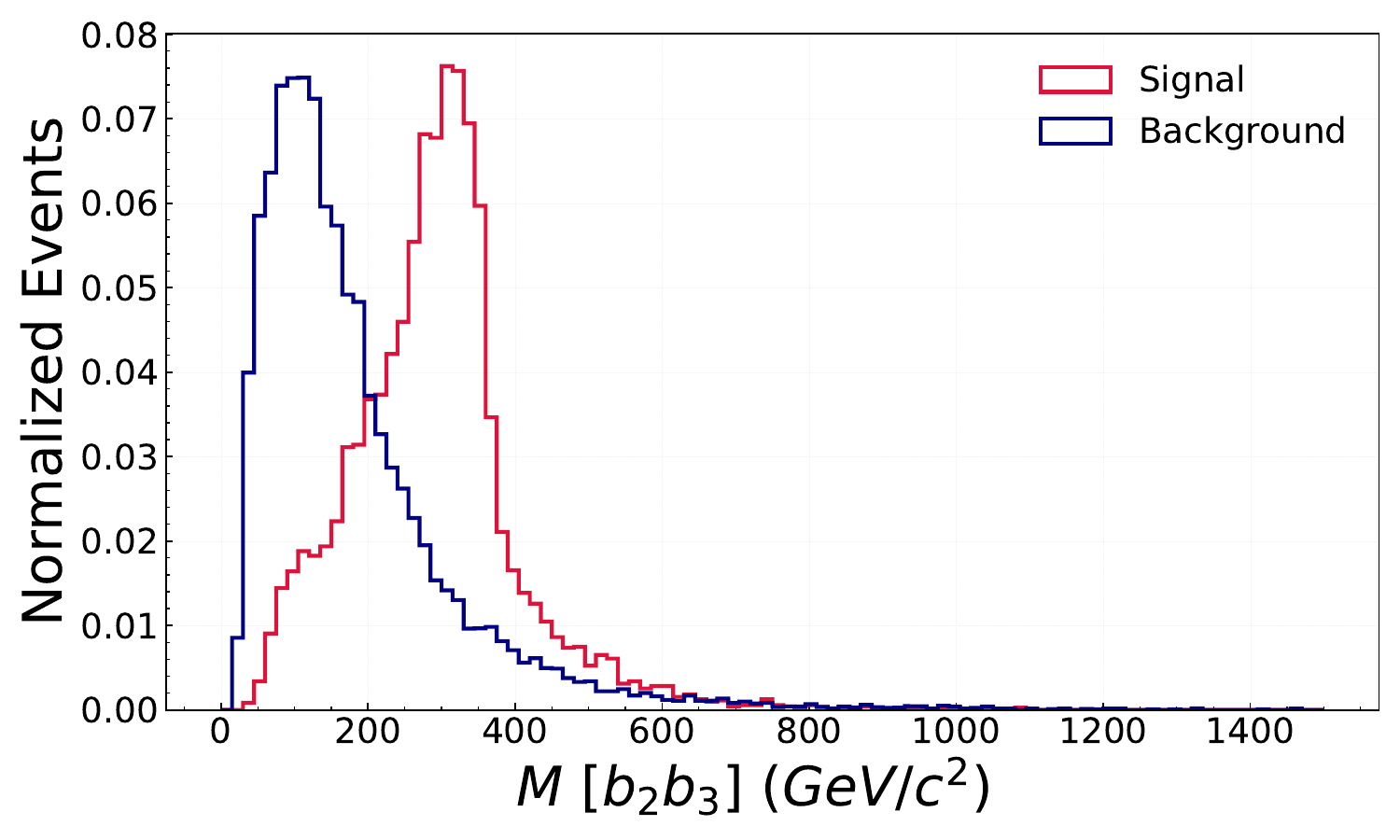}
\includegraphics[width=0.19\textwidth]{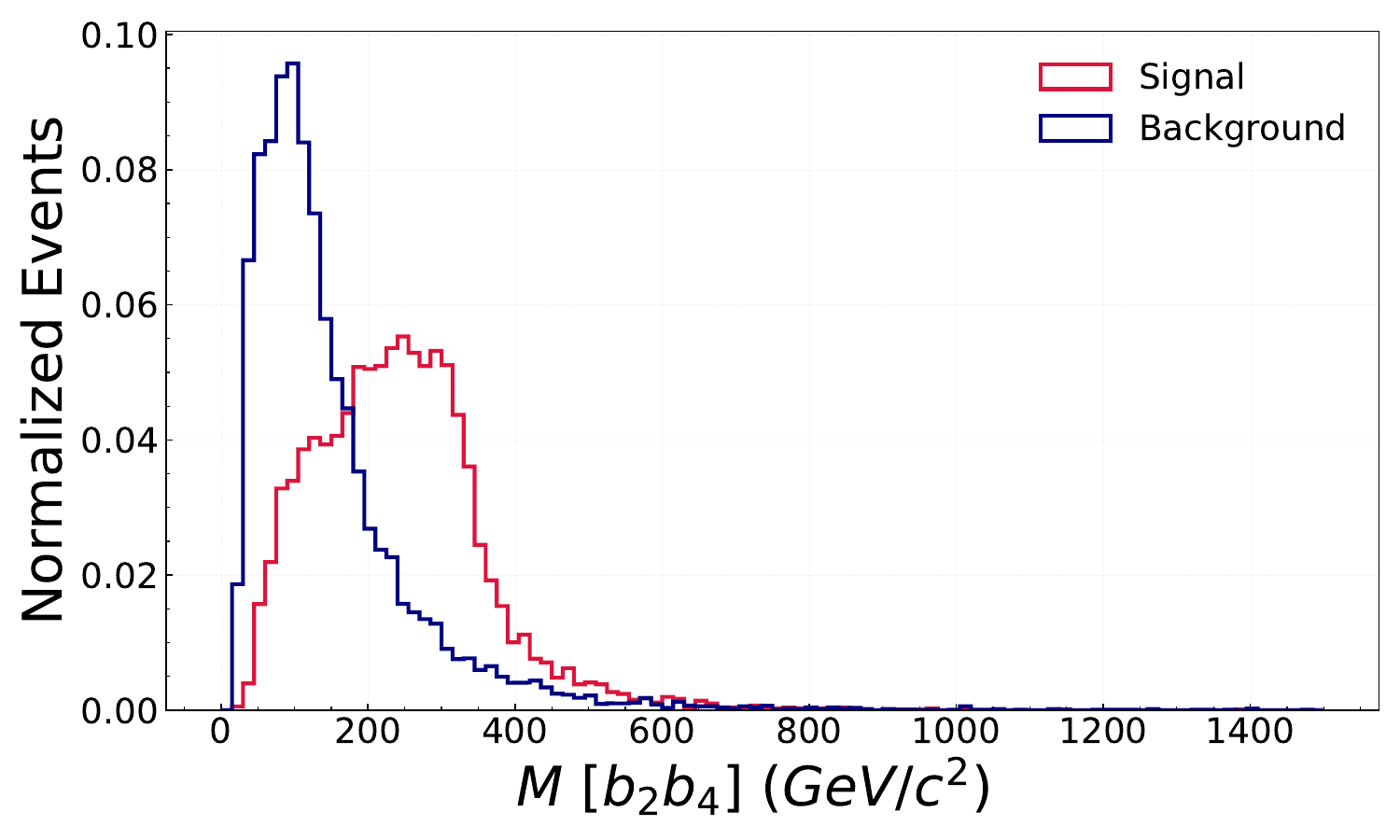}
\includegraphics[width=0.19\textwidth]{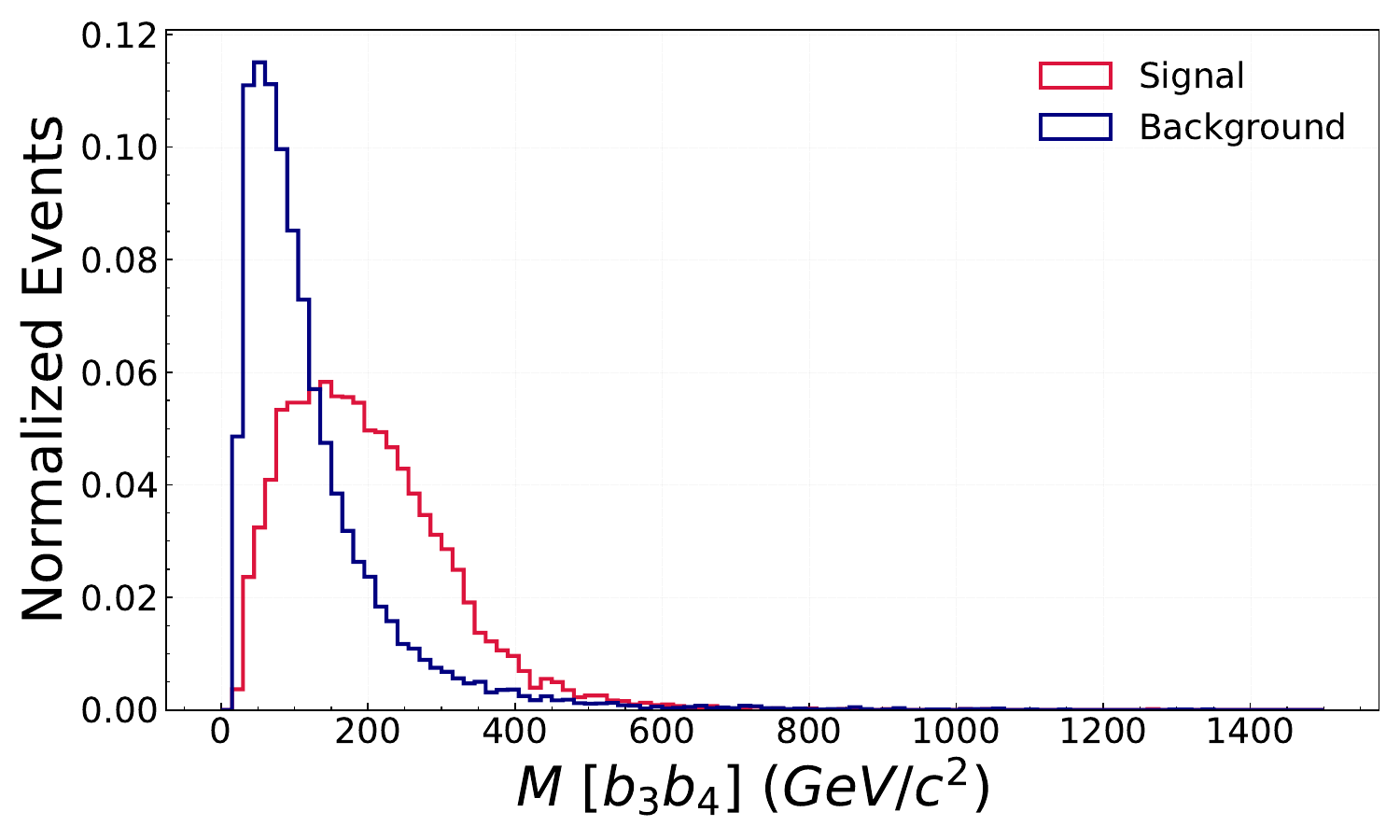}
\includegraphics[width=0.19\textwidth]{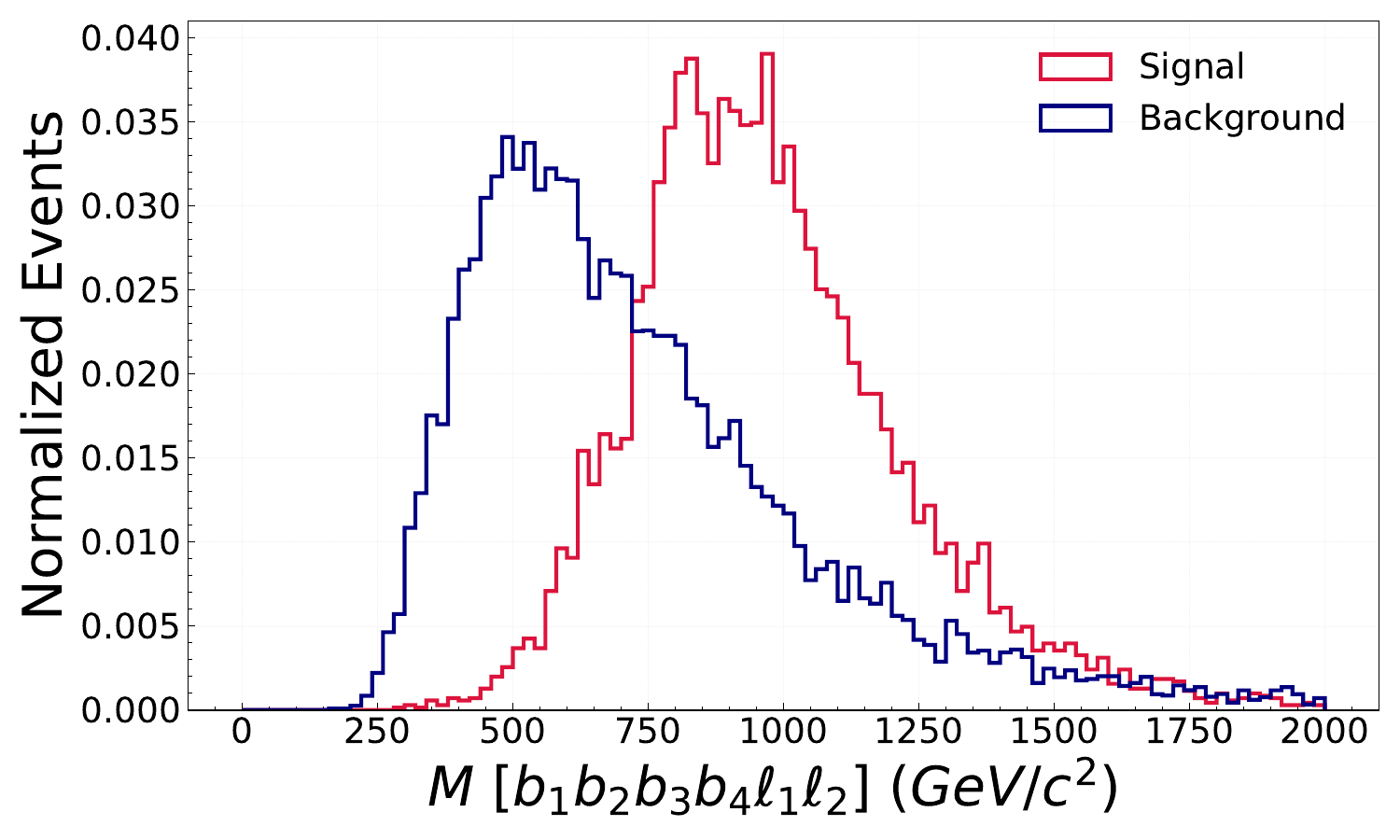}

   \caption{The normalized distributions of the 29  observables mentioned in the text after the pre-selection cuts.}
   \label{tab:Figure1}
   
\end{figure*}

The initial distributions acquired following the application of preselection cuts are depicted in \autoref{tab:Figure1}. By examining these distributions, one can straightforwardly determine the selection cuts that maximizes the signal-to-background ratio. We have established a set of selection criteria from intution gained from the initial observable distribution and the result of this traditional cut and count analysis is presented in \autoref{tab:together} - as one can see, this methodology does not produce a large enough significance (defined consistently throughout the text as $\sigma=\frac{S}{\sqrt{S+B}}$ where $S$ and $B$ refer to the number of signal and background events respectively) because one cannot eliminate the background effectively without incurring the large cost of diminishing the signal cross-section as well.\footnote{Introducing additional cuts in this example does not help much, and the significance hovers around the $3\,\sigma$-$4\,\sigma$ mark. This can also be verified later by applying a series of sequential cuts, selected based on the ranking scheme and the vertical line test(\autoref{tab:alter}), which will be explained in the upcoming section.} 

In addition to purely statistical uncertainties, there could be systematic uncertainties as well. We compute the significance calculation in the presence of systematic uncertainties (computed as $x\%$ of the background cross-section) as in~\autoref{Eq:syst} \cite{Cowan:2010js,Xia:2018kgd} denoted as $Z$. In this formula, \( S \) and \( B \) represent the number of signal and background events at a given instance. We have taken \( x = 0.1 \), corresponding to a \( 10\% \) systematic uncertainty. This quantity has also been evaluated at the end of all the cutflow charts obtained in our study and displayed in ~\autoref{tab:together}, \ref{tab:final_cutflow} and \ref{tab:alter}.

\begin{equation}
 Z = \sqrt{2}\left[(S + B)\text{ln}\left[\frac{(S + B)(B + \Delta^{2}_{B})}{B^{2} + (S + B)\Delta^{2}_{B}}\right] - \frac{B^{2}}{\Delta^{2}_{B}}\text{ln}\left[1 + \frac{\Delta^{2}_{B}S}{B(B + \Delta^{2}_{B})}\right]\right]^{\frac{1}{2}} ~~~ \text{with} ~~~ \Delta_{B} = xB.
  \label{Eq:syst}
\end{equation} 
 We note here that statistical significance is calculated either as $\sigma=\frac{S}{\sqrt{S+B}}$ (as we have done here), or in the limit that the number of background events is larger than the signal (which is often the case in particle physics experiments), as $\sigma=\frac{S}{\sqrt{B}}$. The connection between this significance with purely statistical errors, and the one in~\autoref{Eq:syst} with the systematics accounted for can be understood as follows: in the limit that there are no systematic uncertainties, i.e., $\Delta_{B}=0$,~\autoref{Eq:syst} reduces to $Z=\sqrt{2}\left[(S+B)\text{ln}\left(\frac{S+B}{B}\right)-S \right]$. In the limit that $S\ll B$, we can expand the logarithm and see that this reduces to $S/\sqrt{B}$. Thus, the formula $\sigma=S/\sqrt{B}$ is strictly only true if there are no systematic uncertainties, and the background cross-section is larger than that of the signal. We have displayed our results both for this limit (denoted as $\sigma$) and for the significance estimated in the more general way including both statistical and systematic uncertainties (denoted as $Z$).

Since we defined the objects that calculate significance and uncertainty, the observables identified as promising candidates for selection cuts, based on our inspection, are $P_T(b_1)$, $P_T(b_2)$, $H_T$, and $M(\ell_1, \ell_2, b_1, b_2, b_3, b_4)$. These observables exhibit better signal-to-background separation compared to others. Following the conventional cut-and-count strategy, we have implemented the set of selection cuts listed in \autoref{tab:together}. 

Having set up the premise and the model framework within which we will work, we now move on to the design of the algorithm and discuss various aspects of it.

\begin{center}
\begin{table}
\begin{tabular}{c c c c c } 
 \hline 
 - & $N_{signal}$ & $N_{BG}$ & $\sigma$ &  $Z$\\ [0.5ex] 
 \hline\hline
 Preselection & 677.09 & 222336 & 1.433 &  0.031\\ 
 \hline
 $PT(b_1) > 175 $ & 615.74 & 59315 & 2.515 &  0.103 \\
 \hline
 $PT(b_2) > 125 $ & 564.61 & 30644 & 3.196 &  0.188 \\
 \hline
  $H_T > 600 $ & 561.69 & 29699 & 3.229 &  0.188 \\
 \hline
 $M (\ell_1,\ell_2,b_1,b_2,b_3,b_4) > 600 $ & 558.41 & 29071 & 3.244 & 0.191 \\
 \hline

  \end{tabular}
\caption{Cut Flow chart implemented by inspection of the distributions in \autoref{tab:Figure1}.} 

\label{tab:together}
\end{table}

\end{center}

\section{Development of the Algorithm}
\label{sec:sec4}
The technique we will develop in this section is a coalition of a few simple concepts. The first step in any conventional analysis is to begin by choosing the most ``obvious" variable to put a cut on - oftentimes, this happens to be one the $p_T$'s of the jets or leptons. The subsequent steps might identify more variables that could aid in the isolation of the signal from the background usually culminating in a cut on the invariant mass of the heavy particle whose discovery is sought. While this is by no means standard, it does form a reasonable representation of the cut and count method. What we aim to do here is to optimize this method by answering the following questions in a quantitative manner:
\begin{enumerate}
\item How does one choose the kinematic variable that will maximally aid S vs B?
\item Having identified the variable, how does one choose the exact cut that will maximally isolate the signal?
\item How does one continue the process taking care to ensure the significance increases with every step?
\end{enumerate}

We will begin by introducing relevant quantities of interest that will simultaneously answer these questions. We point out at this stage that various experimental groups already have their own optimization schemes that might involve, among other things, correlations between different kinematic observables. Our goal in presenting the present algorithm is to present something, that would hopefully, standardize practices between different groups.

\subsection{The Area Parameter and Ranking}
\label{sec4:subsec1}

\begin{figure}
 \centering
  \includegraphics[width=0.5\textwidth]{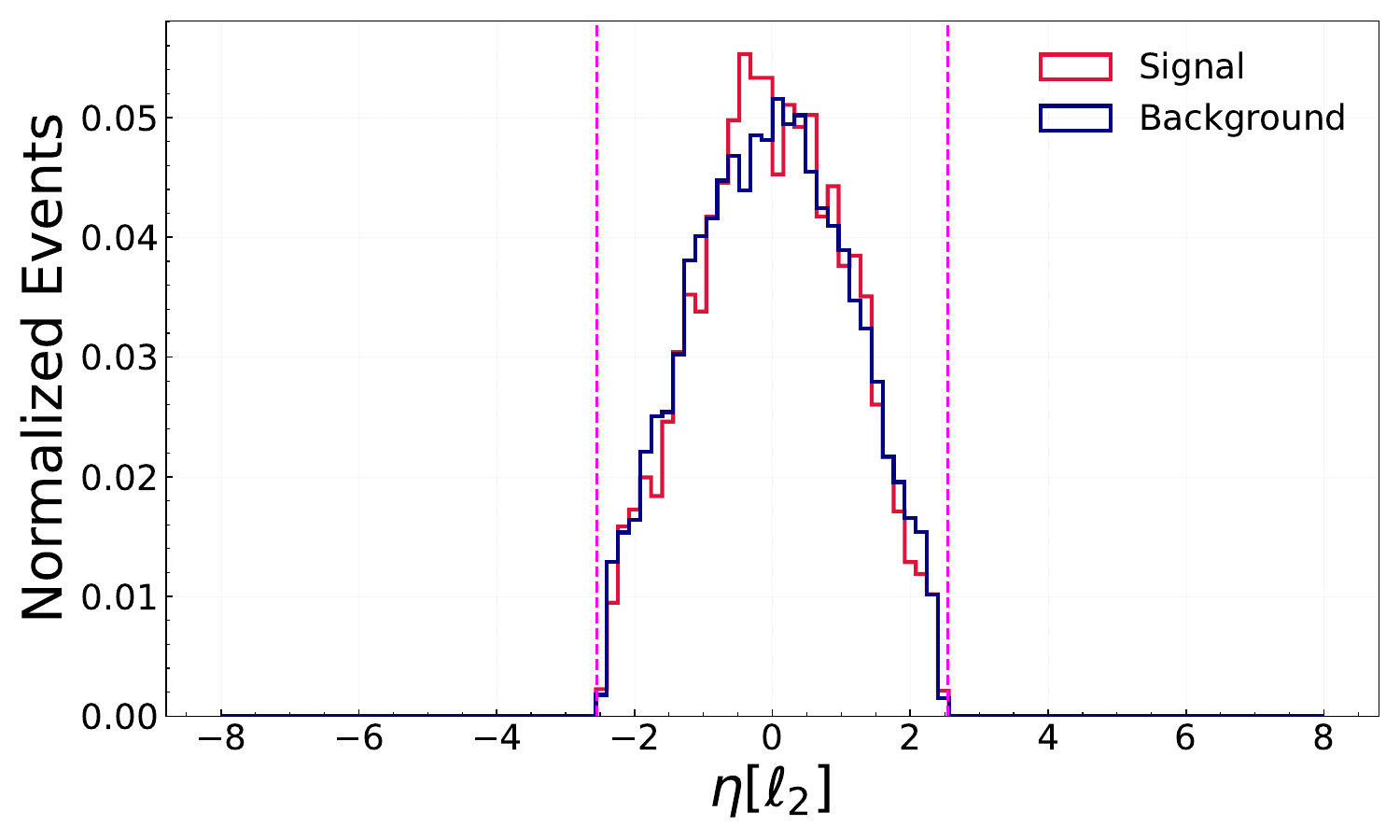}
  \includegraphics[width=0.45\textwidth]{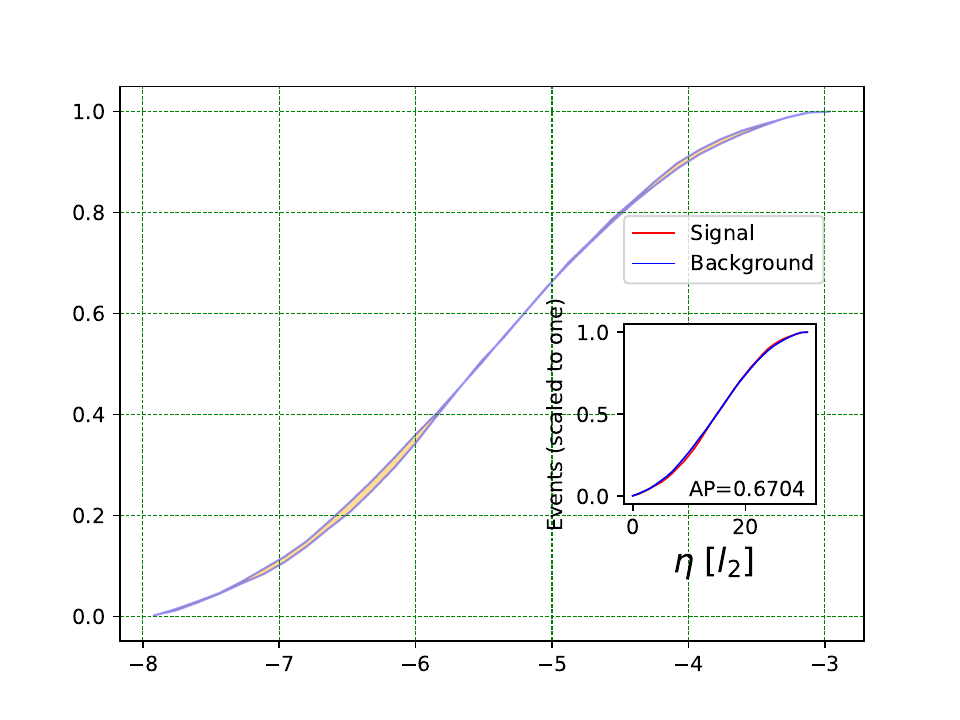}
\caption{(Left): The $\eta[\ell_2]$ distribution for the signal and background. (Right): The CDF representation of $\eta[\ell_2]$ observable after redefining the canvas of the $\eta[\ell_2]$ observable before measuring the area parameter by the omitting bins on the left and right beyond $|\eta|>2.55$. In the inset of the right panel, the x-axis corresponds to the bin index of the redefined canvas.} 
\label{tab:Figure2}      
\end{figure}

As we mentioned, we start our analysis from the normalized distributions obtained from the MadAnalysis5 - after the preselection cuts mentioned earlier - to initiate the process of identifying variables. Even though there are numerous statistical metrics available for evaluating the separation between two normalized distributions (specifically, the signal and background distributions in our case), commonly used in various applications such as Bhattacharyya Distance \cite{3142ae09-8e70-3b5c-a340-fa8eafc77ee5} and Kullback-Leibler Divergence \cite{1320776d-9e76-337e-a755-73010b6e4b64} among others, we introduce a novel parameter termed the \textbf{Area Parameter} (AP). The AP serves as a metric to quantify the extent of separation between the signal and background distributions. To compute the AP, we initially transform the normalized distribution, which is a probability distribution, into a cumulative distribution function (CDF). The CDF of a real variable $Y$ is defined as follows:

\begin{equation}
F_Y(y)=P(Y\leq y),
\label{eq:CDF_def}
\end{equation}
i.e., the CDF is a measure of the probability that the value of the variable $Y$ is less than or equal to $y$. Thus the CDF is, by definition, a curve that starts from zero and asymptotes to one, i.e., since the total probability is bounded between $0$ and $1$, the CDF is a curve whose endpoints are $0$ and $1$. The path which connects the endpoints solely depends on the probability distributions of the observable in question. Of course, if the signal and background share the exact same distribution, then their respective CDF curves will overlap, and any deviation between these curves is attributable to the difference between the signal and the background as they pertain to the observable in question. This informs us that the area enclosed by the region \emph{between} the CDF curves of the signal and the background gives a reliable measure of their separation. Calculating the AP in the occupancy percentage in the observable canvas is important. The canvas for the observable starts from the bin with the first nonzero value and ends in the bin where both signal and background CDFs converge to one. For example, in \autoref{tab:Figure2}, we display the probability distribution of the pseudorapidity $\eta$ of one of the leptons for both the signal and the background on the left and the corresponding CDF on the right. While the domain assigned to the probability distribution is $-8$ to $+8$, in the calculation of the AP, we only consider the range where the distribution is non-zero which in this case corresponds to $-2.55\leq\eta[\ell_2]\leq +2.55$. Thus, even though an overall 100 bins exist between $-8$ to $+8$ in \autoref{tab:Figure2}, only 32 of these are effective in discriminating signal and background. Thus, it is important to redefine the canvas of the observable before calculating the AP in order not to underestimate it.

The AP is calculated as the summation of the absolute difference between the signal CDF curve and the background CDF curve over all the histogram bins in the redefined canvas in percentage, i.e.,

\begin{equation}
    \textbf{AP} = \sum_{i=\rm initial \; \rm bin}^{\rm final \; \rm bin} \frac{  | \rm signal\,(i)- \rm background\,(i)|}{\rm Number \; of \; effective \; bins} * 100,
\end{equation}
where signal (i) and background (i) are the CDF values of the signal and background curves respectively in the $\rm i^{th}$ bin. 

\begin{figure}
 \centering
 \includegraphics[width=0.50\textwidth]{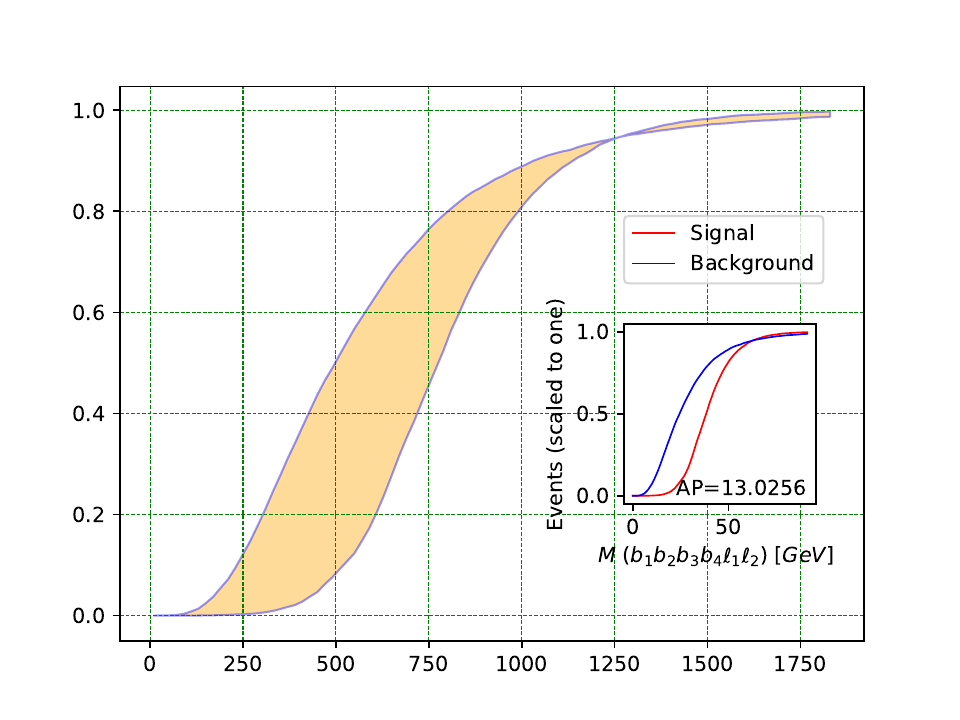}
\caption{The representation of the area parameter for the invariant mass $M (b_1,b_2,b_3,b_4,\ell_1,\ell_2)$ observable - the percentage of area enclosed by the shaded region is found to be 13.0256.}
\label{tab:invariant_mass}      
\end{figure}

In the example of $\eta[\ell_2]$ observable distribution in \autoref{tab:Figure2}, the right hand side plot shows the CDF path of signal and background in the inset. The AP is the percentage of area occupied by the shaded region between the signal and background CDFs and is found to be 0.6704 percent for this observable. Having thus obtained this metric for all possible observables listed in \autoref{sec:sec2}, our next step is to rank them based on the obtained AP values. The AP values obtained for the observables and their respective ranking over the iterations (this process will be explained shortly) are given in \autoref{tab:ranks}. It is found that observable distribution $P_T(b_2)$ has the highest AP value during the first iteration, and it is ranked 1. This observable distribution is thus highly qualified as a discriminant between signal and background and we can thus impose a cut on this variable. While this fact is indeed evident even from the normalized distribution shown in \autoref{tab:Figure1}, the AP allows us to quantify the selection based on a number, which could be more meaningful in some non-trivial selections which we will encounter in the upcoming iterations during which the signal and background curves start getting closer to each other. As another example, the AP corresponding to the observable $M (b_1,b_2,b_3,b_4,\ell_1,\ell_2)$ is calculated and displayed in  \autoref{tab:invariant_mass} - we can contrast this with \autoref{tab:Figure2} to understand that this is a much better variable to put a cut on compared to $\eta[\ell_2]$. The AP in both cases is a measure of the shaded area between the two curves.

Having thus ranked all the observables, we will select the rank 1 observable for imposing the cut and keep the remaining observables on hold - those on hold could still be considered in the subsequent steps.  At this juncture, we need to determine what exactly are the cuts to be imposed on this variable before moving to the next stage of the analysis.

\definecolor{LightCyan}{rgb}{0.88,1,1}

\begin{table*}[t]
\begin{adjustbox}{width=1\textwidth}
  \centering
  \begin{tabular}{ccccccccccccccccccccccccc}
 Variable &  I1:AP & I1:Rank & I2:AP & I2:Rank & I3:AP & I3:Rank & I4:AP & I4:Rank & I5:AP & I5:Rank & I6:AP & I6:Rank & I7:AP & I7:Rank &I8:AP & I8:Rank &I9:AP & I9:Rank &I10:AP & I10:Rank &I11:AP & I11:Rank &I12:AP & I12:Rank  \\ [0.1ex] 
 \hline\hline

$P_T(\ell_1)$  & 3.3617 & 18  & 5.9894 & 6 & 5.9864 & 7 & 6.3205 & 7 &  5.9395 &  5 & 6.1900 & 3 & 13.7353 & 1 & -- & -- & -- & -- & -- & -- & -- & -- & -- & -- \\
\hline

$P_T(\ell_2)$  & 2.1206 & 24 & 3.7197 & 19 & 3.7019 & 17 & 3.7105 & 16 & 3.8176 & 12 & 3.4450 & 13 & 13.1806 & 2 & 5.1624 & 8 & 4.1243 & 9 & 3.1524 & 12 & 3.9961 & 13 & 2.8905 & 15  \\
\hline

$P_T(b_1)$  & 16.0185 & 4 & 7.6891 & 3 & 6.7966 & 4 & 6.7422 & 5 & 10.1955 & 1 & -- & -- & -- & -- & -- & -- & -- & -- & -- & -- & -- & -- & -- & --  \\
\hline

$P_T(b_2)$  & 17.7135 & 1 & -- & -- & -- & -- & -- & -- & -- & -- & -- & -- & -- & -- & -- & -- & -- & -- & -- & -- & -- & -- & -- & --  \\
\hline

$P_T(b_3)$  & 16.5329 & 2 & 10.2161 & 1 & -- & -- & -- & -- & -- & -- & -- & -- & -- & -- & -- & -- & -- & -- & -- & -- & -- & -- & -- & -- \\
\hline

$P_T(b_4)$  & 13.2858 & 6 & 10.0154 & 2 & 8.5847 & 1 & -- & -- & -- & -- & -- & -- & -- & -- & -- & -- & -- & -- & -- & -- & -- & -- & -- & -- \\
\hline

$\eta(\ell_1)$  & 1.1868 & 28 & 0.2464 & 28 & 0.4149 & 27 & 0.7917 & 26 & 0.8998 & 25 & 0.8591 & 24 & 4.8328 & 19 & 0.9350 & 21 & 1.7119 & 17 & 1.9367 & 19 & 3.0777 & 14 & 2.3573 & 18  \\
\hline

$\eta(\ell_2)$  & 0.6704 & 29 & 0.5063 & 27 & 1.2176 & 26 & 1.9345 & 24 & 1.7759 & 20 & 0.9485 & 23 & 3.4901 & 23 & 1.0943 & 20 & 2.8702 & 11 & 4.0199 & 6 & 4.7926 & 8 & 4.9241 & 6  \\
\hline

$\eta(b_1)$  & 2.3921 & 21 & 2.6296 & 22 & 2.7370 & 20 & 2.8195 & 19 & 1.4721 & 21 & 2.1279 & 19 & 6.9513 & 13 & 2.1377 & 16 & 2.1910 & 14 & 1.9653 & 18 & 4.7043 & 10 & 4.0144 & 11  \\
\hline

$\eta(b_2)$  & 2.2355 & 23 & 1.7256 & 25 & 1.8816 & 24 & 2.2384 & 23 & 1.2321 & 22 & 1.4690 & 21 & 6.1015 & 16 & 2.0969 & 17 & 1.6298 & 19 & 2.9772 & 14 & 2.3579 & 18 & 4.2431 & 10  \\
\hline

$\eta(b_3)$  & 1.8940 & 25 & 1.8969 & 24 & 1.9303 & 23 & 2.2504 & 22 & 1.0828 & 24 & 1.0468 & 22 & 4.5020 & 20 & 1.3057 & 19 & 1.7532 & 16 & 2.0066 & 17 & 2.8563 & 15 & 3.3417 & 13  \\
\hline

$\eta(b_4)$  & 1.5189 & 27 & 1.3898 & 26 & 1.4532 & 25 & 1.6287 & 25 & 1.1174 & 23 & 1.7669 & 20 & 3.5423 & 22 & 0.8084 & 22 & 1.2895 & 20 & 2.0379 & 16 & 5.7876 & 4 & 6.1115 & 3  \\
\hline

$\Delta R (\ell_1,\ell_2)$  & 2.3455 & 22 & 1.9777 & 23 & 2.5584 & 21 & 2.8273 & 18 & 3.2236 & 15 & 3.4891 & 12 & 4.4359 & 21 & 3.4875 & 11 &7.3403 & 2 & 5.0143 & 5 & 5.1973 & 6 & 3.2046 & 14  \\
\hline

$\Delta R (b_1,b_2)$  & 4.5723 & 14 & 3.6682 & 20 & 3.8640 & 16 & 3.9813 & 14 & 3.0608 & 17 & 2.4316 & 16 & 5.1544 & 18 & 2.2061 & 14 & 2.8254 & 12 & 3.3804 & 11 & 4.6761 & 11 & 4.3513 & 9  \\
\hline

$\Delta R (b_1,b_3)$  & 4.2727 & 15 & 5.4252 & 9 & 6.3944 & 6 & 6.7504 & 4 & 5.3403 & 7 & 5.0791 & 8 & 9.5012 & 9 & 5.2139 & 7 & 4.7692 & 7 & 3.4846 & 10 & 2.7880 & 16 & 2.8328 & 16  \\
\hline

$\Delta R (b_1,b_4)$  & 3.8570 & 17 & 4.7790 & 12 & 5.7117 & 9 & 6.7196 & 6 & 5.2007 & 10 & 5.3312 & 6 & 10.2961 & 5 & 6.6470 & 3 & 6.7573 & 3 & 7.8569 & 2 & 7.1543 & 2 & 8.6001 & 1  \\
\hline

$\Delta R (b_2,b_3)$  & 5.7210 & 13 & 7.1856 & 4 & 7.9829 & 3 & 8.1413 & 2 & 6.8821 & 3 & 7.4668 & 2 & 11.8727 & 3 & 7.5893 & 1 & -- & -- & -- & -- & -- & -- & -- & -- \\
\hline

$\Delta R (b_2,b_4)$  & 3.8667 & 16 & 4.5096 & 15 & 5.0776 & 11 & 5.6453 & 10 & 5.2806 & 9 & 6.0250 & 4 & 10.0965 & 7 & 5.6649 & 4 & 4.9146 & 6 & 5.9851 & 3 & 4.7311 & 9 & 2.7679 & 17  \\
\hline

$\Delta R (b_3,b_4)$  & 2.7495 & 19 & 2.9393 & 21 & 3.4689 & 18 & 3.7238 & 15 & 2.6281 & 18 & 2.2612 & 17 & 6.9512 & 14 & 2.3649 & 13 & 1.6574 & 18 & 3.6952 & 8 & 2.2928 & 19 & 3.5267 & 12 \\
\hline

$THT$  & 16.1644 & 3 & 3.9193 & 18 & 2.4791 & 22 & 2.3578 & 21 & 1.8862 & 19 & 2.1712 & 18 & 7.4464 & 11 & 2.1789 & 15 & 2.5088 & 13 & 2.6384 & 15 & 4.2212 & 12 & 4.6349 & 7 \\
\hline

$\cancel{E}_T$  & 1.7355 & 26 & 5.7884 & 7 & 6.6667 & 5 & 7.1572 & 3 & 7.2595 & 2 & 10.2216 & 1 & -- & -- & -- & -- & -- & -- & -- & -- & -- & -- & -- & -- \\
\hline

$M (\ell_1,\ell_2)$  & 2.6089 & 20 & 4.2209 & 17 & 4.1303 & 14 & 4.4462 & 12 & 3.3314 & 14 & 2.6968 & 14 & 11.3749 & 4 & 2.6294 & 12 & 1.2051 & 21 & 1.6969 & 20 & 2.4345 & 17 & 4.3776 & 8  \\
\hline

$M (b_1,b_2)$  & 14.7516 & 5 & 4.5795 & 14 & 4.2270 & 13 & 4.2277 & 13 & 6.7322 & 4 & 5.0024 & 9 & 5.8740 & 17 & 4.4535 & 9 & 3.9830 & 10 & 3.9448 & 7 & 5.8002 & 3 & 5.5309 & 5  \\
\hline

$M (b_1,b_3)$  & 9.9225 & 8 & 4.6414 & 13 & 4.3594 & 12 & 4.7321 & 11 & 3.0780 & 16 & 2.6530 & 15 & 6.9374 & 15 & 1.9113 & 18 & 2.1463 & 15 & 3.0944 & 13 & 4.9090 & 7 & 6.5611 & 2  \\
\hline

$M (b_1,b_4)$  & 7.8832 & 10 & 5.5992 & 8 & 5.8642 & 8 & 6.1065 & 8 & 5.9306 & 6 & 5.3838 & 5 & 10.1575 & 6 & 5.4703 & 6 & 6.5688 & 4 & 8.4909 & 1 & -- & -- & -- & --  \\
\hline

$M (b_2,b_3)$  & 9.0154 & 9 & 5.2798 & 10 & 5.3685 & 10 & 5.7117 & 9 & 5.2833 & 8 & 5.1407 & 7 & 9.8008 & 8 & 5.6536 & 5 & 7.3876 & 1 & -- & -- & -- & -- & -- & -- \\
\hline

$M (b_2,b_4)$  & 7.6765 & 11 & 4.3867 & 16 & 4.0921 & 15 & 3.4960 & 17 & 3.9514 & 11 & 3.7447 & 11 & 7.1702 & 12 & 4.1541 & 10 & 5.1734 & 5 & 5.3406 & 4 & 8.2886 & 1 & -- & -- \\
\hline

$M (b_3,b_4)$  & 7.2371 & 12 & 5.1073 & 11 & 3.3353 & 19 & 2.5889 & 20 & 3.7302 & 13 & 3.8932 & 10 & 8.0344 & 10 & 7.3550  & 2 & 4.6694 & 8 & 3.4975 & 9 & 5.4720 & 5 & 5.8167 & 4   \\
\hline

$M(bbbb\ell\ell)$  & 13.0256 & 7 & 6.4208 & 5 & 8.5071 & 2 & 9.7585 & 1 & -- & -- & -- & -- & -- & -- & -- & -- & -- & -- & -- & -- & -- & -- & -- & -- \\
\hline

\end{tabular}
\end{adjustbox}
\caption{The AP values of the listed 29 variables and their rankings at each iteration.}
\label{tab:ranks}
\end{table*}

\subsection{Vertical Line Test and The Cut Choice}
\label{sec4:subsec2}

\begin{figure}[H]
 \centering
  \includegraphics[width=0.60\textwidth]{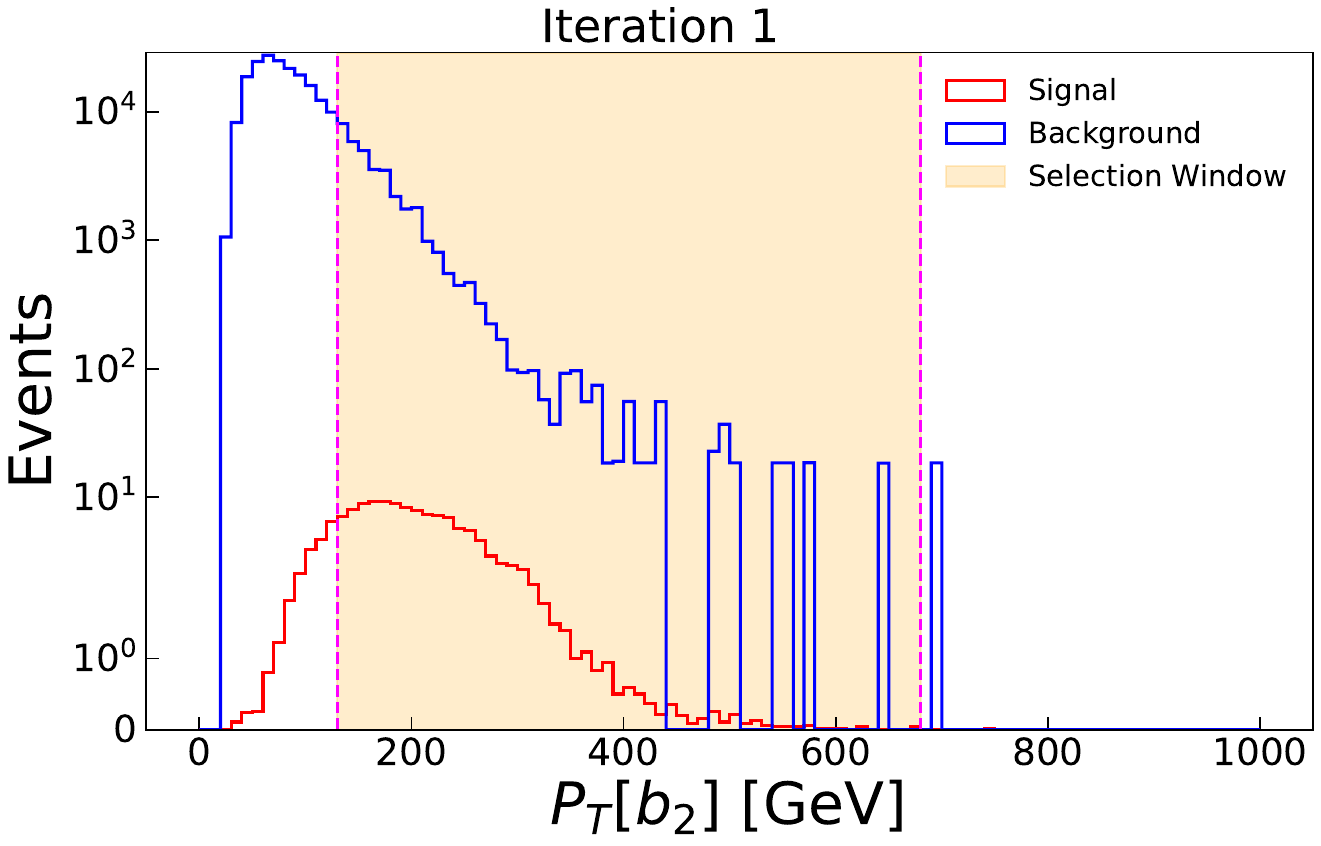}
\caption{Vertical line test carried over the rank 1 $P_T(b_2)$ observable during the first iteration - the suggested selection cut is shown between the two dashed lines.}
\label{tab:optimisation}      
\end{figure}

The conventional way of imposing cuts on an observable involves an educated guess based on what the underlying model is and where one expects the signal and background to be maximal and then by simply looking at the distributions and deciding based on their shapes. While this process certainly works, it is not guaranteed that one can identify the cut (or a set of cuts) that will certainly maximize S vs B. Herein, we seek to simply extend this method in a democratic fashion by introducing the Vertical Line Test - this involves fixing two vertical lines at the ends of the distribution and scanning the entire parameter space by changing the positions of these lines while calculating $S/\sqrt{S+B}$ for all possible cases by simply counting the signal and background events between these two vertical lines. If there are 100 bins in a histogram, then there are 4950 possible configurations of vertical lines to consider and out of these many regions, the one with the highest significance will be considered the optimal choice. While this process is certainly more computationally intensive than simply choosing one cut, it has the advantage that it can isolate the most interesting region of phase space in cases where one cannot isolate obvious bumps or other features in the distributions thus providing a neat handle on the hunt for new physics. While doing the test, it is also important to keep sufficient number of signal events while focusing on maximum significance - we have imposed the condition that the number of signal events should be a minimum of 10.\footnote {This number can, of course be modified depending on the nature of the phenomenological search.}  In the present case, the observable $P_T(b_2)$ turns out to have rank one during Iteration 1, and the suggestion from the optimization method is to apply a selection cut on the region between the vertical lines at 130 and 990 - this is displayed in \autoref{tab:optimisation}. Having thus imposed a cut on this kinematic variable, we will drop it from further analysis - as can be seen from \autoref{tab:ranks}, $P_T(b_2)$ has served its purpose at the stage of the first iteration.

Given that one would typically need multiple cuts on the events to isolate the signal from the background to effect a discovery process, the next step is to continue this process. Herein, one needs to address the question raised in the introduction to this article: in general, cutting portions of phase space will affect the distributions of the other kinematic variables as well and one needs to be cognizant of the fact that decisions on suitable cuts of these variables based on the shapes of original distributions might not be optimal anymore.\footnote{We stress here that these cuts might still \emph{work}, i.e., they might further refine the $S/\sqrt{B}$, however our point here is that they might not necessarily be the \emph{optimal} ones to choose.} While a detailed study of the correlations between all possible observables is beyond the scope of this article, we adopt the simple solution of recalculating the distributions and the associated AP of all the variables after the first cut has been imposed. This is reflected in the column ``Iteration 2" in \autoref{tab:ranks} for the example at hand. It is seen from that table that the variable exhibiting the next highest degree of separation as dictated by the area parameter is $P_T(b_3)$. We then run the vertical line test for this observable and design a suitable cut on it. In the next section, we summarize the methodology of this iterative process.

\subsection{Iterations and The Cut Flowchart}
\label{subsec:Iteration}

Having thus imposed a cut on the second observable, our significance is likely to change. While we could simply progress to the next step, herein we introduce a choice: the decision on whether or not to include this observable in our phenomenological studies depends on how much the significance has changed. If, for example, the significance comes down, this step is not of much use. If the significance does increase but not by a lot, we again have the freedom to either use or disregard this observable. In this work, we follow the (admittedly subjective) criterion that if the significance at any particular step, say $\sigma(k)$, is greater than the previous step $\sigma(k-1) $ by 0.10, i.e., if  $\sigma(k) > \sigma(k-1) + 0.10\, \sigma $, then that particular selection cut is accepted and we drop that observable from further analysis. This condition is adaptive to the analysis of our interest. In general, the condition can be formulated as $\sigma(k) > \sigma(k-1) + \Delta\sigma$. This condition is imposed to ensure a bare minimum improvement in each step of the algorithm for faster convergence towards a $5\sigma$ discovery potential. For this analysis, we have taken $\Delta\sigma = 0.10$ and applied a selection cut based on the satisfaction of this condition. Any other value could have been chosen - however, there is a risk that the algorithm may get stuck if none of the observables in the ranking order provide a selection cut suggestion that satisfies the bare minimum condition if it is chosen too high. In such a scenario, it is adaptive to choose a lower value of $\Delta\sigma$ to continue the progression of the methodology. In addition, we also require that if  $\sigma(k) \geq 5 \sigma$ for any $k$, the analysis stops at that point as the desired significance is reached (we call this the Lower Limiting Condition or the LL condition in what follows). 

\begin{figure*}
\centering
\includegraphics[width=1\textwidth]{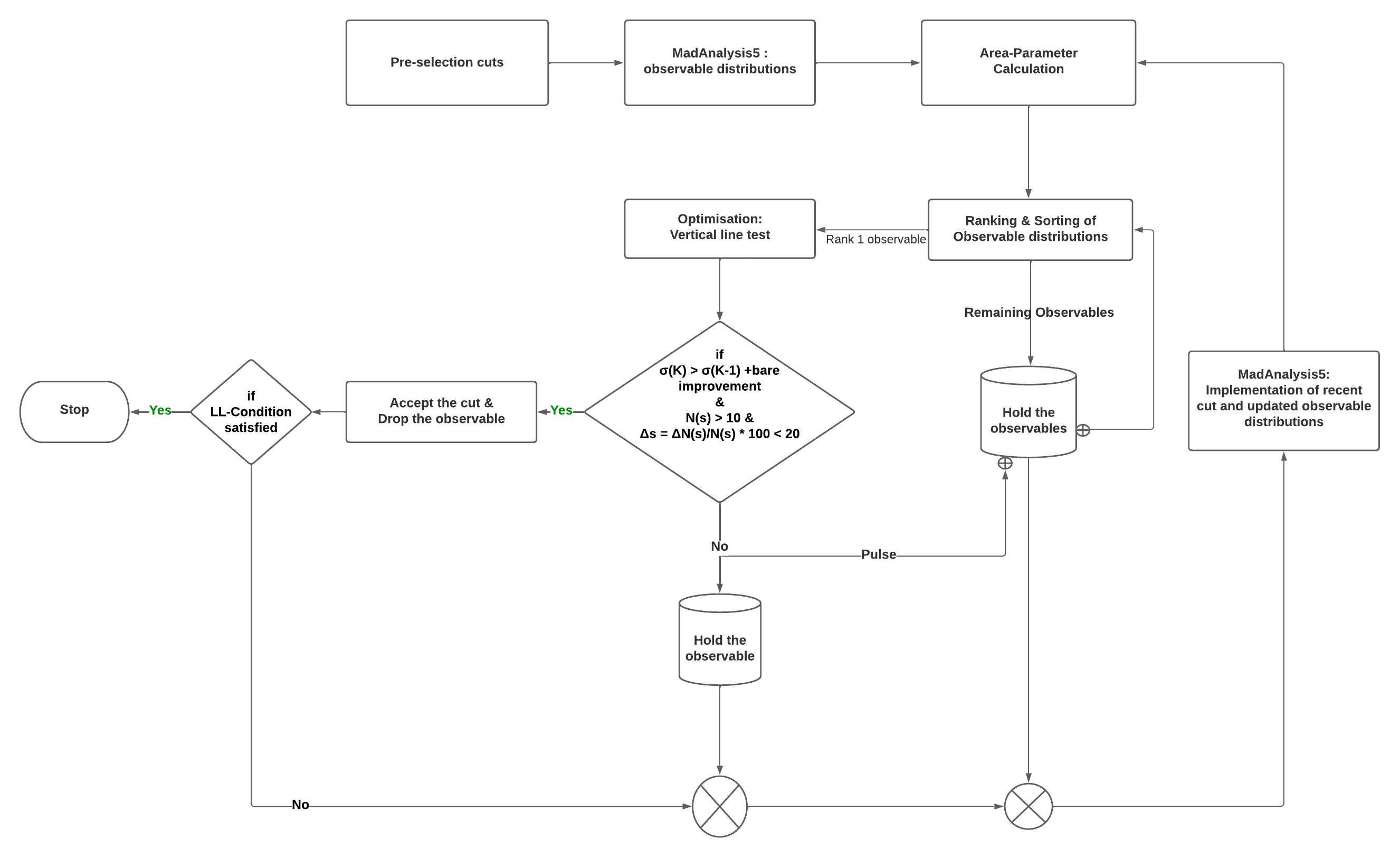}
\caption{ A flowchart representation of the proposed algorithm.}
\label{tab:flowchart} 
\end{figure*}

\begin{figure}
 \centering
  \includegraphics[width=0.17\textwidth]{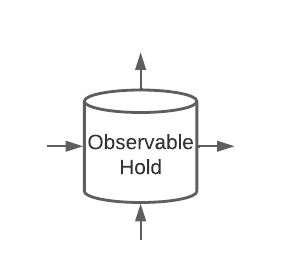}
  \includegraphics[width=0.17\textwidth]{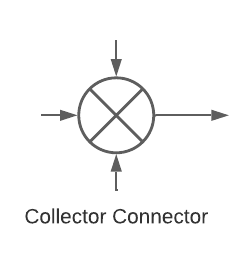}
  \includegraphics[width=0.17\textwidth]{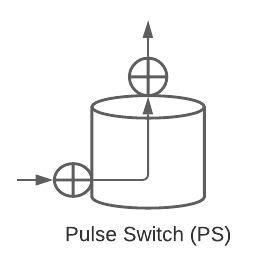}
\caption{ The three important components in the flowchart.}
\label{tab:componets}      
\end{figure}

If, on the other hand, the significance at step $k$ is such that $\sigma(k) < \sigma(k-1) + 0.10$, then we conclude that the significance has not improved much from the suggested cut from the vertical line test and so we keep that particular observable in hold (as it might well become important down the line in the analysis). We then proceed with the analysis picking the next variable with the highest AP. To understand the algorithm visually, it might be useful to represent the methodology as a flowchart that summarizes it in its entirety - this is shown in \autoref{tab:flowchart}. The different components in the flowchart are magnified in \autoref{tab:componets}.

Below, we summarize the features of this flowchart as a way to recap of the essential features of the proposed algorithm.

\begin{figure*}
 \centering
 \includegraphics[width=0.30\textwidth]{./Plots/Iteration1.pdf}
 \includegraphics[width=0.30\textwidth]{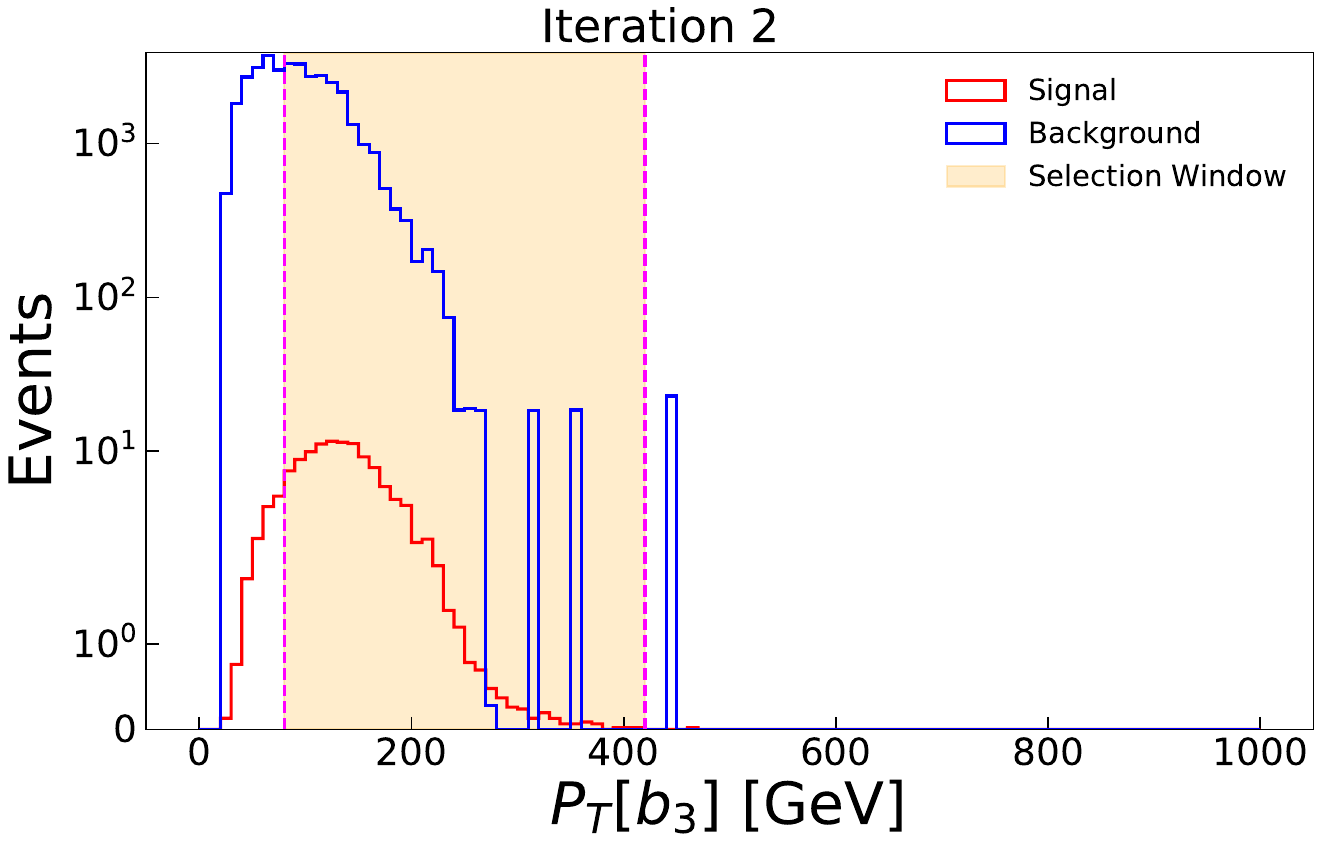}
 \includegraphics[width=0.30\textwidth]{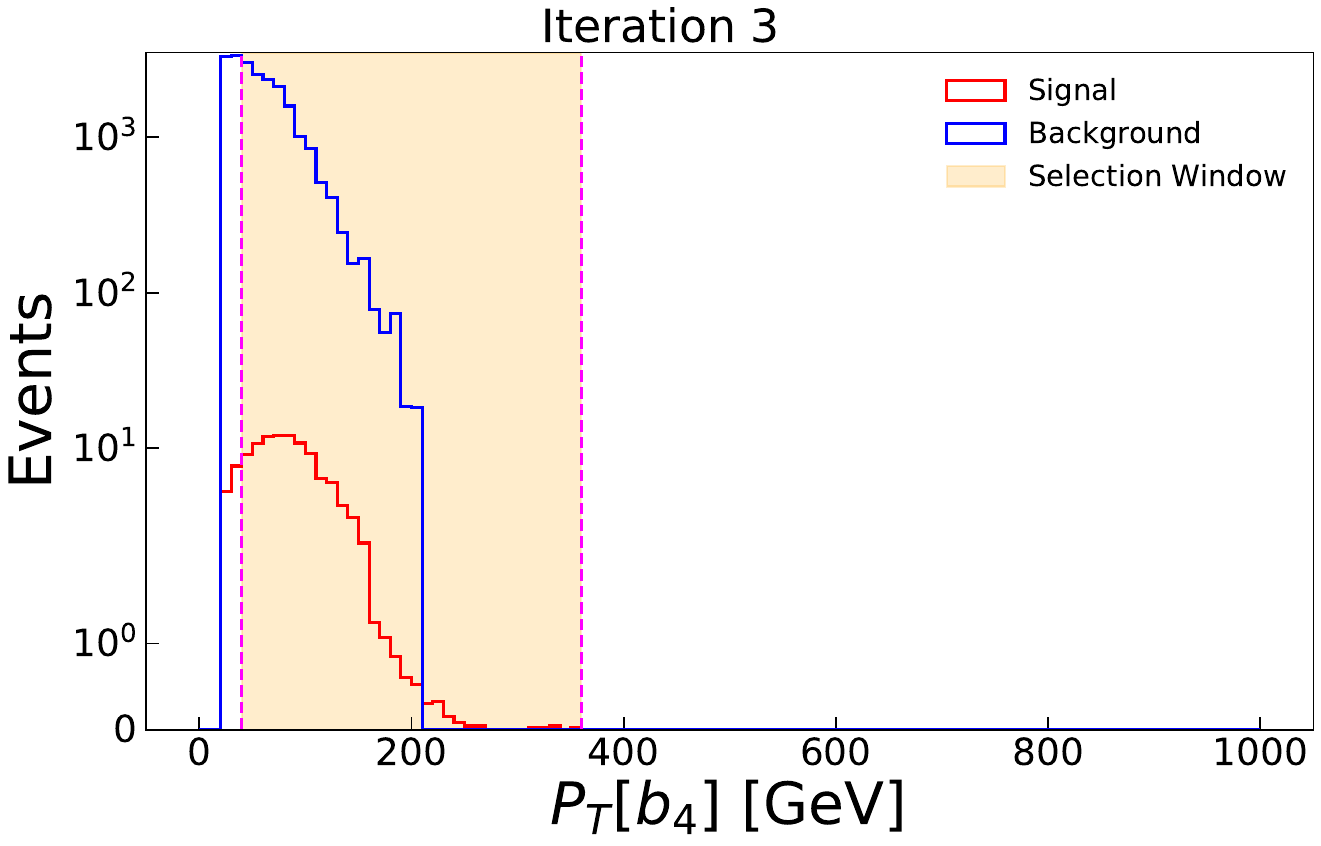}
 \includegraphics[width=0.30\textwidth]{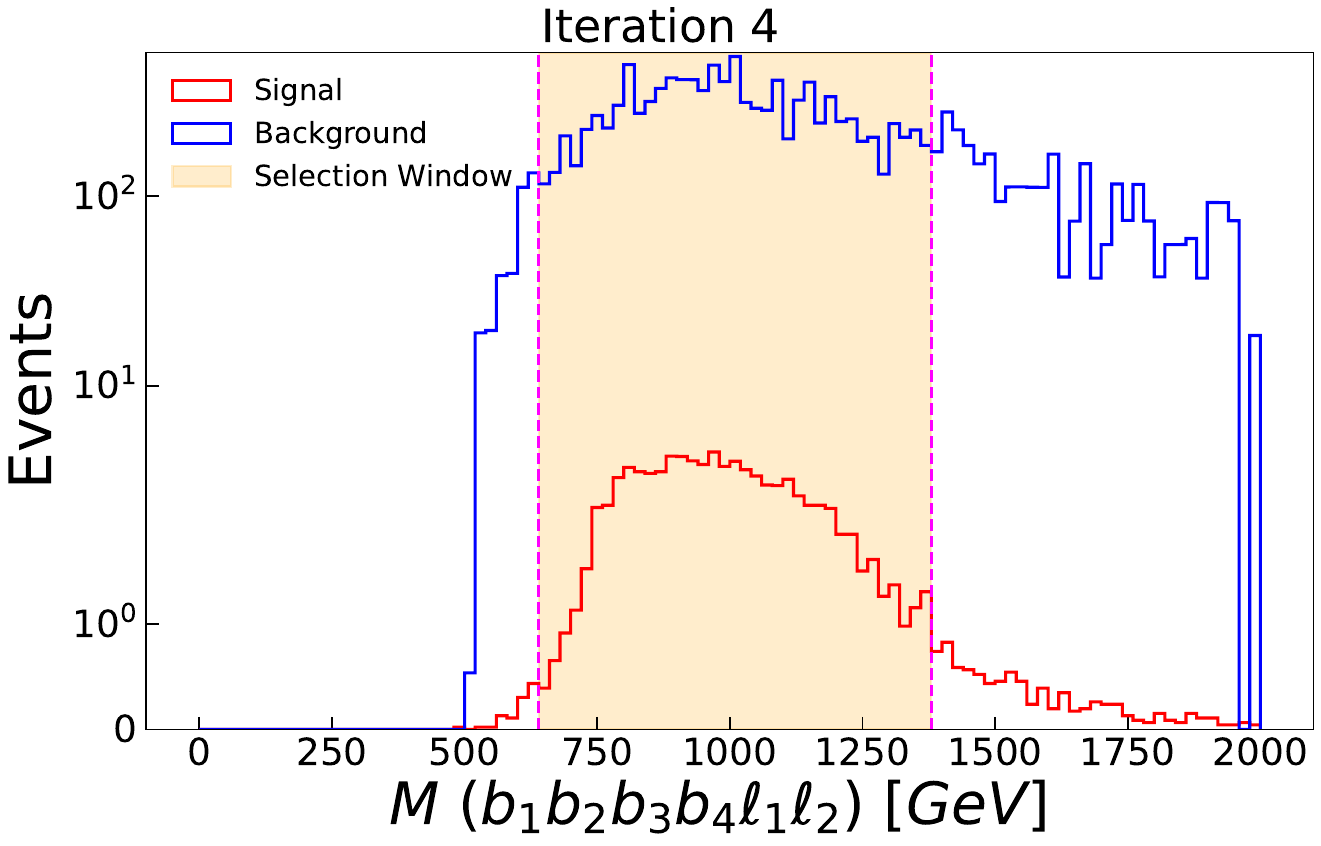}
 \includegraphics[width=0.30\textwidth]{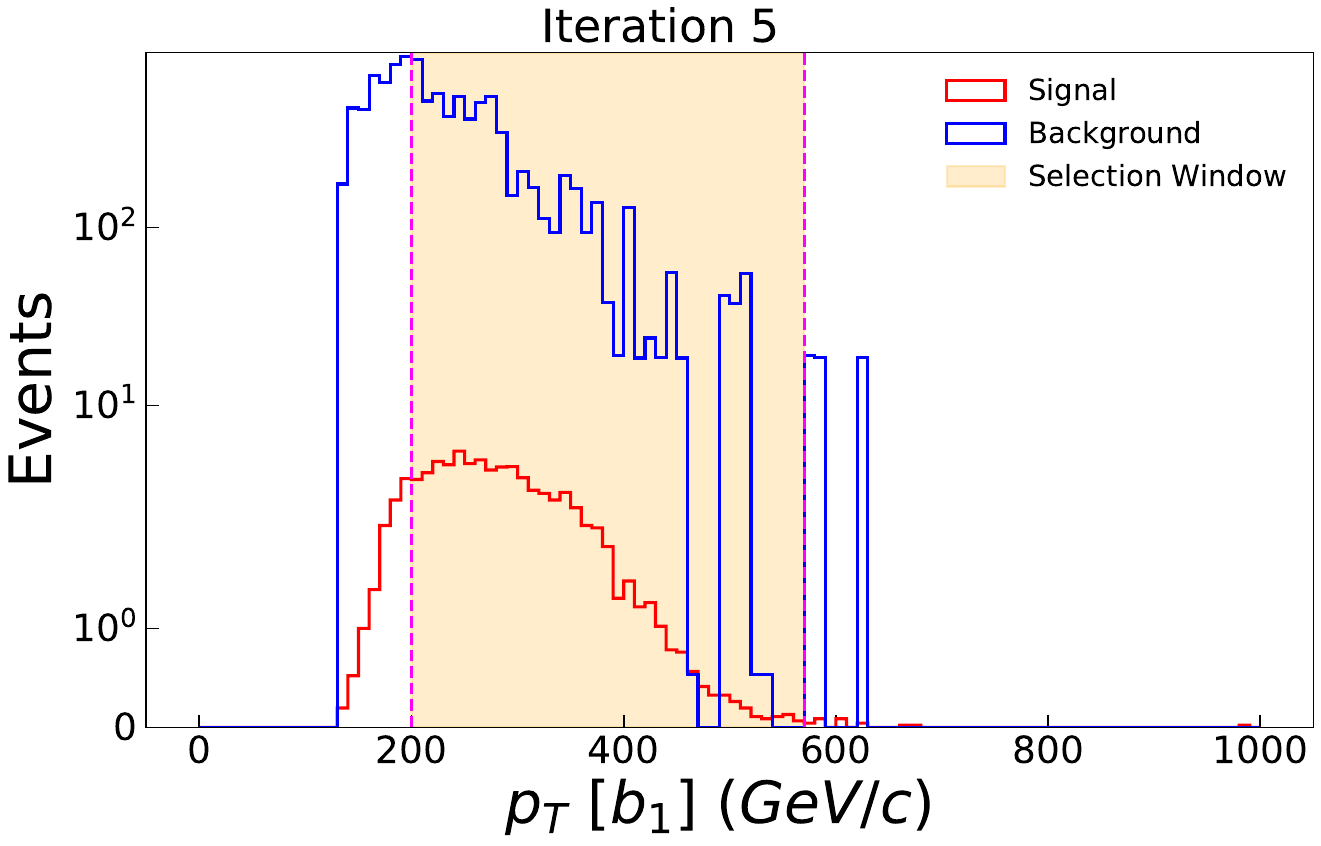}
 \includegraphics[width=0.30\textwidth]{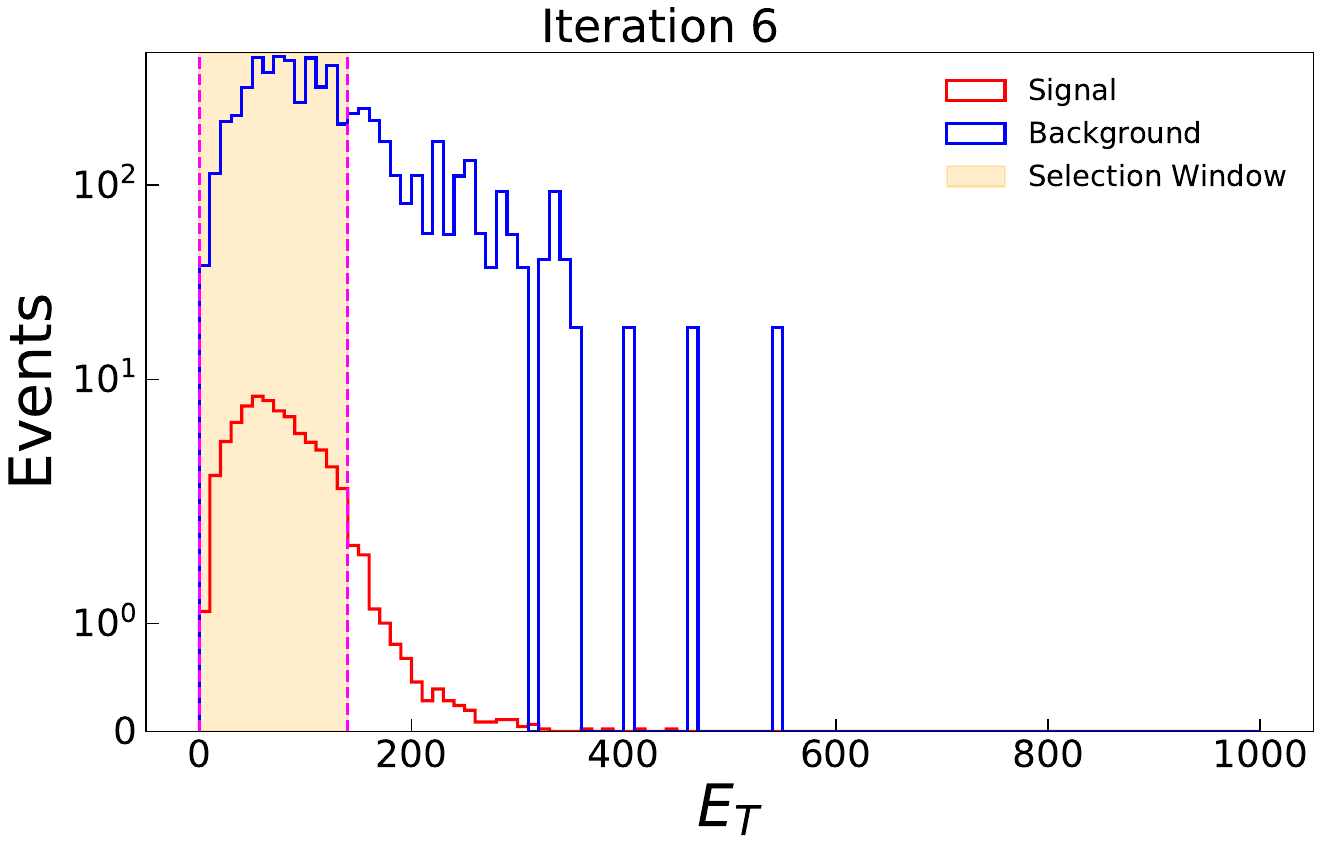}
 \includegraphics[width=0.30\textwidth]{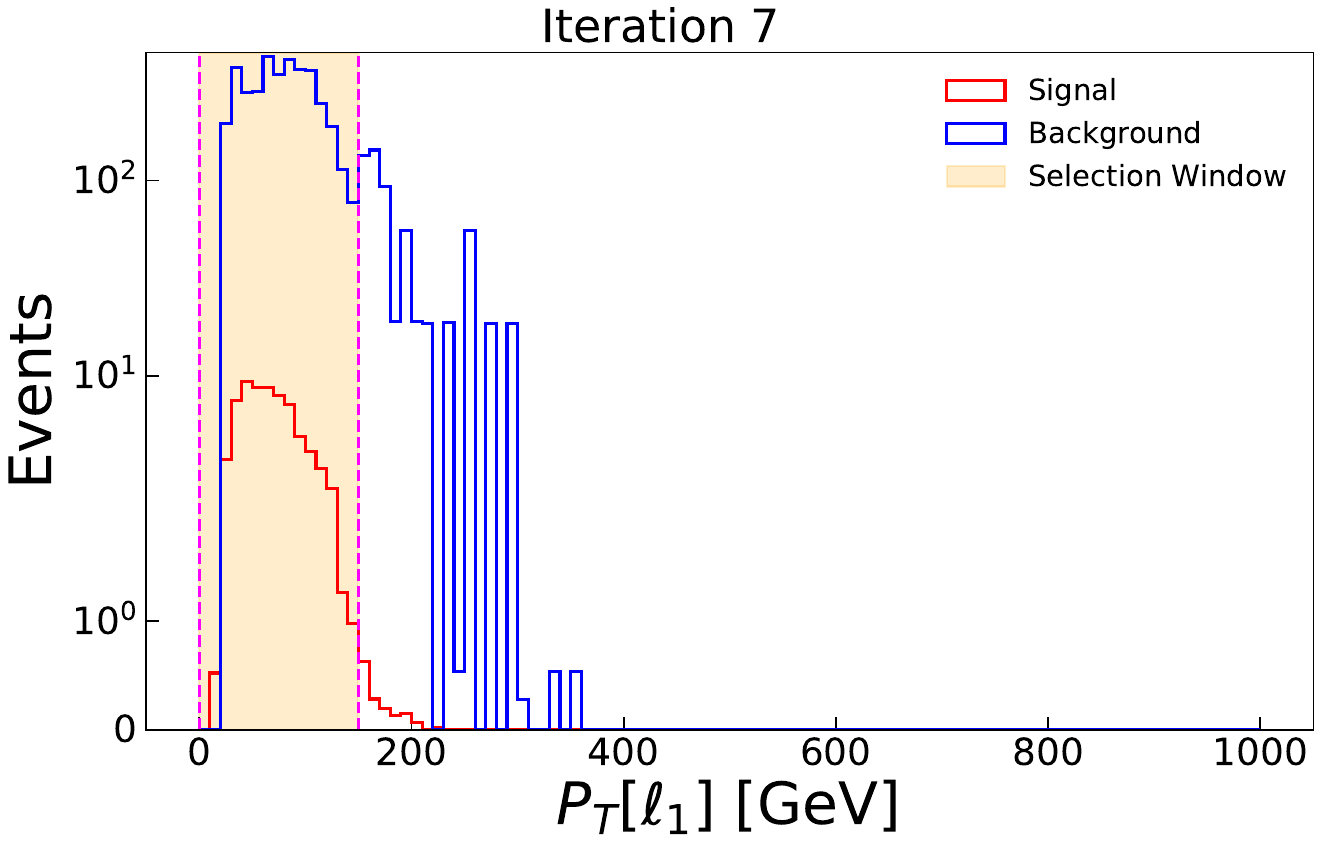}
  \includegraphics[width=0.30\textwidth]{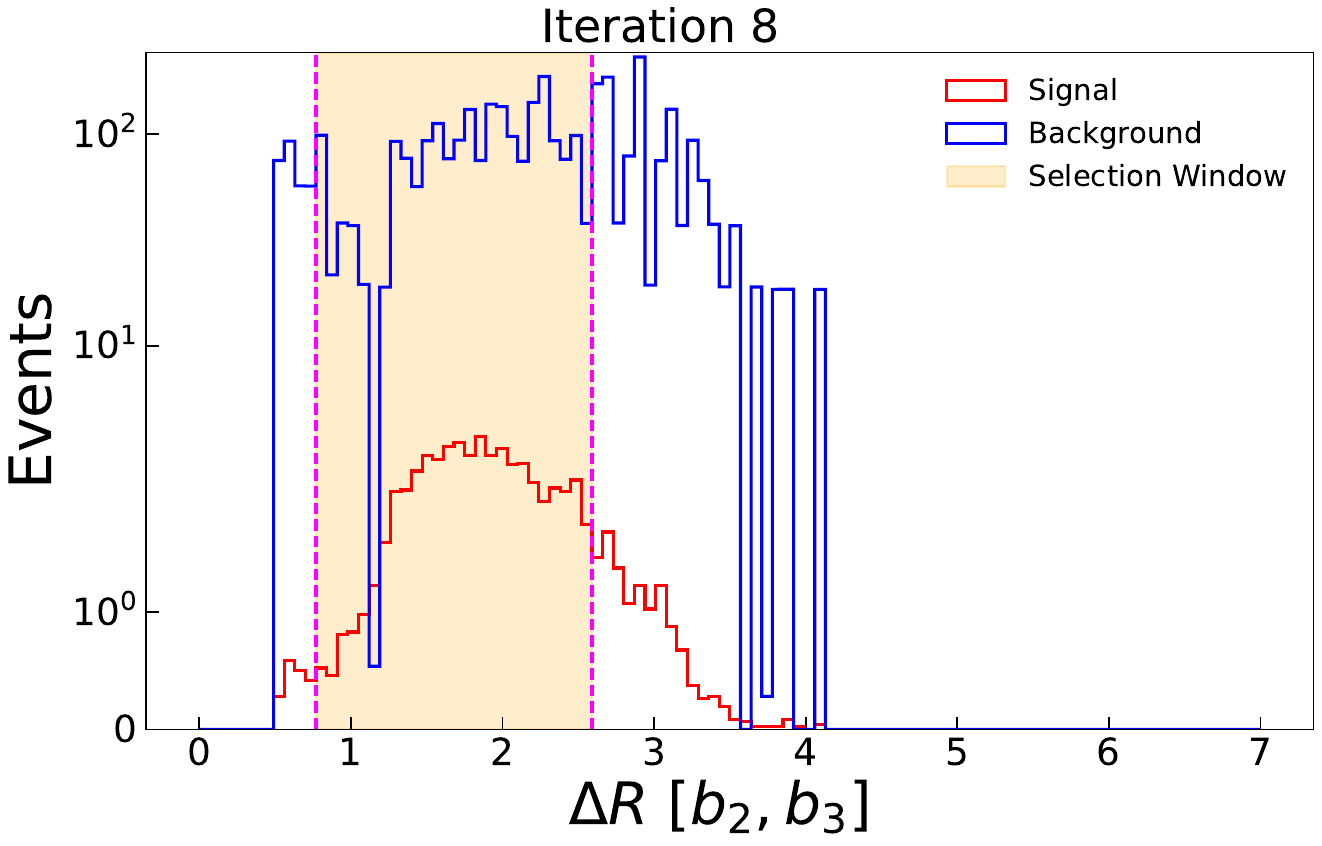}
  \includegraphics[width=0.30\textwidth]{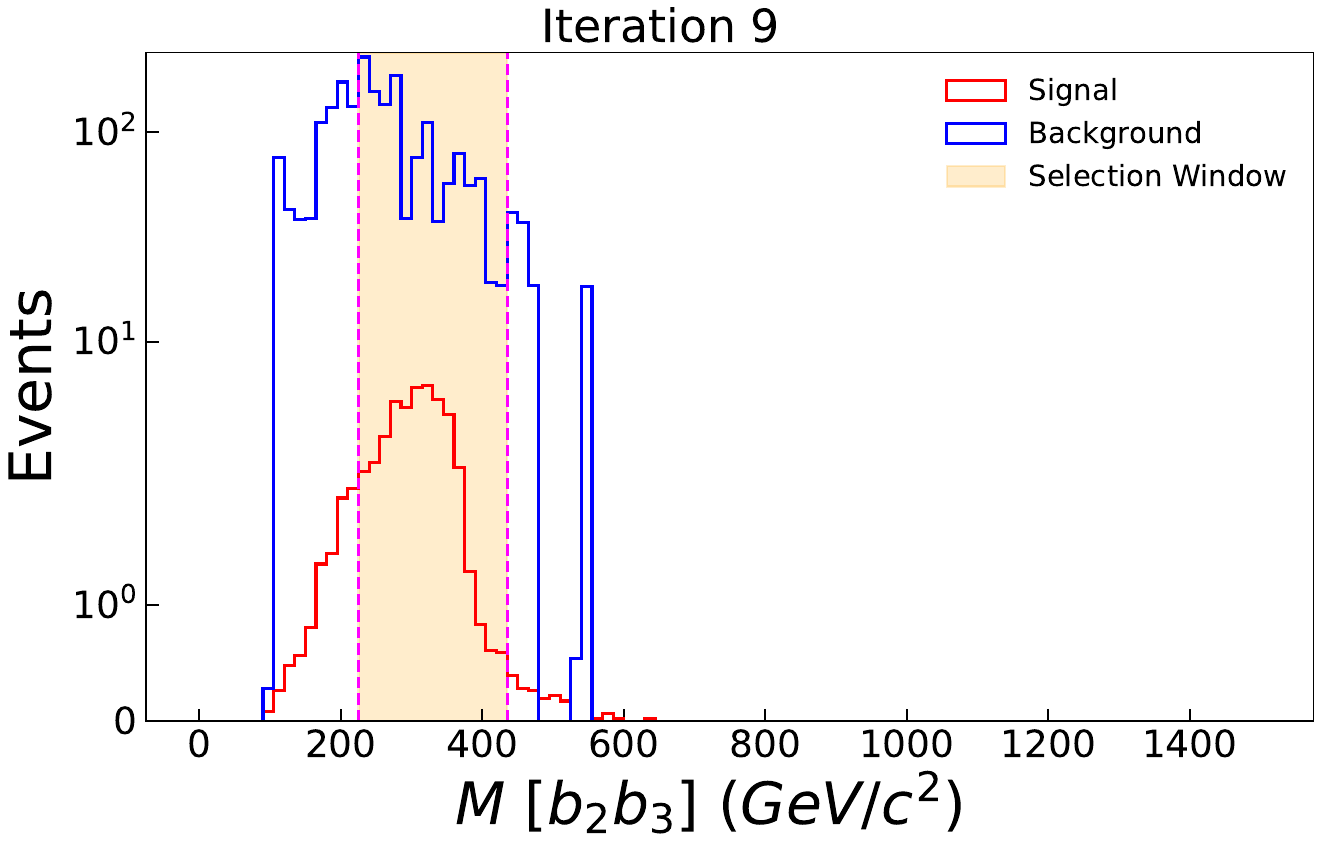}
 \includegraphics[width=0.30\textwidth]{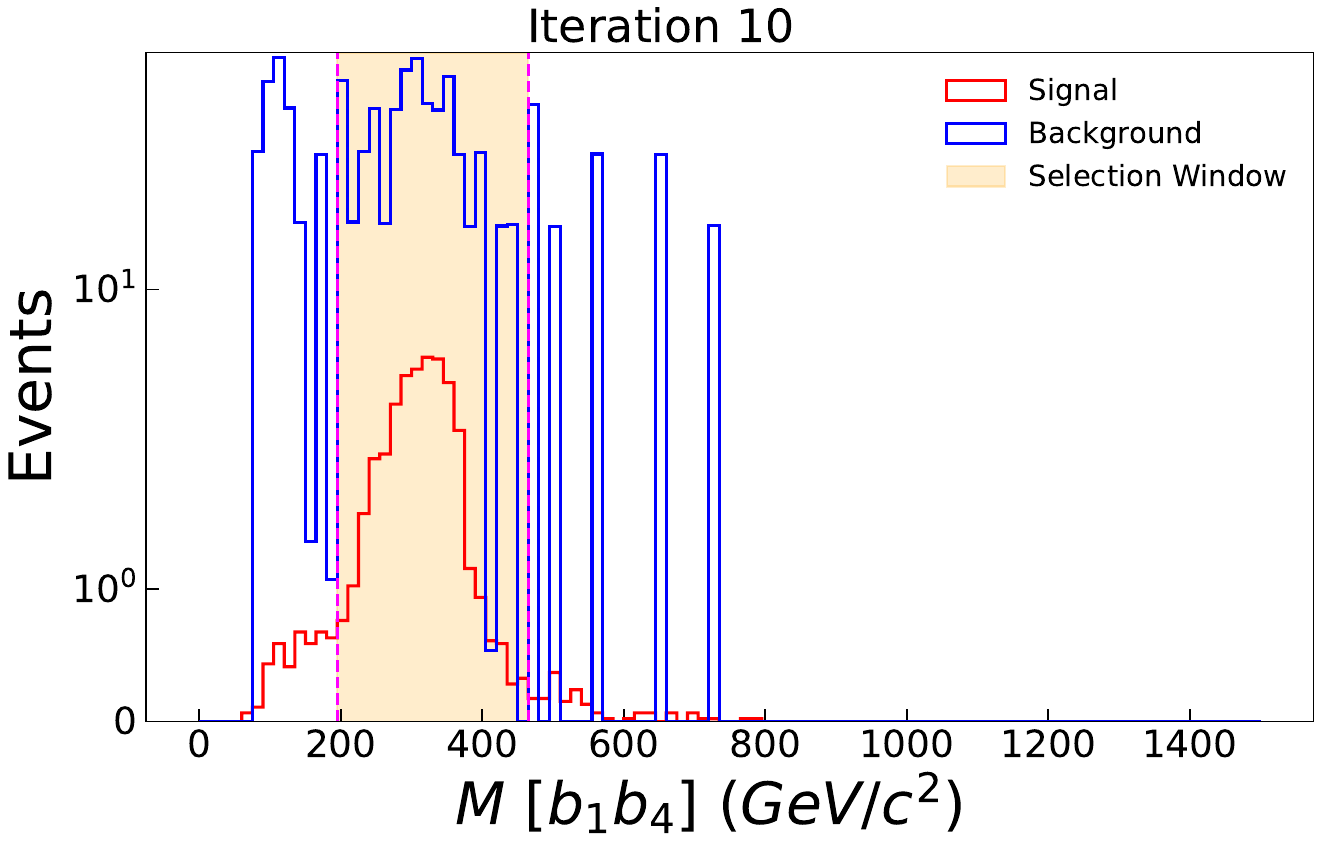}
  \includegraphics[width=0.30\textwidth]{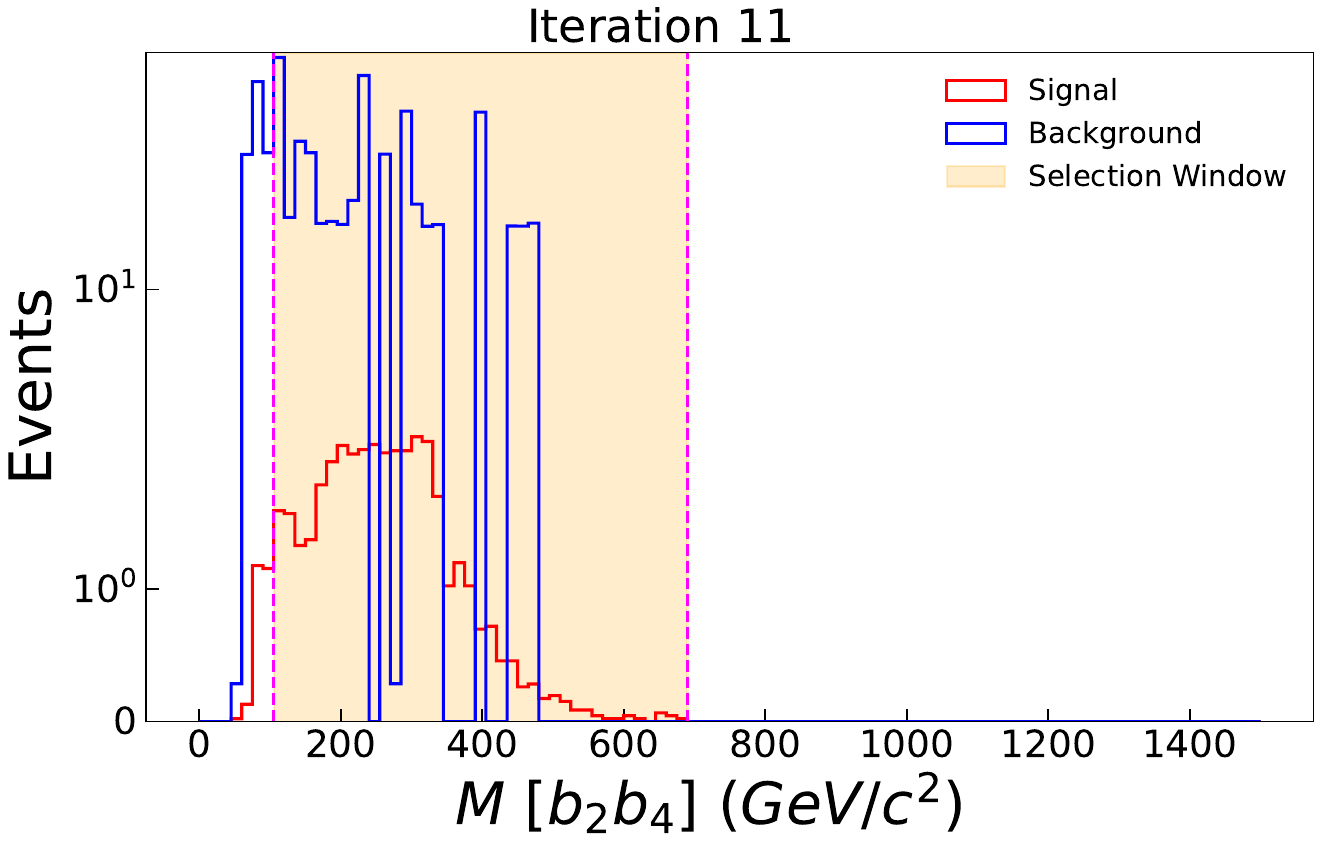}
  \includegraphics[width=0.30\textwidth]{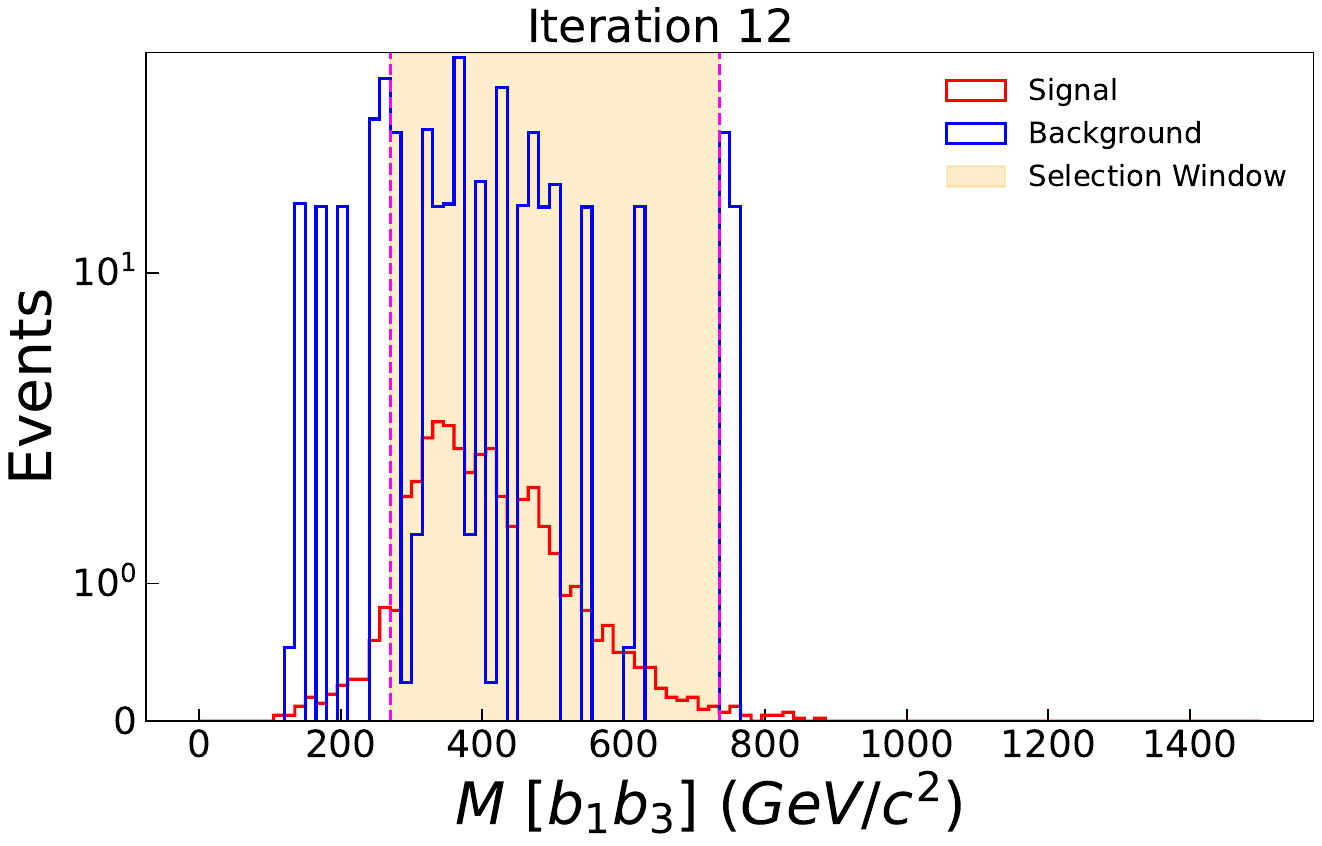}
\caption{ The cutflow suggestions from each iteration as dicated by the vertical line test.}
\label{fig:suggestions}      
\end{figure*}

\begin{enumerate}
    \item The generated signal and background events (after the imposition of the preselection cuts) are fed into MadAnalysis 5 and the distributions of the various kinematic obsevables are obtained.
    \item The Area Parameter is computed for all observables and used to rank them accordingly. The top-ranked observable is then passed to the vertical line test stage, which determines the optimal cut to maximize the significance, subject to the constraint that the signal yield does not drop by more than $20\%$ compared to the previous iteration. A condition is imposed such that $\Delta S = \frac{\Delta N(s)}{N(s)}\times 100 < 20$, where $\Delta N(s)$ denotes the reduction in the signal yield relative to the current signal yield $N(s)$.
    \item \textbf{Observable hold:}  This representation showcases where we hold the remaining observable distributions for future iterations.
    \item If, after the imposition of the cut, enough signal events remain ($N_s>$ 10) and if an improvement in significance is obtained, the cut is accepted and the observable is dropped from further consideration. In addition, if after this cut, we satisfy the LL condition (i.e., significance becomes $5\sigma$ or higher), the process terminates.
    \item If, after the imposition of the cut the LL condition is not satisfied, we pass on to the \textbf{Collector Connector (CC):}  It takes the observable distributions from the hold and pushes it to the next step. Once a proper instruction (indicated by an arrow) hits the CC, it will collect all the observable distribution sets from the hold connected to it and then recalculate all the distributions using MadAnalysis 5, calculate the AP and then sort the observables as explained previously. 
    \item \textbf{Pulse Switch (PS):} This is an instantaneous switch that triggers the execution of an instruction when a specific condition is met (typically an `if' condition in the program to check the minimum significance criterion). Otherwise, it remains inactive. Specifically, if it turns out that $\sigma(k) < \sigma(k-1) + 0.10$, then that particular observable will be sent to hold, and a pulse will activate the switch, which is connected to another hold with remaining variables. This action will make all the remaining observables in the hold go through the sorting based on ranking, followed by optimization via vertical line test on the rank-1 observable.\footnote {This particular scenario was encountered in Iteration-12 in our example.}
    \item The same steps continue until we either run out of observables for ranking or when the LL conditions are satisfied, i.e., the significance improved beyond $5\sigma$\footnote{Here, a $5\sigma$ significance was already achieved at iteration 8. However, we proceeded with a few additional iterations to allow for further optimization, thereby demonstrating the robustness and effectiveness of the approach. While it would have been reasonable to stop at that point, we continued in order to study the performance trend over successive iterations.}. The suggested cuts on following the algorithm at each iteration are shown in \autoref{fig:suggestions} and the final cut flow chart is shown in \autoref{tab:final_cutflow}.
\end{enumerate}

At first glance, our proposed methodology may appear similar to the Boosted Decision Tree (BDT) approach. However, BDTs are ensemble-based algorithms designed primarily for classification tasks and rely on data-driven training procedures. In contrast, the proposed methodology is a guided optimization technique that yields interpretable outputs, corresponding to meaningful physical constraints. A detailed clarification highlighting the distinction between the proposed approach and BDTs is provided in \autoref{tab:A1} of the Appendix.

\begin{center}
\begin{table}
\small
\centering 
\begin{tabular}{ccccccccc} 
 \hline
- & Rank &- & $N_{signal}$ & $N_{BG}$ & $\sigma$ & $\Delta\sigma$ &  Accepted  & $Z$\\ [0.3ex] 
 \hline\hline
- &  &initial & 677.09 & 222336 & 1.433 & - & - & 0.030 \\ 
 \hline
 I-1& 1&$130 < PT(b_2) < 680 $ & 579.95 & 36746 & 3.002 & 1.569 & $\checkmark$ &  0.157 \\
 \hline
 I-2&1&$ 80 < PT(b_3) > 420 $ & 514.58 & 21859 & 3.440 & 0.438 & $\checkmark$ &   0.233 \\
 \hline
 I-3&1&$40 < PT(b_4) < 360 $ & 457.97 & 15212 & 3.659 & 0.219 & $\checkmark$ &  0.297 \\
 \hline
 I-4&1&$640 < M (\ell,\ell,b,b,b,b) < 1380 $ & 413.78 & 10771 & 3.913 & 0.254 & $\checkmark$ &  0.378 \\
 \hline
  I-5 &1&$200 < PT(b_1) < 570 $ & 359.48 & 6502.6 & 4.340 & 0.427 & $\checkmark$ &  0.539 \\
 \hline
  I-6&1&$ 0 < \cancel{E}_T < 140 $ & 318.02 & 4284.8 & 4.688 & 0.348 & $\checkmark$ & 0.716 \\
 \hline
  
 I-7 &1&$0 < PT(\ell_1) < 150 $ & 311.49 & 3689.2 & 4.925 & 0.237 & $\checkmark$ & 0.811 \\
 \hline

 I-8 &1&$ 0.7700< \Delta R(b_2,b_3) < 2.5900 $ & 250.61 & 2118.5 & 5.149 & 0.224 & $\checkmark$ &  1.114 \\
 \hline
 \midrule
 \hline
 \hline
 \midrule
 \hline
I-9 &1&$225 < M (b_2b_3) < 435 $ & 204.15 & 1258.9 & 5.337 & 0.188 & $\checkmark$ &  1.486 \\
 \hline

I-10 &1&$195 < M (b_1b_4) < 465 $ & 178.30 & 766.4 & 5.801 & 0.464 & $\checkmark$ &  2.046 \\
 \hline
I-11 &1&$105 < M (b_2b_4) < 690 $ & 167.38 & 617.4 & 5.975 & 0.174 & $\checkmark$ &  2.330 \\
 \hline
 I-12 &1&$ 0.8400< \Delta R(b_1,b_4) < 3.8150 $ & 167.38 & 617.4 & 5.975 & 0.0 & $\times$ &  -- \\
 \hline
 I-12 &2&$270 < M (b_1b_3) < 735 $ & 152.44 & 402.8 & 6.469 & 0.494 & $\checkmark$ &  3.065 \\
 \hline

\end{tabular}
\caption{The final cutflowchart from the analysis as dictated by the algorithm.}
\label{tab:final_cutflow}
\end{table}
\end{center}

\begin{figure}
    \centering
 \includegraphics[width=0.55\linewidth]{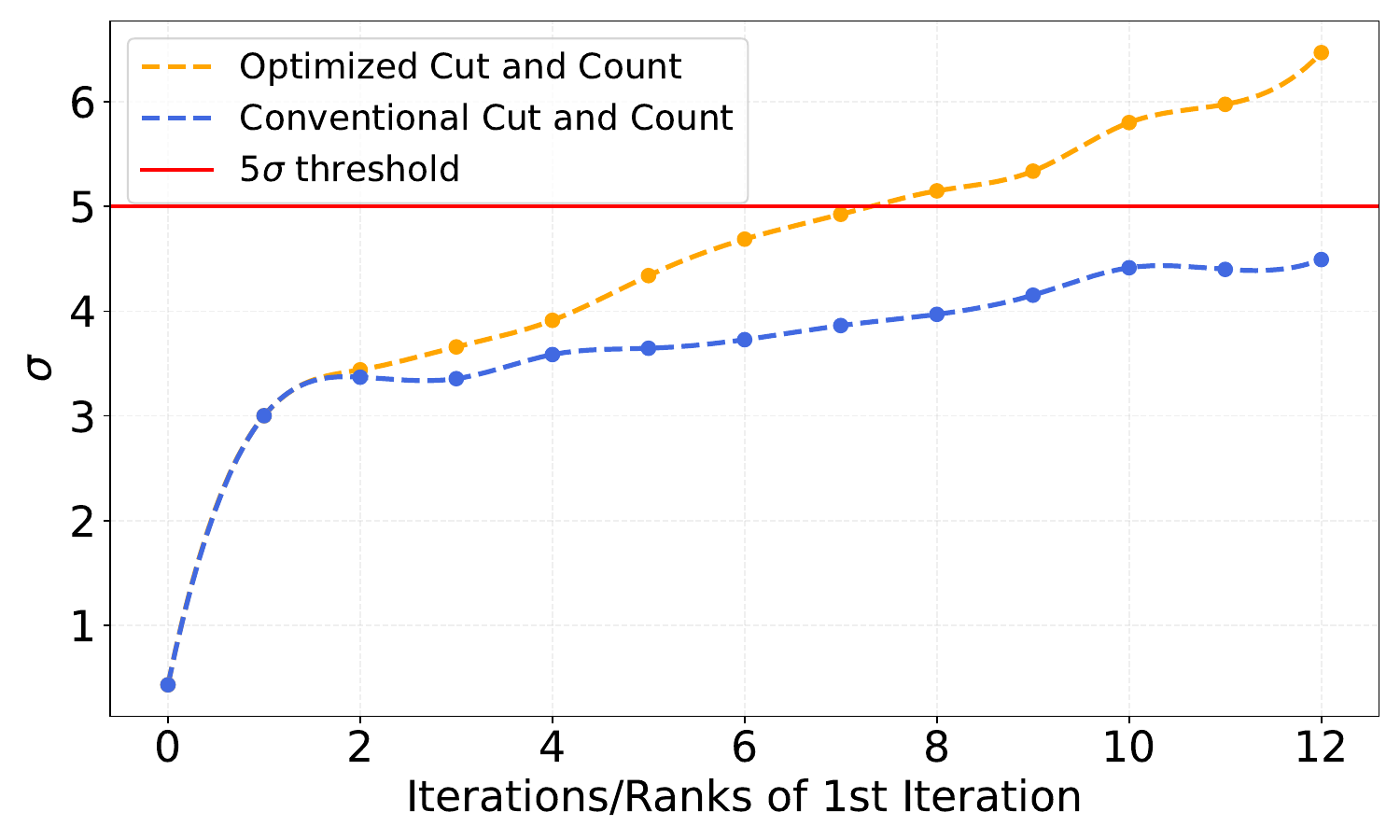}
    \caption{Comparison of the improvement in significance between the conventional cut-and-count method applied sequentially to the top-ranked observables (\autoref{tab:alter}) and the optimized cut-and-count results obtained via the designed algorithm (\autoref{tab:final_cutflow}).}
    \label{fig:compare}
\end{figure}

At this point, it would be reasonable to ask if an iterative process is indeed needed, and whether one can just use the AP designed initially and rank the variables and put in cuts as suggested by the vertical line test. This would certainly be computationally less extensive. In \autoref{tab:alter}, we display the cut flow chart after applying the cuts suggested by the vertical lines test to the top 12 ranked variables from the first iteration. It is evident that the desired level of significance or effective rejection of background events has not yet been achieved (while simultaneously maintaining the essential conservation of the signal events). In contrast, using the iterative method clearly demonstrates significant improvements in the analysis, facilitating both signal conservation and background rejection concurrently as shown in \autoref{tab:final_cutflow}. The main reason for the superiority of the cutflow chart shown in \autoref{tab:final_cutflow}, which is obtained from the algorithm, is that at each iteration, MadAnalysis 5 takes into account the distributions of every observable that is modified due to an imposed threshold or selection cut on a particular observable suggested at that point in the iteration. Thus, MadAnalysis 5 reconsiders the correlation of observables at each step\footnote{The evolution of variable ranking is decided by the interplay between signal and background distributions under repeated choices, rather than pairwise correlations. Although correlations show statistical relationships between observables, their effects vary for signal and background samples and and their subsequent values are also calculated after the optimization decision made by the finding significance $S/\sqrt{S+B}$. As a result, the discriminating power of an observable can only be determined after earlier cuts have reshaped the accessible phase space, making ranking evolution intrinsically non-linear.~\autoref{tab:A1} includes Pearson correlation matrices for completeness.}, recalculates distributions, and adjusts for the effects of the imposed cuts. On the other hand, \autoref{tab:alter}, although it considers the most important observables at the start of the first iteration, cannot account for the correlations among observables. As a result, many of the subsequent cuts either become redundant or negatively affect the significance. On comparing~\autoref{tab:final_cutflow}, which presents the final cutflow chart obtained from the algorithm, with~\autoref{tab:alter}, which follows a sophisticated conventional cut-and-count approach, the improvement in significance over iterations demonstrates that the cut-and-count scheme applied by the algorithm is far superior. It is found that a $5\sigma$ discovery potential is achievable by following the algorithm, whereas the significance of the conventional method hovers around $4\sigma$. A graphical representation of this comparison is provided in~\autoref{fig:compare}.

\begin{center}
\begin{table}
\small
\centering
\begin{tabular}{cccccc} 
 \hline
- & - & $N_{signal}$ & $N_{BG}$ & $\sigma$ &  $Z$\\ [0.3ex] 
 \hline\hline
- &  initial & 677.09 & 222336 & 1.433 &  0.030 \\ 
 \hline
 R-1&$130 < PT(b_2) < 680 $ & 579.95 & 36746 & 3.002 &  0.157 \\
 \hline
 R-2&$ 70 < PT(b_3) > 420 $ & 537.22 & 24868 & 3.370 &  0.214 \\
 \hline
 
 R-3&$ 720 < H_T  < 2000 $ & 484.63 & 20374 & 3.356 &  0.235 \\
 \hline
 R-4&$ 190 < PT(b_1) < 700 $ & 457.24 & 15799 & 3.586 &  0.286 \\
 \hline
 R-5&$ 315 < M (b_1,b_2) < 1275 $ & 432.41 & 13631 & 3.646 &  0.310\\
 \hline
 R-6&$ 30 < PT(b_4) < 360 $ & 411.59 & 11779 & 3.728 &  0.344 \\
 \hline
 
 R-7& $ 700 <  M (\ell,\ell,b,b,b,b) < 1740 $ & 402.83 & 10470 & 3.863 & 0.378 \\
 \hline
 
 R-8&$ 240 < M (b_1,b_3) < 900 $ & 377.26 & 8654.6 & 3.970 & 0.427\\
 \hline
 R-9&$ 180 < M (b_2,b_3) < 630 $ & 344.36 & 6531.7 & 4.153 & 0.514\\
 \hline
 R-10&$ 180 < M (b_1,b_4) < 690 $ & 312.84 & 4707.8 & 4.415 & 0.644\\
 \hline
 R-11&$ 120 < M (b_2,b_4) < 975 $ & 284.21 & 3889.9 & 4.399 & 0.705\\
 \hline
 R-12&$ 75 < M (b_3,b_4) < 675 $ & 269.38 & 3325.9 & 4.493 & 0.778\\
 \hline
 
\end{tabular}
\caption{The cutflow chart for the same process following a sophisticated conventional cut and count method - we simply choose the observables that can best alienate the signal from BG from Ranking scheme and apply cuts in a sequential fashion using vertical line test.}
\label{tab:alter}
\end{table}
\end{center}

\begin{figure*}
 \centering

\includegraphics[scale=0.22]{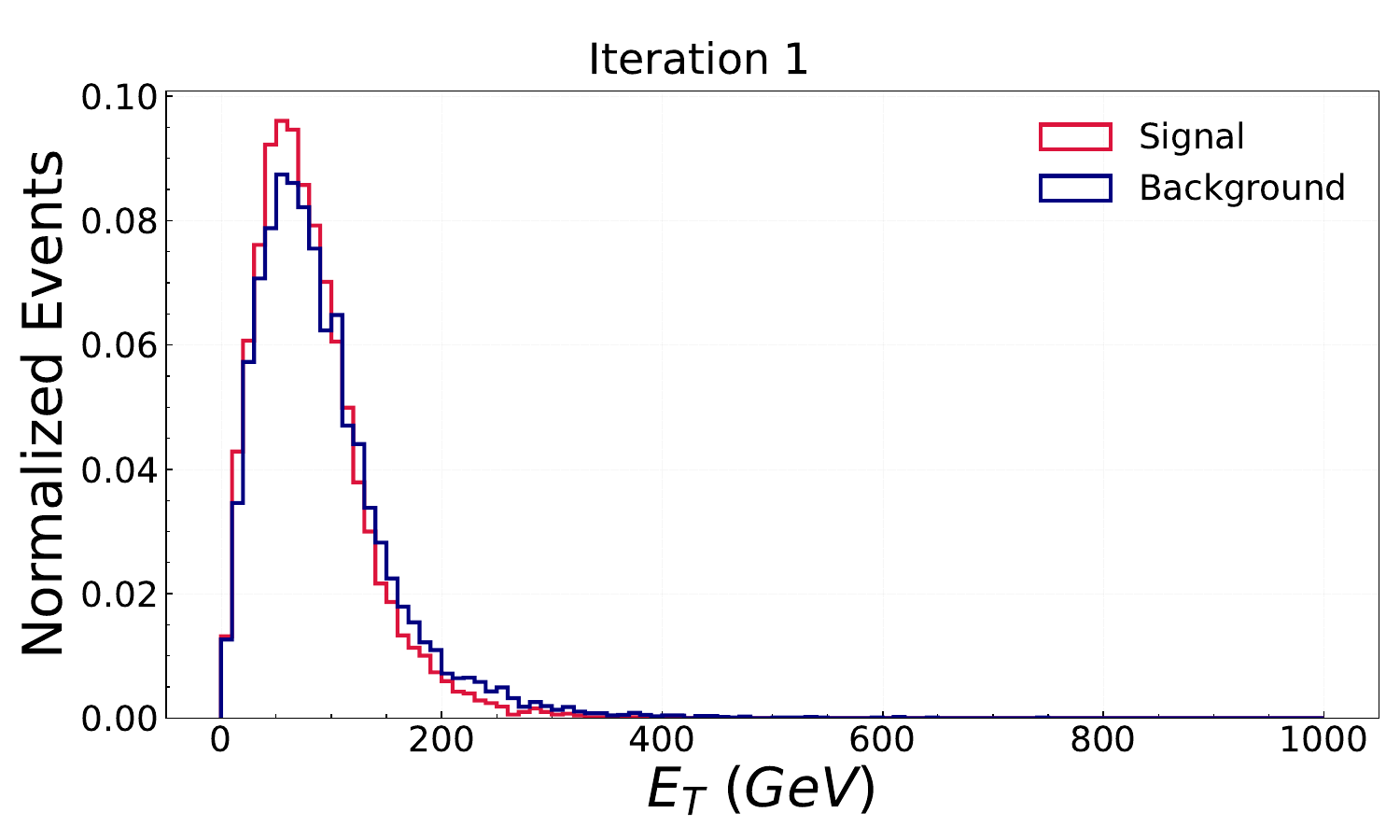} \includegraphics[scale=0.22]{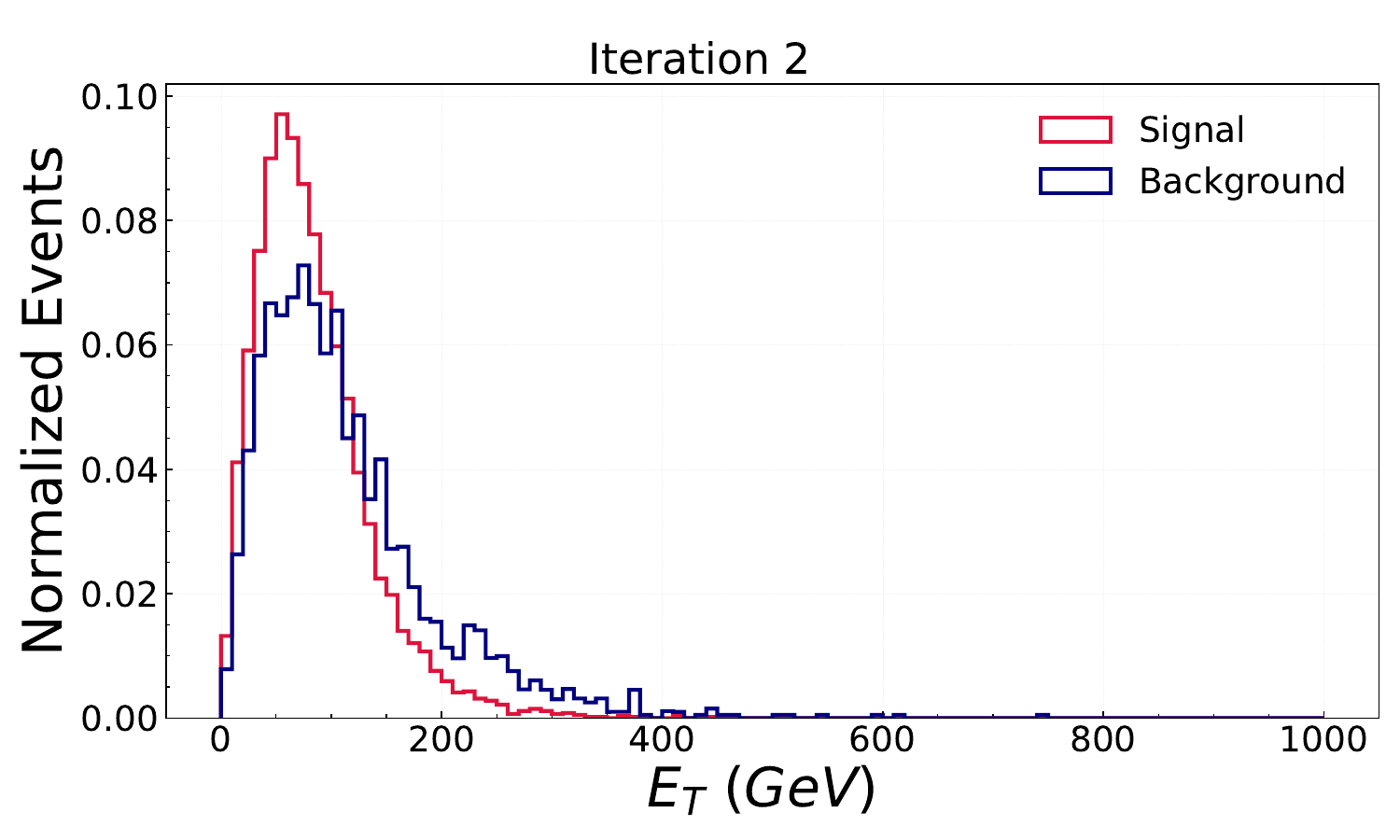}\includegraphics[scale=0.22]{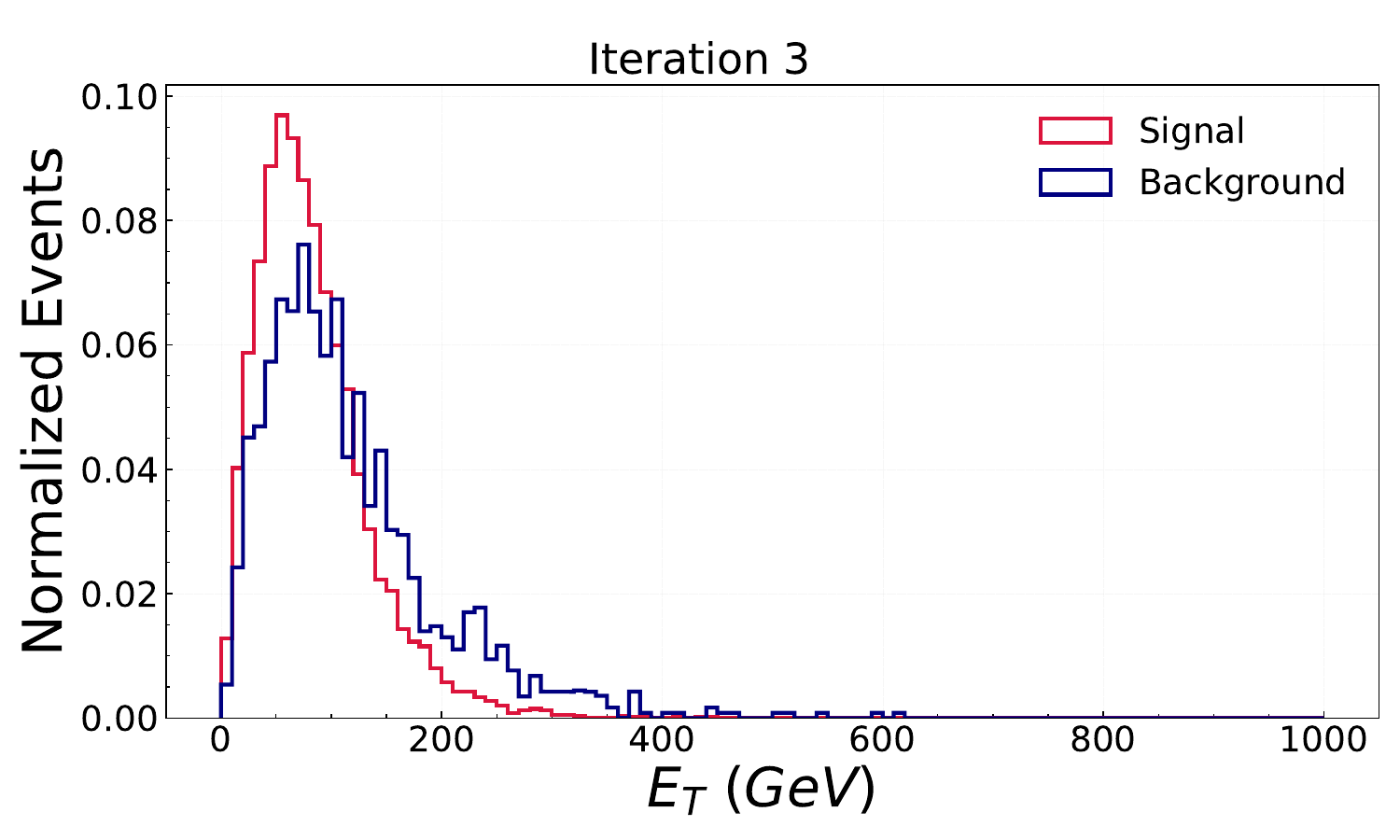}  \includegraphics[scale=0.22]{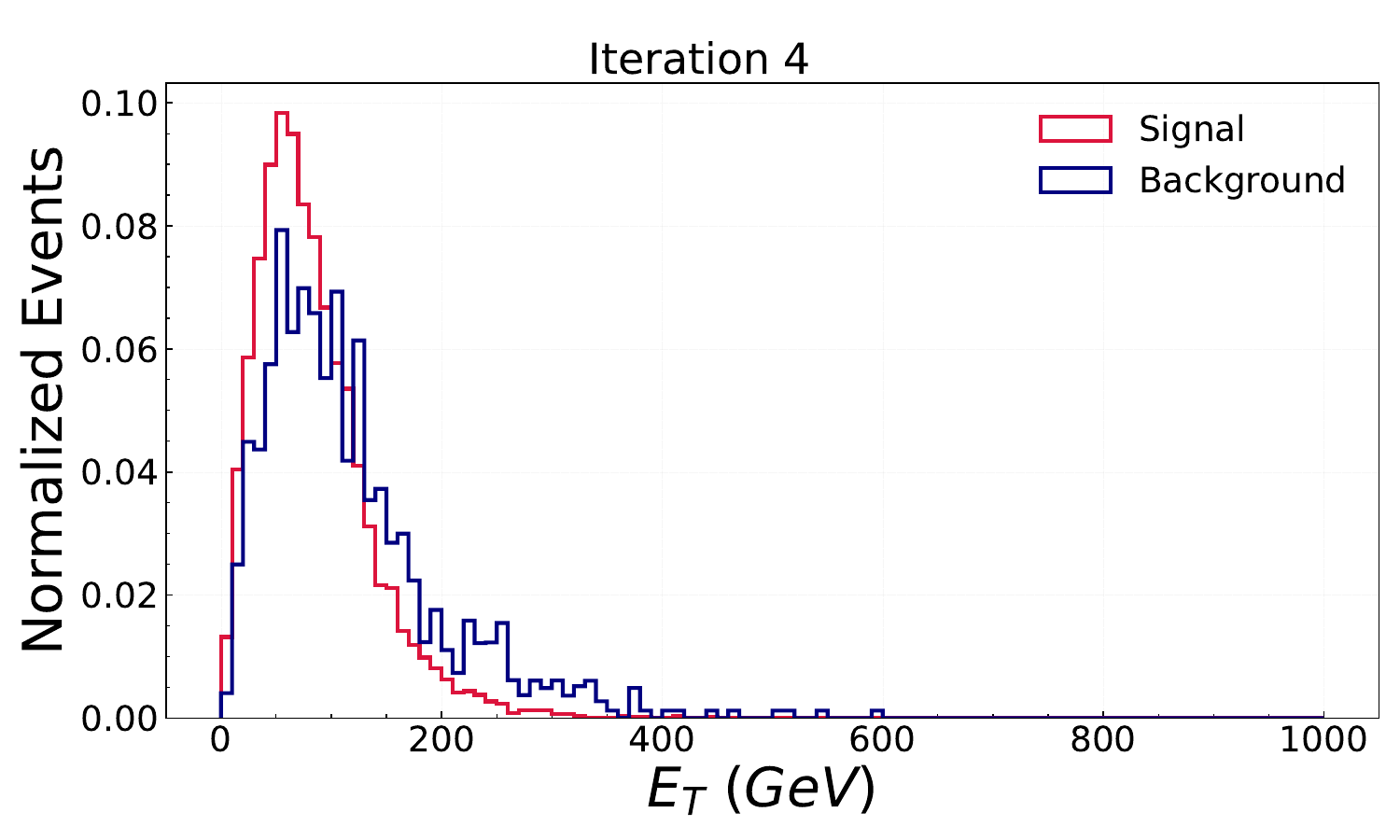}\includegraphics[scale=0.22]{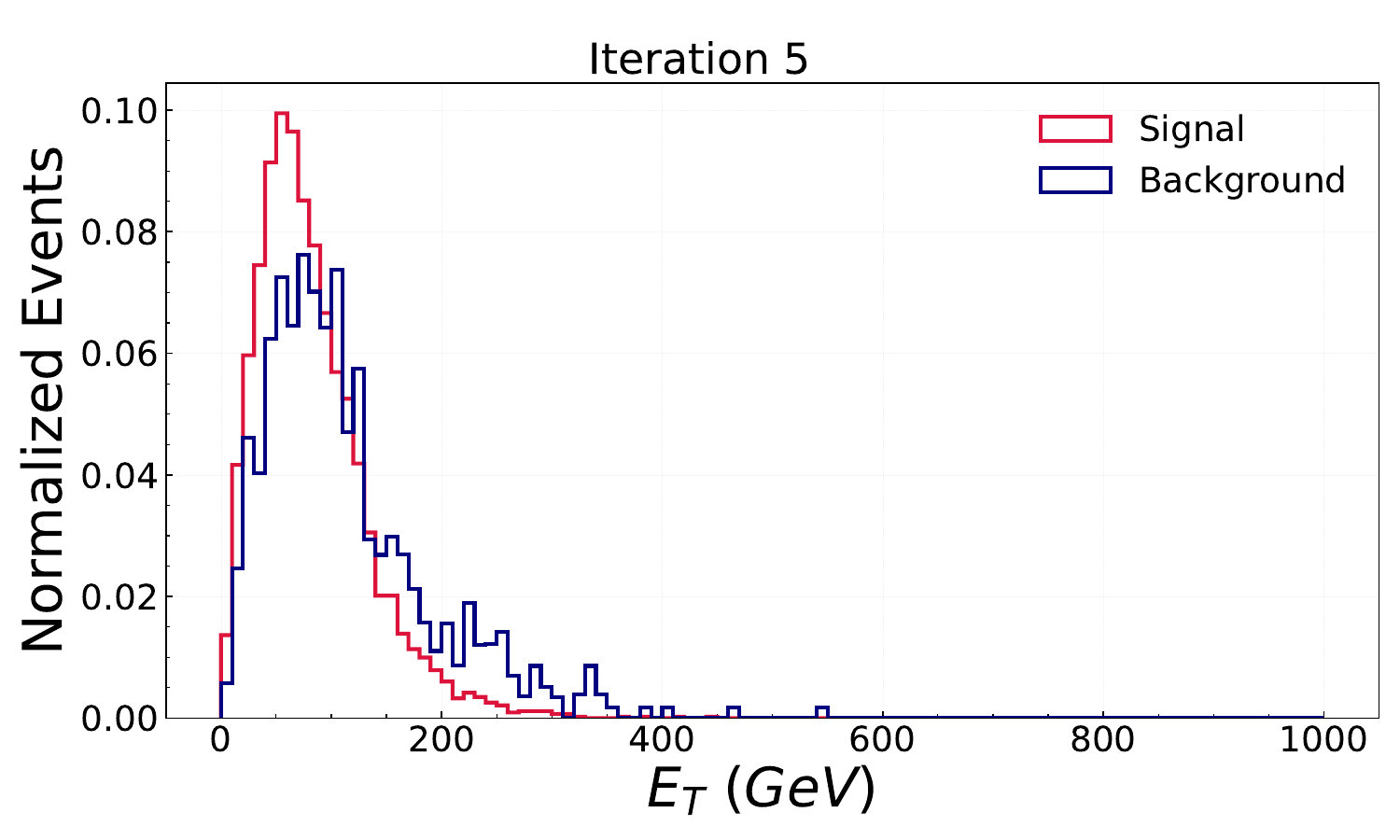} \includegraphics[scale=0.22]{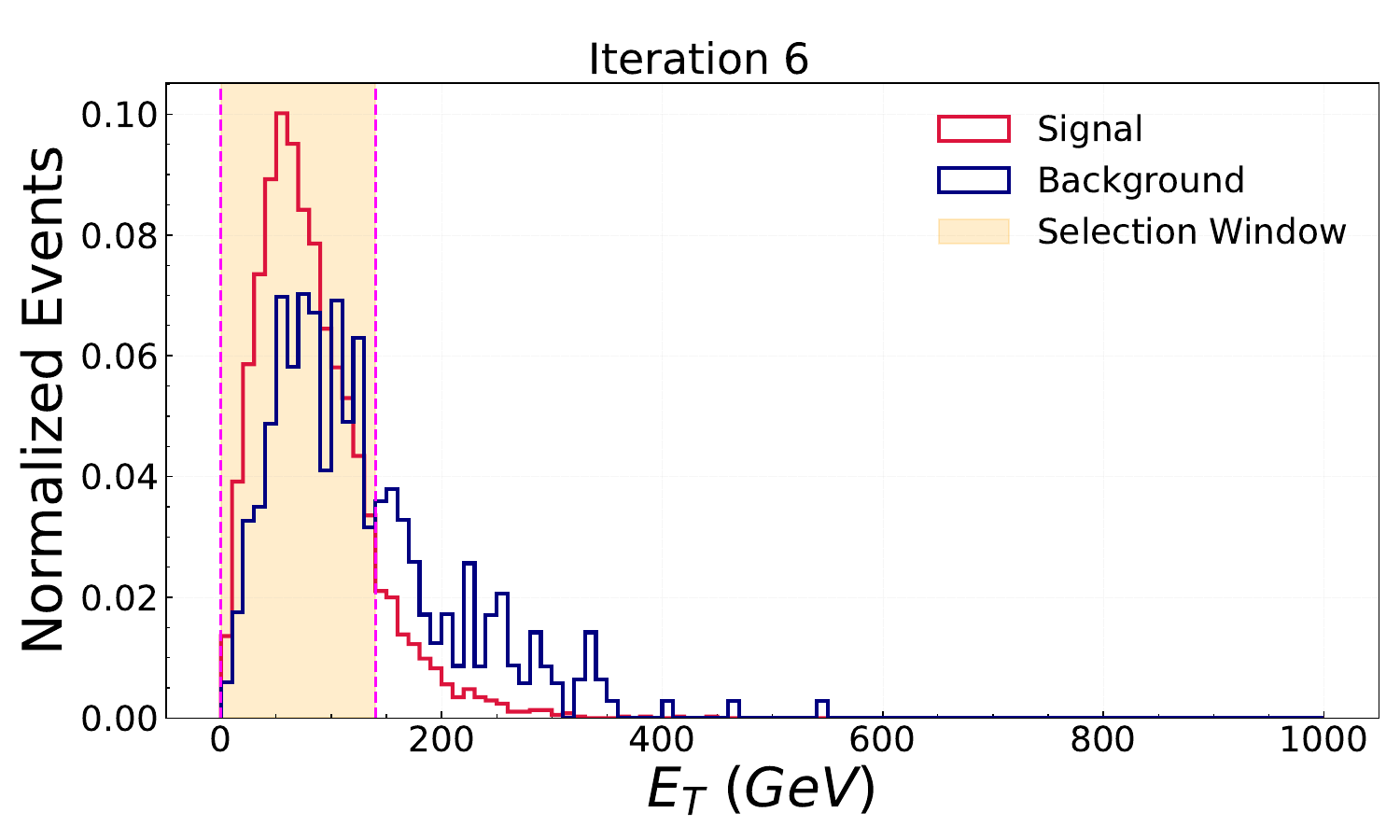}  \\

\caption{Evolution of the normalised $\cancel{E}_T$ distribution from Iteration-1 to Iteration-6, illustrating how the observable gradually becomes more effective in distinguishing signal from background, thereby emerging as a promising candidate for cut-based selection.}
\label{tab:gt}      
\end{figure*}

Additionally, an important revelation from the methodology discussed is that, during the initial iteration, the observable $\cancel{E}_T$ was positioned at the 26th out of 29 observables, making its status as one of the least preferred observables which is to be selected for vertical line test. This aligns with its low preference in the first intuition within a conventional cut and count methodology. As illustrated in \autoref{tab:ranks}, the rank of $\cancel{E}_T$ based on \textbf{AP} improves with each iteration, and in its selection eventually as the top-ranked observable in the sixth iteration. Subsequently, the vertical line test was applied to propose a selection cut. This exemplifies how constraining one observable can impact on the other observables. The selection cuts preceding $\cancel{E}_T$ have exhibited positive effects, leading to incremental improvements in its rank based on \textbf{AP} at each step. The evolution of observable $\cancel{E}_T$ across iterations is detailed in \autoref{tab:gt}, highlighting its constraint by the vertical line test in the sixth iteration -an achievement not attainable in a typical conventional cut-based study.

\section{BDT Analysis}
\label{sec:comparision}
One machine learning algorithm that has been quite similar to the intended method is BDT. In contrast to BDT, which is nonlinear by construction and interpretability using the physical observables is opaque, the optimization schema of the designed algorithm has been driven by the $argmax(\frac{s}{\sqrt{s+b}})$, the significance formula and the cutflow suggestions themselves are linear in fashion, where one cut follows another. Although it is challenging to provide an interpretability in terms of physical observables for such a BDT method, we attempt to compare our approach in this part by doing the analysis using BDT. However, the BDT's individual components are a collection of several weak decision trees, which are then combined using a technique known as boosting to create a powerful classifier. Therefore, we will demonstrate a first single decision tree up to layer 4 in order to determine whether BDT is learning similar characteristics learnt by such an algorithm. After that, we will use an adaptive boost approach to improve the classifier. The step by step implementation and further details are as follows. 

\subsection{Preparation of the Dataset for the BDT Analysis}

The tools and framework described in the \autoref{sec:intro} were utilized to produce the signal and background events needed for the analysis. To minimize the phase space and limit background contribution, the data were extracted using a set of baseline selection criteria. The event selection requirement is the same as it was prior to the algorithm's implementation. Based on the signal's final state, we imposed number cuts as $N_\ell > 1 $ and $N_b > 3$ with $\Delta R > 0.4$ for all jet and lepton combinations and the event by event data are recorded for all 29 observables which described in the~\autoref{sec:sec2}. The background and signal samples were merged into a single dataset to create the final labeled dataset for training the model. The dataset was constructed by taking into account their respective production cross sections and scaled event yields. The relative fractions were determined accordingly, and events were randomly sampled from each background dataset in these proportions. This ensures that the combined background sample faithfully represents the expected physical composition of the total SM background. The dataset was randomly separated into training and testing subsamples, which accounted for $80\%$ and $20\%$ of the total occurrences, respectively.\\
In order to gain a better understanding of the data, we calculated the Pearson linear correlation coefficient matrix, which is $\rho = \frac{\langle xy \rangle-\langle x \rangle \langle y \rangle}{\sigma_x \sigma_y}$. The expectation value and standard deviation of $x$ are $\langle x \rangle$ and $\sigma_x$, respectively, and are calculated independently for the signal and background samples. The results are shown in the~\autoref{tab:A1}

\subsection{Decision Tree and BDT Analysis Setup and Results}

\subsubsection{Single Decision Tree Analysis}
\label{DT}

\begin{figure}
    \centering
    \includegraphics[width=1\linewidth]{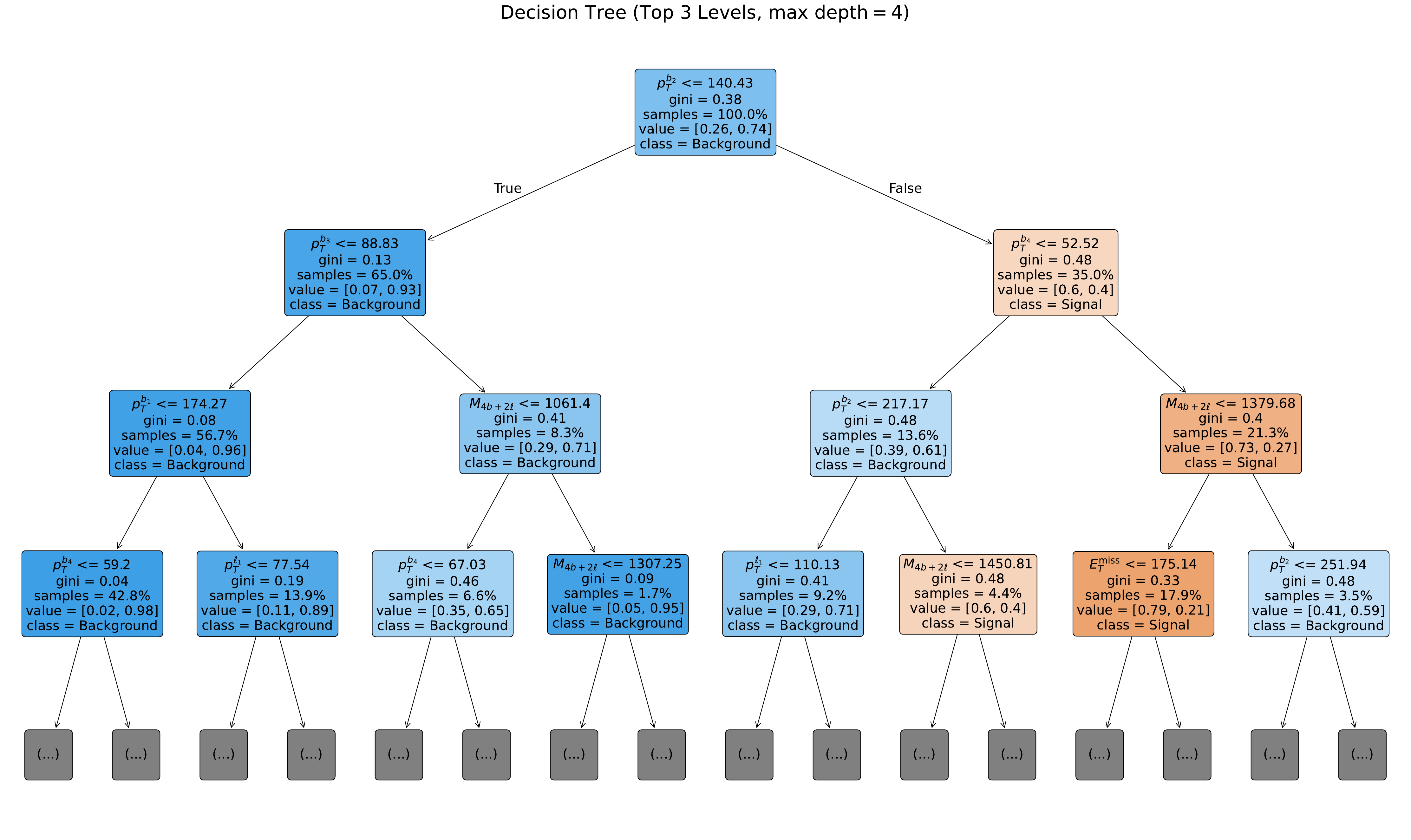}
    \caption{Visualization of the trained decision tree classifier with a maximum depth of four. Each node corresponds to a kinematic selection, and the color shading represents the relative purity of signal and background events. The structure illustrates how the algorithm hierarchically separates the two classes using the most discriminating observables.}
    \label{fig:DT}
\end{figure}

To investigate the discriminating capacity and produce an interpretable multivariate selection, a single DT is initially trained. The \texttt{scikit-learn} framework was used to create the classifier, and the Gini impurity criteria~\cite{Coadou:2022nsh} for node splitting was used for training. The tree was limited to four levels of maximum depth in order to prevent overfitting and maintain interpretability. The purpose of demonstrating this particular DT was to investigate the learning process of the kinematic variable in relation to the algorithm that was devised. The~\autoref{fig:DT} represents the DT that was trained using the dataset. Every DT node shows a number of information that describe the classification process. At the top of each node is a threshold applied to a particular observable, followed by a quantity labeled gini, which stands for the Gini impurity. This quantifies the degree of mixing between signal and background events in that node; a lower value indicates greater purity. The percentage of events that reach that node following further choices is indicated by the entrance samples. The dominant class assigned to that node is indicated by the class label, while the value field displays the relative proportions of signal and background events. Together, these elements provide an interpretable representation of how the classifier learns a hierarchical sequence of kinematic selections to discriminate between signal and background.

\subsubsection{Boosted Decision Tree Analysis}
\label{bdt}

\begin{figure}
    \centering
    \includegraphics[width=0.49\linewidth]{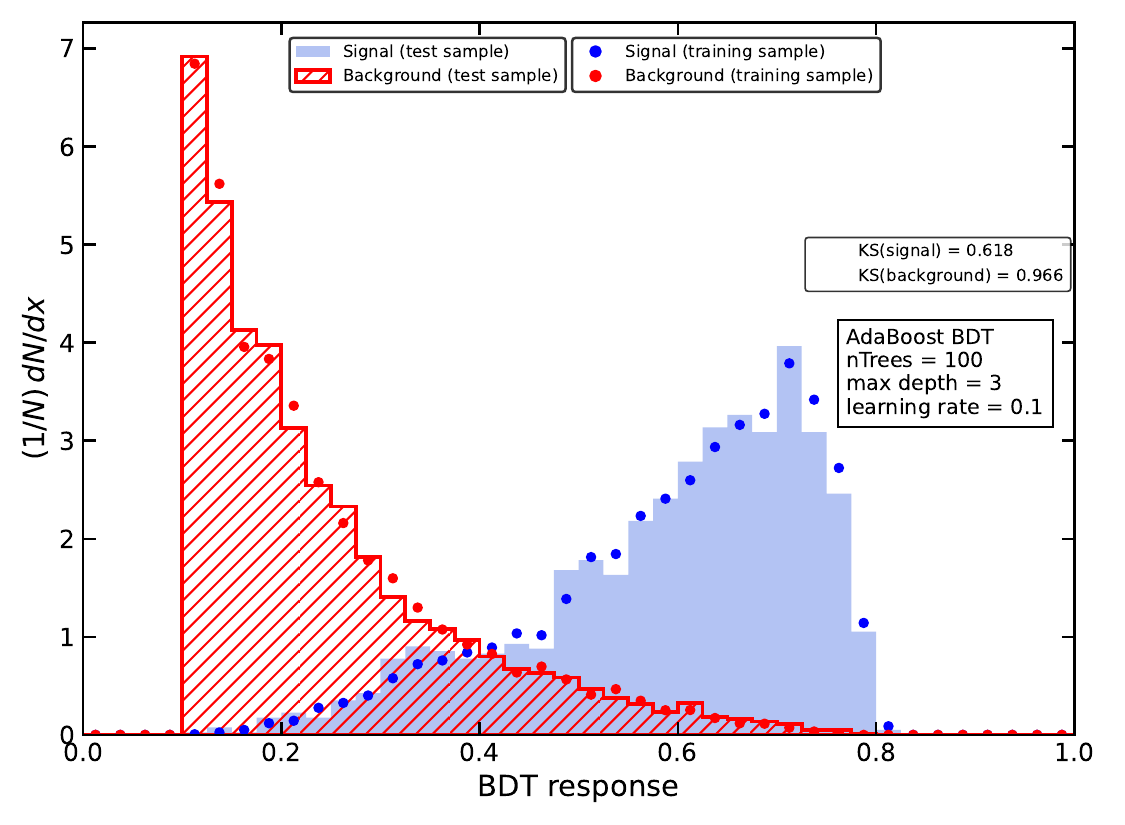}
    \includegraphics[width=0.49\linewidth]{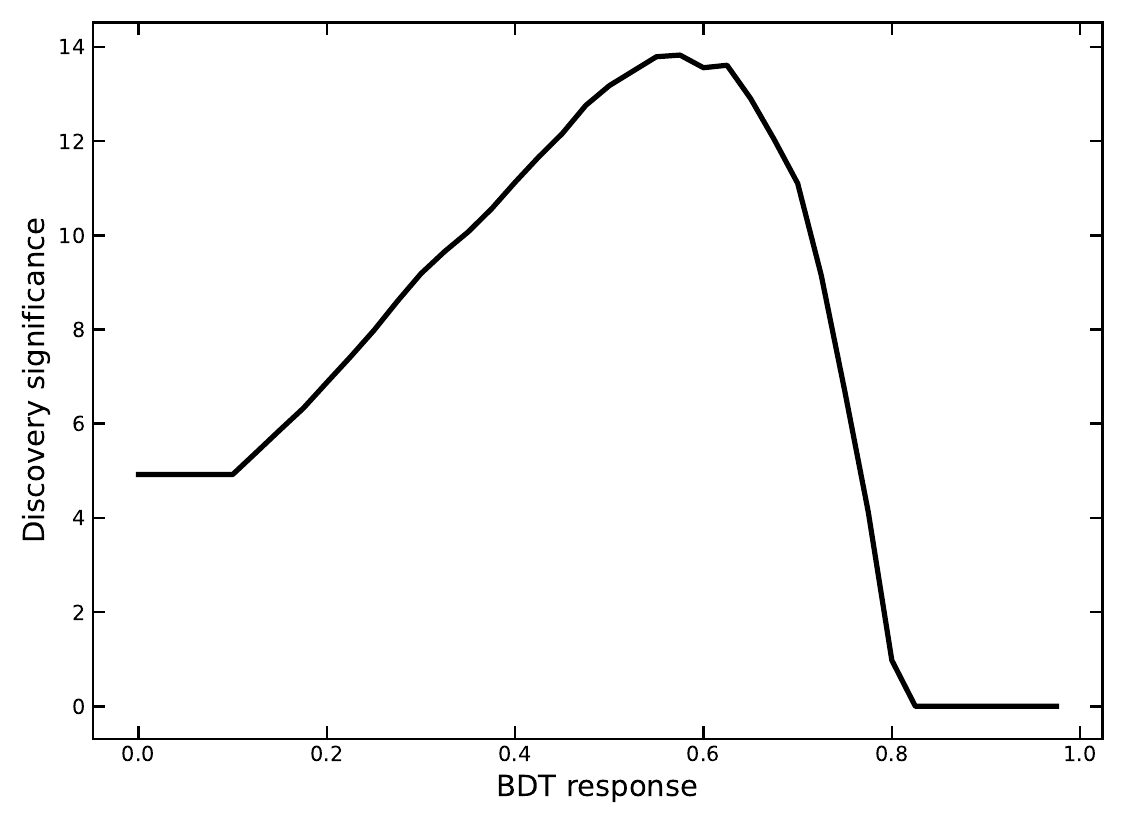}
    \caption{BDT response curve for the training and testing samples (left) and discovery significance as a function of the BDT response (right).}
    \label{fig:BDT_significance}
\end{figure}

To enhance discrimination beyond a single DT classifier, a BDT algorithm based on the AdaBoost method~\cite{Choudhury:2024crp} was used. In contrast to a single tree, which depends on a single hierarchical series of kinematic choices, the BDT builds a more reliable and stable classifier by combining an ensemble of weak decision trees. By iteratively reweighting incorrectly categorized events and concentrating on challenging-to-classify areas of phase space, this method improves the separation between signal and background. 
Decision trees with a maximum depth of three were used as weak learners in the construction of the BDT. The classifier is checked for overfitting using the p-value of the Kolmogorov–Smirnov(KS) test, which compares the BDT response curves for the training and testing subsamples. Additionally, unlike by employing the receiver-operator-characteristics (ROC) curve, which quantifies the BDT performance, we attempt to interpret it using the variation of discovery significance as a function of BDT response. The ensemble consisted of 100 trees and was trained with a learning rate of 0.1.~\autoref{fig:BDT_significance} displays the BDT response curve for the training and testing subsamples as well as the discovery significance as a function of the BDT response.  

\subsubsection{Comparative Insights}
\label{compare}

We find a clear correspondence between the variables selected by the trained DT~(\autoref{DT}) and those identified by our iterative cut-based algorithm. In particular, the observables selected by our algorithm in the first three iterations, namely $p_T(b_2)$, $p_T(b_3)$, and $p_T(b_4)$, appear already within the first two layers of the trained decision tree, as shown in~\autoref{fig:DT}. Proceeding to deeper layers of the tree, additional observables such as $M_{4\ell+2b}$ and $p_T(b_1)$ emerge, which correspond precisely to the variables suggested by our algorithm in iterations four and five. At the next layer, the tree introduces $\cancel{E}_T$ and $p_T(\ell_1)$, which align with the variables identified in iterations six and seven of our method. This striking agreement demonstrates that the hierarchical structure learned by the decision tree mirrors the linear, step-by-step ranking produced by our transparent optimisation algorithm, indicating that both approaches are sensitive to similar underlying physical features of the signal. Also, a full BDT analysis is performed using the AdaBoost method, in which the classifier is constructed as an ensemble of weak decision trees. Using appropriately chosen BDT hyperparameters (as detailed in~\autoref{bdt}), we evaluate the discovery significance as a function of the BDT response. We find that the BDT achieves a higher signal significance compared to the purely cut-based algorithm presented. While the proposed algorithm does not reach the performance of sophisticated machine-learning techniques, it significantly outperforms the conventional cut-and-count approach while preserving interpretability, thereby providing a meaningful addition to existing analysis strategies.

\section{Discussion and Conclusion}
\label{sec:Conclusion}

Particle physics is in the midst of an exciting time - while the searches at the LHC for hints of BSM physics have not borne fruit thus far, it is nevertheless a very interesting question to ask how much and how far one could probe models with more refined and sophisticated techniques in order to help discover the expected new TeV scale physics. While the ML and other computational techniques are apt for this purpose, our motivation here is slightly different - the traditional cut and count method of doing phenomenological analysis has been enormously successful and has stood the test of time thus far. Thus, it is interesting to ask if one could keep the spirit of this approach while trying to optimize this strategy for better results. This can also yield results that are phenomenologically interpretable, unlike deep neural networks, which cannot explain why their parameters can perfectly execute the assigned task, leading to what we refer to as the black-box nature. 

In this spirit, we demonstrated in this paper that one could use a cut-based strategy to uncover new physics but with a more systematic and streamlined approach. Our analysis starts by ranking all observables associated with any given new physics process on the basis of the newly introduced Area Parameter - a metric that looks for maximum signal-background separation. Having ranked the observables, we employ the Vertical Line Test to systematically isolate the region of phase space that yields the maximum significance. The process is then iterated over the other observables yielding a final significance that can supercede the one obtained by conventional techniques.

In our analysis, we considered a process featuring a pair of charged Higgs particles decaying into $W^+W^-$, accompanied by a pair of two pseudoscalar particles undergoing subsequent decays into $b\bar{b}$. This results in a final state topology that is rather complex and challenging, thus providing an excellent platform for evaluating the effectiveness of the proposed algorithm. Contrasted with the traditional cut-and-count method, we demonstrated that our current algorithm performs significantly better in terms of isolating the signal from the background, giving us a significance of more than $5\sigma$ while the traditional cut and count methodology for this same final state yields only around $4\sigma$. However, this does come at a price - it entails a notably higher level of computational complexity and thus, total computational time. This increased complexity stems primarily from the dynamic nature of observable rankings, which change with each iteration as we isolate different regions of phase space. As a consequence, the background execution of MadAnalysis 5 is necessary to compute the CDFs related to the observables after each iteration. Nonetheless, we expect that this might be less important than achieving a superior searching technique, particularly for new physics signals that might involve complicated final states.

We also conducted a comparative study of the designed algorithm with the single DT and BDT and found that both strategies are sensitive to the same set of observables identified for optimising signal–background discrimination. On performing a thorough BDT analysis using the AdaBoost approach, which builds the classifier as an ensemble of weak decision trees. The discovery significance was evaluated as a function of the BDT response. We find that the BDT achieves a higher signal significance compared to the purely cut-based algorithm presented. While the suggested algorithm does not equal the performance of more advanced machine-learning approaches, it greatly outperforms the traditional cut-and-count strategy while retaining interpretability, thereby providing a meaningful addition to existing analysis strategies.

\section{Acknowledgements}

BC acknowledges support from the Department of Science and Technology, India, under Grant CRG/2020/004171. GBK is supported by the Prime Minister's Research Fellowship (Grant No: PMRF-192002-1802) from the Department of Science and Technology, India. AS acknowledges the monetary support which is received from Anusandhan National Research Foundation (ANRF) through the SERB-NPDF grant (Ref No: PDF/2023/002572). The work of S.S. is supported by Funda\c{c}\~ao de Amparo \`a Pesquisa do Estadode S\~ao Paulo (FAPESP) grant 2021/09547-9.

\bibliography{reference}

\section{APPENDIX}
\label{tab:Appendix}

\subsection{Correlation Matrix of the observables}
\label{tab:A1}

\begin{figure}[H]
    \centering
    \includegraphics[width=0.49\linewidth]{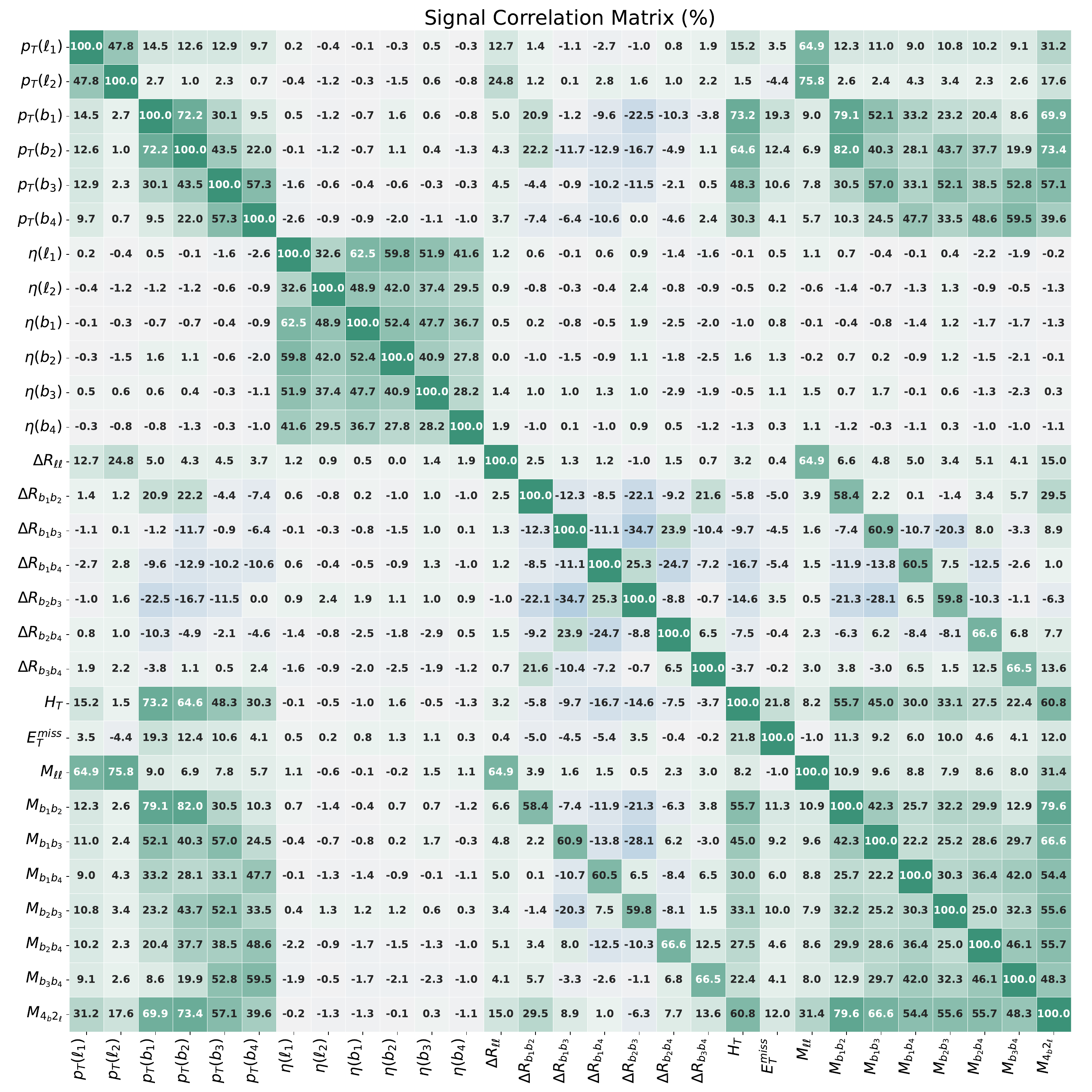}
    \includegraphics[width=0.49\linewidth]{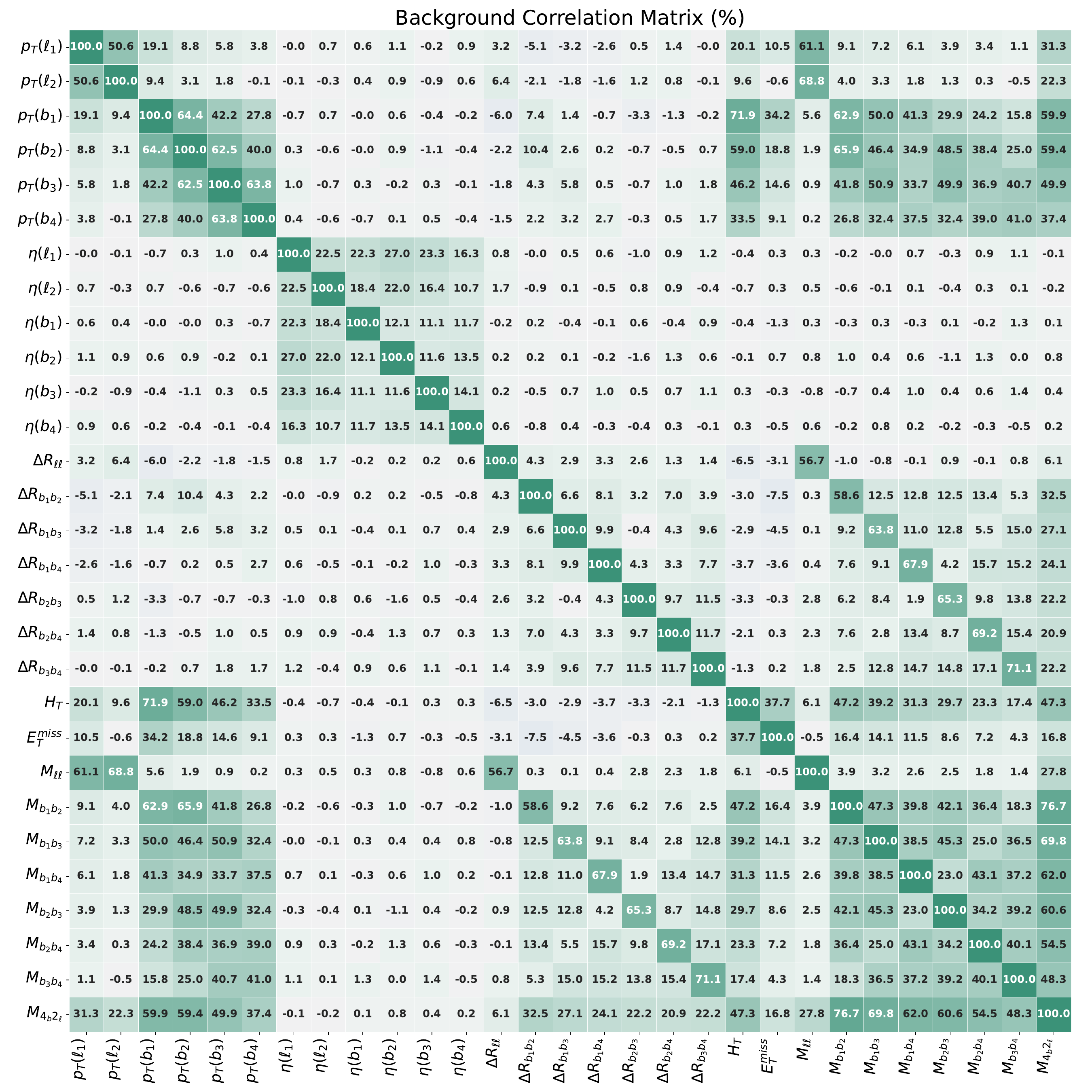}
    \caption{Correlation matrix (in $\%$) of the input kinematic variables for the signal sample(left) and background samples(right) used in the BDT training.}
\end{figure}

\end{document}